\documentclass{lmcs}
\pdfoutput=1

\usepackage{lastpage}
\lmcsdoi{16}{3}{1}
\lmcsheading{}{\pageref{LastPage}}{}{}%
{Oct.~19,~2018}{Jul.~03,~2020}{}

\usepackage[utf8]{inputenc}
\usepackage{amsthm,amsmath,amssymb,xspace,MnSymbol}
\usepackage{tikz}
\usepackage{verbatim}
\usepackage{booktabs}


\newcommand{\ie}{\emph{i.e.}}
\newcommand{\eg}{\emph{e.g.}}
\newcommand{\cf}{\emph{cf.}}

\renewcommand{\phi}{\varphi}
\newcommand{\Nat}{\ensuremath{\mathbb{N}}}
\newcommand{\PNat}{\ensuremath{\mathbb{N}_{+}}}
\newcommand{\Int}{\ensuremath{\mathbb{Z}}}
\newcommand{\Domain}{\ensuremath{\mathbb{D}}}

\newcommand{\set}[1]{\{#1\}}

\newcommand{\pref}[1]{\upharpoonright#1}
\newcommand{\restr}[1]{\upharpoonright#1}

\newcommand{\pwrset}{\mathcal{P}}
\newcommand{\nepwrset}{\mathcal{P}^{+}}
\newcommand{\closure}{\mathit{cl}}

\newcommand{\sidediego}[1]{}
\newcommand{\sidepraveen}[1]{}
\newcommand{\review}[1]{#1}

\newcommand{\pmainlogic}{\textup{LRV}}

\newcommand{\sys}{{\tt system}\xspace}
\newcommand{\Sys}{{\tt System}\xspace}
\newcommand{\env}{{\tt environment}\xspace}
\newcommand{\Env}{{\tt Environment}\xspace}

\newcommand{\tup}[1]{\ensuremath{\langle #1 \rangle}}

\newcommand{\ptup}[1]{\ensuremath{\langle #1 \rangle^{-1}}}

\newcommand{\oblieq}[3]{\ensuremath{#1 \approx \tup{#3 ?}#2}}

\newcommand{\oblineq}[3]{\ensuremath{#1 \not\approx \tup{#3 ?}#2}}
\newcommand{\oblieqp}[3]{\ensuremath{#1 \approx \ptup{#3 ?} #2}}
\newcommand{\oblineqp}[3]{\ensuremath{#1 \not\approx \ptup{#3 ?} #2}}
\newcommand{\toblieq}[2]{\ensuremath{#1 \approx \Diamond #2}}
\newcommand{\toblieqp}[2]{\ensuremath{#1 \approx \Diamond^{-1} #2}}

\newcommand{\toblineqp}[2]{\ensuremath{#1 \not\approx \Diamond^{-1} #2}}

\newcommand{\oblieqlocal}{\approx}
\newcommand{\oblineqlocal}{\not\approx}

\newcommand{\mynext}{{\sf X}}
\newcommand{\nxt}{\mynext}
\newcommand{\previous}{{\sf X}^{-1}}
\newcommand{\until}{{\sf U}}
\newcommand{\since}{{\sf S}}
\newcommand{\future}{{\sf F}}
\newcommand{\always}{{\sf G}}

\newcommand{\symbvals}{{\sf FR}}
\newcommand{\symbmodels}{\models_{\mathit{symb}}}
\newcommand{\pobli}{ {\sf PO}}
\newcommand{\inc}{\mathit{inc}}
\newcommand{\dec}{\mathit{dec}}
\newcommand{\nci}{\mathit{nc}}

\newcommand{\bvars}{\textit{BVARS}}
\newcommand{\dvars}{\textit{DVARS}}

\newcommand{\val}{\upsilon}
\newcommand{\vals}{\Upsilon}
\newcommand{\bvale}{\mathit{b\val}^{e}}
\newcommand{\dvale}{\mathit{d\val}^{e}}
\newcommand{\bvals}{\mathit{b\val}^{s}}
\newcommand{\dvals}{\mathit{d\val}^{s}}
\newcommand{\bvalse}{\mathit{B\vals}^{e}}
\newcommand{\dvalse}{\mathit{D\vals}^{e}}
\newcommand{\bvalss}{\mathit{B\vals}^{s}}
\newcommand{\dvalss}{\mathit{D\vals}^{s}}

\newcommand{\te}{\mathit{te}}
\newcommand{\ts}{\mathit{ts}}

\newcommand{\curtrs}{\ensuremath{\mathit ct_{s}}}

\newcommand{\act}[1]{\xrightarrow{#1}}
\newcommand{\init}{\mathit{init}}
\newcommand{\fin}{\mathit{fin}}

\newcommand{\op}{\mathit{op}}
\newcommand{\nop}{\mathit{nop}}
\newcommand{\se}{\mathit{se}}
\newcommand{\ssys}{\mathit{ss}}

\newcommand{\symbval}{\mathit{fr}}
\newcommand{\RH}{\mathit{RH}}
\newcommand{\rh}{\mathit{H}}
\newcommand{\po}{\mathit{O}}

\begin{document}
\title{Playing with Repetitions in Data Words\texorpdfstring{\\}{} Using Energy Games}
\author{Diego Figueira\rsuper{a}}
\address{\lsuper{a}Univ. Bordeaux, CNRS,  Bordeaux INP, LaBRI, UMR 5800, F-33400, Talence, France}
\thanks{Partially supported by ANR project BraVAS (grant ANR-17-CE40-0028), ANR project DeLTA (ANR-16-CE40-0007)} 

\author{Anirban Majumdar\rsuper{b}}
\address{\lsuper{b}Chennai Mathematical Institute, India and UMI ReLaX}
\thanks{Partially supported by a grant from the Infosys foundation}

\author{M.~Praveen\rsuper{b}}
\thanks{Partially supported by a grant from the Infosys foundation}
\begin{abstract}
    We introduce two-player games which build words over infinite
    alphabets, and we study the problem of checking the existence of winning strategies. These games are played by two players, who take turns in choosing valuations for
    variables ranging over an infinite data domain, thus generating multi-attributed \emph{data words}. The winner of the  game is specified by formulas in the Logic of Repeating Values, which
    can reason about repetitions of data values in infinite
    data words. We prove that it is undecidable to check if one of the players
    has a winning strategy, even in very restrictive settings.  However, we prove that if
    one of the players is restricted to choose valuations ranging over
    the Boolean domain, the games
    are effectively equivalent to \emph{single-sided} games on vector
    addition systems with states (in which one
    of the players can change control states but cannot change
    counter values), known to be decidable and effectively equivalent
    to energy games.

Previous works have shown that the satisfiability
    problem for various variants of the logic of repeating values is equivalent to the
    reachability and coverability problems in vector addition systems. Our results raise
    this connection to the level of games, augmenting further the
    associations between logics on data words and counter systems.
\end{abstract}

\maketitle

\section{Introduction}

Words over an unbounded domain ---known as \emph{data words}--- is a
structure that appears in many scenarios, as abstractions of timed
words, runs of counter automata, runs of concurrent programs with an
unbounded number of processes, traces of reactive systems, and more
broadly as abstractions of any record of the run of processes handling
unbounded resources. Here, we understand \review{a} data word as a (possibly infinite) word in which every position carries a vector of elements from a possibly infinite domain (\eg, a vector of numbers).

Many specification languages have been proposed
to specify properties of data words, both in terms of automata~\cite{Neven&Schwentick&Vianu04,Segoufin06} and logics~\cite{BDMSS06:journal,DL-tocl08,Kara&Schwentick&Zeume10,Figueira11}.
One of the most basic mechanisms for expressing properties on these
structures is based on whether a data value at a
given position is repeated either \emph{locally} (\eg, in the
$2$\textsuperscript{nd} component of the vector at the
$4$\textsuperscript{th} future position),
or \emph{remotely} (\eg, in the $1$\textsuperscript{st} component of a
vector at some position in the past). This
has led to the study of linear temporal logic extended with these
kind of tests, called \emph{Logic of Repeating Values} (LRV)~\cite{DDG2012}.
The satisfiability problem for LRV is inter-reducible with
the reachability problem for Vector Addition Systems with States
(VASS), and when the logic is
restricted to testing remote repetitions only in the future, it is
inter-reducible with the coverability problem for VASS~\cite{DDG2012,DFP2016}. These connections also
extend to data trees and branching VASS~\cite{AbriolaFF17}.

Previous works on data words have been centered around the
satisfiability, containment, or model checking problems. Here, we
initiate the study of two-player \emph{games} on such structures,
motivated by the realizability problem of reactive systems (hardware,
operating systems, communication protocols). A reactive system keeps
interacting with the environment in which it is functioning, and a data
word can be seen as a trace of this interaction. The values of some
variables are decided by the system and some by the environment. The
reactive system has to satisfy a specified property, given as a
logical formula over data words. The \emph{realizability} problem
asks whether it is possible that there exists a system that always
satisfies the specified property, irrespective of what the environment
does. This can be formalized as the existence of a winning strategy
for a two-player game that is defined to this end. In this game, there
are two sets of variables. Valuations for one
set of variables are decided by the system player (representing the
reactive system) and for the other set of variables, valuations are
decided by the environment player (representing the environment in
which the reactive system is functioning). The two players take turns
giving valuations to their respective variables and build an infinite
sequence of valuations. The system player wins a game if the resulting
sequence satisfies the specified logical formula. Motivated by the
\emph{realizability problem} of Church~\cite{church1962logic}, the
question of existence of winning strategies in such games are studied
extensively (starting from~\cite{pnueli1989synthesis}) for the case
where variables are Boolean and the logic used is propositional linear
temporal logic. To the best of our knowledge there have been no works
on the more general setup of infinite domains. This work can be seen
as a first step towards considering richer structures, this being the
case of an infinite set with an equivalence relation.

\subsection*{Contributions} By combining known relations  between
satisfiability of (fragments of) LRV and (control state) reachability
in VASS~\cite{DDG2012,DFP2016} with existing knowledge  about
realizability games (\cite{pnueli1989synthesis} and numerous papers
expanding on it), it is not difficult to show that realizability games
for LRV are related to games on VASS\@. Using known results about
undecidability of games on VASS, it is again not difficult to show
that realizability games for LRV are undecidable. Among others, one
way to get decidable games on VASS is to make the game asymmetric,
letting one player only change control states, while the other player
can additionally change values in counters, resulting in the so called
single-sided VASS games~\cite{AMSS2013}. Our first contribution in
this paper is to identify that the corresponding asymmetry in LRV
realizability is to give only Boolean variables to one of the players
and let the logic test only for remote repetitions in the past (and
disallow testing for remote repetitions in the future). Once this
identification of the fragment is made, the proof of its
inter-reducibility with single-sided VASS games follows more or less
along expected lines by adapting techniques developed in~\cite{DDG2012,DFP2016}.

To obtain the fragment mentioned in the previous paragraph, we impose
two restrictions; one is to restrict one of the players to Boolean
variables and the other is to dis-allow testing for remote repetitions
in the future. Our next contribution in this paper is to prove that
lifting either of these restrictions lead to undecidability. A common
feature in similar undecidability proofs (e.g., undecidability of VASS
games~\cite{ABD2008}) is a reduction from the reachability problem for
$2$-counter machines (details follow in the next section) in which one
of the players emulates the moves of the
counter machine while the other player catches the first player in
case of cheating. Our first undecidability proof uses a new technique
where the two players cooperate to emulate the moves of the counter
machine and one of the players has the additional task of detecting
cheating. Another common feature of similar undecidability proofs is
that emulating zero testing transitions of the counter machine is
difficult while emulating incrementing and decrementing transitions
are easy. Our second undecidability proof uses another new technique
in which even emulating decrementing transitions is difficult and
requires specific moves by the two players.
%

\subsection*{Related works} The relations between satisfiability of various logics over data
words and the problem of language emptiness for automata models have been explored
before. In~\cite{BDMSS06:journal}, satisfiability of the two variable
fragment of first-order logic on data words is related to reachability
in VASS\@. In~\cite{DL-tocl08}, satisfiability of LTL extended with
freeze quantifiers is related to register automata.

A general framework
for games over infinite-state systems with a well-quasi ordering is
introduced in~\cite{ABD2008} and the restriction of downward closure
is imposed to get decidability. In~\cite{RSV2005}, the two players
follow different rules, making the abilities of the
two players asymmetric and leading to decidability. A possibly
infinitely branching version of VASS is studied in~\cite{BJK2010}, where decidability is obtained in the restricted case
when the goal of the game is to reach a configuration in which one of the
counters has the value zero. Games on VASS with inhibitor
arcs are studied in~\cite{BHSS-concur12} and decidability is obtained
in the case where one of the players can only increment counters and
the other player can not test for zero value in counters. In~\cite{CRR2014}, energy games are studied, which are games on counter
systems and the goal of the game is to play for ever without any
counter going below zero in addition to satisfying parity conditions on the control
states that are visited infinitely often. Energy games are further studied in~\cite{AMSS2013}, where they are related to single-sided VASS games,
which restrict one of the players to not make any changes to the
counters. Closely related perfect half-space games are studied in~\cite{CJLS2017}, where it is shown that optimal complexity upper
bounds can be obtained for energy games by using perfect half space
games.

This paper is an extended version of a preliminary version~\cite{ConfVersion}. Results about nested formulas in sections~\ref{dec-singlesided-nestedpast} and~\ref{undec-singlesided-nestedpast} are new in this version.


\subsection*{Organization} In Section~\ref{prelims} we define the logic LRV, counter machines, and  VASS games. In Section~\ref{games} we introduce
LRV games. Section~\ref{undec-plrvgame} shows undecidability results
for the fragment of LRV with data repetition tests restricted to past.
Section~\ref{single-past-decidable} shows the decidability result of
past-looking single-sided LRV games. Section~\ref{undec-singlesided-plrvgame} shows undecidability of
future-looking single-sided LRV games, showing that in some sense the
decidability result is maximal. In
Section~\ref{dec-singlesided-nestedpast}, we show that decidability is
preserved for past looking single-sided games if we allow nested
formulas that can only use past LTL modalities. We show in
Section~\ref{undec-singlesided-nestedpast} that even past looking
single-sided games are undecidable if we allow nested formulas to use
future LTL modalities.
We conclude in Section~\ref{conclusion}.


\section{Preliminaries}%
\label{prelims}
We denote by $\Int$ the set of integers and by $\Nat$ the set of
non-negative integers. We let $\PNat$ denote the set of integers that
are strictly greater than $0$. For any set $S$, we denote by $S^{*}$ (resp.\ $S^\omega$) the set
of all finite (resp.\ countably infinite) sequences of elements in $S$. For a sequence $\sigma \in
S^{*}$, we denote its length by $|\sigma|$. We denote by
$\pwrset(S)$ (resp.~$\nepwrset(S)$) the set of all subsets
(resp.~non-empty subsets) of $S$.

\subsection*{Logic of repeating values} We recall the syntax and
semantics of the logic of repeating values from~\cite{DDG2012,
DFP2016}. This logic extends the usual propositional linear
temporal logic with the ability to reason about repetitions of data
values from an infinite domain. We let this logic use both Boolean
variables (\ie, propositions) and data variables ranging over an infinite data domain
$\Domain$. The Boolean variables can be simulated by data
variables. However, we need to consider fragments of the logic, for
which explicitly having Boolean variables is convenient. Let
$\bvars=\set{q,t,\ldots}$ be a countably infinite set of Boolean
variables ranging over $\set{\top,\bot}$, and let $\dvars=\set{x,y,\ldots}$  be a countably
infinite set of `data' variables ranging over $\Domain$. We denote by
\pmainlogic{} the logic whose formulas are defined as follows:\footnote{In a previous work~\cite{DFP2016} this logic was denoted by PLRV (LRV + Past).}%
\begin{align*}
\phi ::= & ~ q ~|~ x \oblieqlocal \mynext^{j}y ~|~ \oblieq{x}{y}{\phi} ~|~
\oblineq{x}{y}{\phi} ~|~ \oblieqp{x}{y}{\phi} \\ & ~|~ \oblineqp{x}{y}{\phi}
~|~ \review{\phi \land \psi} ~|~ \lnot \phi ~|~ \mynext \phi ~|~ \review{\phi \until
\psi} ~|~ \previous \phi \\ & ~|~ \review{\phi \since \psi} ~,~ \text{where } q \in \bvars,~
x,y \in \dvars, ~ j \in \Int
\end{align*}

A \emph{valuation} is the union of a mapping from $\bvars$ to
$\set{\top, \bot}$ and a mapping from $\dvars$ to $\Domain$. A
\emph{model} is a finite or infinite sequence of valuations. We use
$\sigma$ to denote models and $\sigma(i)$ denotes the
$i$\textsuperscript{th} valuation in $\sigma$, where $i \in \PNat$.
For any model $\sigma$ and position $i \in \PNat$,
the satisfaction relation $\models$ is defined
inductively as shown in Table~\ref{tab:semantics}.
%
\review{The semantics of temporal operators next ($\mynext$),
previous ($\previous$), until ($\until$), since ($\since$) and the
Boolean connectives are defined in the usual way, but for the sake of
completeness we provide
their formal definitions.} In Table~\ref{tab:semantics}, $q \in \bvars$, $x,y \in \dvars$.
\begin{table}
\begin{align*}
    \sigma, i \models q  & \text { iff } \sigma(i)(q)=\top \\
    \sigma, i \models x \oblieqlocal \mynext^{j}y & \text{ iff }
    1 \le i+j \le |\sigma|,~  \sigma(i)(x) = \sigma(i+j)(y)\\
    \sigma, i \models \oblieq{x}{y}{\phi} &  \text{ iff }
    \exists j > i \text{ s.t. } \sigma(i)(x) =
    \sigma(j)(y), ~\sigma, j \models \phi \\
    \sigma, i \models \oblineq{x}{y}{\phi} &  \text{ iff }
    \exists j > i \text{ s.t. } \sigma(i)(x) \ne
    \sigma(j)(y), ~\sigma, j \models \phi \\
    \sigma, i \models \oblieqp{x}{y}{\phi} &  \text{ iff }
    \exists j < i \text{ s.t. } \sigma(i)(x) =
    \sigma(j)(y), ~\sigma, j \models \phi \\
    \sigma, i \models \oblineqp{x}{y}{\phi} &  \text{ iff }
    \exists j < i \text{ s.t. } \sigma(i)(x) \ne
    \sigma(j)(y), ~\sigma, j \models \phi \\
   \sigma, i \models \mynext \phi & \text{ iff }
    \sigma, {i+1} \models \phi \\
    \sigma, i \models \phi \until \psi & \text{ iff }
    \exists j \ge i \text{ s.t. } \sigma, j \models \psi
    \text{ and }
    ~\forall i \le k < j, \ \sigma, k \models \phi \\
    \sigma, i \models \previous \phi & \text{ iff }
    i > 0 \text{ and } ~\sigma, {i-1} \models \phi \\
    \sigma, i \models \phi \since \psi & \text{ iff }
    \exists j \le i \text{ s.t. } \sigma, j \models \psi
    \text{ and }
    ~\forall j < k \le i, \ \sigma, k \models \phi \\
    \sigma, i \models \phi \land \psi & \text{ iff }
    \sigma, i \models \phi \text{ and } ~\sigma,i \models \psi \\
    \sigma, i \models \lnot \phi & \text{ iff }
    \sigma, i \not\models \phi
\end{align*}
\caption{Semantics of \pmainlogic{}.}%
\label{tab:semantics}
\end{table}
Intuitively, the formula $x \oblieqlocal \mynext^{j} y$ tests that the
data value mapped to the variable $x$ at the current position repeats
in the variable $y$ after $j$ positions.
 We use the notation
$\mynext^{i}x \oblieqlocal \mynext^{j}y$ as an abbreviation for the
formula $\mynext^{i} (x \oblieqlocal \mynext^{j-i}y)$ (assuming
without any loss of generality that $i \le j$). The formula
$\oblieq{x}{y}{\phi}$ tests that the data value mapped to $x$ now
repeats in $y$ at a future position that satisfies the nested formula
$\phi$. The formula $\oblineq{x}{y}{\phi}$ is similar but tests for
disequality of data values instead of equality. If a model is being
built sequentially step by step and these formulas are to be satisfied
at a position, they create obligations (for repeating some data
values) to be satisfied in some future step. The formulas
$\oblieqp{x}{y}{\phi}$ and $\oblineqp{x}{y}{\phi}$ are similar but
test for repetitions of data values in past positions.

\review{We consider fragments of LRV in which only past LTL modalities are
allowed. Formally, the grammar is:
\begin{align}
\phi ::= q ~|~ x \oblieqlocal \mynext^{-j}y ~|~
\oblieqp{x}{y}{\phi}  ~|~ \oblineqp{x}{y}{\phi}
~|~ \phi \land \psi ~|~ \lnot \phi ~|~
\previous \phi ~|~ \phi \since \psi ~,~\nonumber\\ \text{where } q \in \bvars,~
x,y \in \dvars, ~ j \in \Nat
    \label{eq:PastLTLodalities}
\end{align}}

We append symbols to \pmainlogic{} for denoting syntactic restrictions
as shown in the following table. For example, \pmainlogic$[\top,
\oblieqlocal, \leftarrow]$ denotes the fragment of \pmainlogic{} in
which nested formulas,
disequality constraints and future obligations
are not allowed. For clarity, we replace $\tup{\top ?}$ with
$\Diamond$ in formulas. E.g., we write $\oblieq{x}{y}{\top}$ as simply
$\toblieq{x}{y}$.
\begin{center}
\begin{tabular}[top]{p{0.09\textwidth}p{0.7\textwidth}}
    \toprule
    Symbol & Meaning\\
    \midrule
    $\top$ & $\phi$ has to be $\top$ in $\oblieq{x}{y}{\phi}$
    (no nested formulas)\\
    $\oblieqlocal$ & disequality constraints ($\oblineq{x}{y}{\phi}$ or
$\oblineqp{x}{y}{\phi}$) are not allowed\\
$\rightarrow$ & past obligations ($\oblieqp{x}{y}{\phi}$ or
$\oblineqp{x}{y}{\phi}$) are not allowed\\
$\leftarrow$ & future obligations ($\oblieq{x}{y}{\phi}$ or
$\oblineq{x}{y}{\phi}$) are not allowed\\
$\langle \previous, \since \rangle$ & $\mynext, \until$ (and
operators derived from them) not allowed in nested formulas \review{(grammar
    in~\eqref{eq:PastLTLodalities})}\\
$\langle \future \rangle$ & $\future$ operator allowed in nested
formulas\\
    \bottomrule
\end{tabular}\\
\end{center}

\subsection*{Parity games on integer vectors} We recall the definition
of games on Vector Addition Systems with States (VASS) from~\cite{AMSS2013}. The game is played between two players: \sys and \env.
A VASS game is a tuple $(Q, C, T, \pi)$ where $Q$ is a
finite set of states, $C$ is a finite set of counters, $T$ is a finite
set of transitions and $\pi: Q \to \set{1, \ldots, p}$, for some integer $p$, is a colouring
function that assigns a number to each state. The set $Q$ is
partitioned into two parts $Q^{e}$ (states of \env)
and $Q^{s}$ (states of \sys). A transition in $T$ is a tuple $(q, \op, q')$ where $q, q'
\in Q$ are the origin and target states and $\op$ is an operation
of the form $x++$, $x--$ or $\nop$, where $x \in C$ is a counter.
We say that a transition of a VASS game belongs to \env if
its origin belongs to \env; similarly for \sys. A VASS game is \emph{single-sided} if every \env
transition is of the form $(q, \nop, q')$. It is assumed that every
state has at least one outgoing transition.

A configuration of the VASS game is an element $(q, \vec{n})$ of $Q
\times \Nat^{C}$, consisting of a state $q$ and a valuation $\vec{n}$
for the counters. A play of the VASS game begins at a designated
initial configuration. The player owning the state of the current
configuration (say $(q, \vec{n})$) chooses an outgoing transition
(say $(q, \op, q')$) and changes the configuration to $(q',\vec{n'})$,
where $\vec{n'}$ is obtained from $\vec{n}$ by incrementing
(resp.~decrementing) the counter $x$ once, if $\op$ is $x++$
(resp.~$x--$). If $\op=\nop$, then $\vec{n'}=\vec{n}$.
We denote this update as $(q,\vec{n}) \act{(q, \op, q')} (q',
\vec{n'})$. The play is then continued similarly by the owner of the
state of the next configuration. If any player wants to take a
transition that decrements some counter, that counter should have a
non-zero value before the transition. Note that in a single-sided VASS game, \env cannot change the value of the counters. The game
continues forever and results in an infinite sequence of
configurations $(q_{0}, \vec{n_{0}}) (q_{1}, \vec{n_{1}}) \cdots$. \Sys wins the game if the maximum colour occurring infinitely
often in $\pi(q_0) \pi(q_1) \pi(q_2) \cdots$ is even. We assume without loss of
generality that from any configuration, at least one transition is
enabled (if this condition is not met, we can add extra states and
transitions to create an infinite loop ensuring that the owner of the
deadlocked configuration loses). In our constructions, we use a
generalized form of transitions $q \act{\vec{u}} q'$ where
$\vec{u} \in \Int^{C}$, to indicate that each counter $c$ should be
updated by adding $\vec{u}(c)$. Such VASS games can be effectively
translated into ones of the form defined in the previous paragraph,
preserving winning regions.

A strategy $\se$ for \env in a VASS game is a
mapping $\se: {(Q \times \Nat^{C})}^{*} \cdot (Q^{e} \times
\Nat^{C}) \to T$ such that for all $\gamma \in {(Q \times
\Nat^{C})}^{*}$, all $q^{e} \in Q^{e}$ and all $\vec{n} \in
\Nat^{C}$, $\se(\gamma\cdot(q^{e},\vec{n}))$ is a transition whose
source state is $q^{e}$. A strategy $\ssys$ for \sys is a
mapping $\ssys: {(Q \times \Nat^{C})}^{*} \cdot (Q^{s} \times
\Nat^{C}) \to T$ satisfying similar conditions. \Env
plays a game according to a strategy $\se$ if the resulting sequence
of configurations $(q_{0}, \vec{n_{0}}) (q_{1}, \vec{n_{1}}) \cdots$
is such that for all $i \in \Nat$, $q_{i} \in Q^{e}$ implies
$(q_{i}, \vec n_{i}) \act{\se( (q_{0}, \vec n_{0}) (q_{1},
\vec n_{1}) \cdots (q_{i}, \vec n_{i}))} (q_{i+1}, \vec n_{i+1})$.
The notion is extended to \sys player similarly. A strategy
$\ssys$ for \sys is winning if \sys wins all
the games that she plays according to $\ssys$, irrespective of the
strategy used by \env. It was shown in~\cite{AMSS2013} that it is decidable to check whether \sys has a winning strategy in a given single-sided VASS game and an
initial configuration. An optimal double exponential upper bound was
shown for this problem in~\cite{CJLS2017}.

\subsection*{Counter machines}
\review{An $n$-counter machine is a tuple $(Q,q_{\init},n,\delta)$ where
$Q$ is a finite set of states,
$q_{\init} \in Q$ is an initial state, $c_{1}, \ldots,
c_{n}$ are $n$ counters and $\delta$ is a finite set of instructions of the form `$(q:
c_{i}:=c_{i}+1; \mathrm{goto}~q')$' or `$(q:\mathrm{If}~c_{i}=0~
\mathrm{then}~
\mathrm{goto}~q'~\mathrm{else}~c_{i}:=c_{i}-1;~\mathrm{goto}~q'')$'
where $i \in [1,n]$ and $q,q',q'' \in Q$.} A configuration of the
machine is described by a tuple $(q,m_{1}, \ldots, m_{n})$ where
$q \in Q$ and $m_{i} \in \Nat$ is the content of the counter
$c_{i}$. The possible computation steps are defined as follows:
\begin{enumerate}
    \item $(q,m_{1}, \ldots, m_{n}) \rightarrow (q',m_{1}, \ldots,
        m_{i}+1, \ldots, m_{n})$ if there is an instruction $(q:
c_{i}:=c_{i}+1; \mathrm{goto}~q')$. This is called an incrementing
transition.
    \item $(q,m_{1}, \ldots, m_{n}) \rightarrow (q',m_{1}, \ldots,
        m_{n})$ if there is an instruction $(q:\mathrm{If}~c_{i}=0~
\mathrm{then}~
\mathrm{goto}~q'~\mathrm{else}\\c_{i}:=c_{i}-1;~\mathrm{goto}~q'')$ and
$m_{i}=0$. This is called a zero testing transition.
    \item $(q,m_{1}, \ldots, m_{n}) \rightarrow (q'',m_{1}, \ldots,
        m_{i}-1, \ldots, m_{n})$ if there is an instruction $(q:\mathrm{If}~c_{i}=0~
\mathrm{then}~
\mathrm{goto}~q'~\mathrm{else}~c_{i}:=c_{i}-1;~\mathrm{goto}~q'')$ and
$m_{i}>0$. This is called a decrementing transition.
\end{enumerate}
\review{A counter machine is \emph{deterministic} if for every state
$q$, there is at most one instruction of the form $(q:
c_{i}:=c_{i}+1; \mathrm{goto}~q')$ or $(q:\mathrm{If}~c_{i}=0~
\mathrm{then}~
\mathrm{goto}~q'~\mathrm{else}~c_{i}:=c_{i}-1;~\mathrm{goto}~q'')$
where $i \in [1,n]$ and $q',q'' \in Q$. This ensures that for every
configuration
$(q,m_{1}, \ldots, m_{n})$ there exists at most one configuration
$(q',m_{1}', \ldots, m_{n}')$ so that $(q,m_{1}, \ldots, m_{n}) \act{}
(q',m_{1}', \ldots, m_{n}')$}. For our undecidability results we will
use deterministic
2-counter machines \review{(\ie, $n=2$)}, henceforward just ``counter machines''.
Given a counter machine $(Q,q_0,2,\delta)$ and two of its states
$q_{\init}, q_{\fin} \in Q$, the reachability problem is to determine
if there is a sequence of transitions of the $2$-counter machine starting
from the configuration $(q_{\init}, 0, 0)$ and ending at the
configuration $(q_{\fin}, n_{1}, n_{2})$ for some $n_{1}, n_{2} \in
\Nat$. It is known that the reachability problem for \review{deterministic} $2$-counter
machines is undecidable~\cite{Minsky1961}. To simplify our
undecidability results we further assume, without any loss of
generality, that there exists an instruction $\hat t = (q_{\fin}:
c_{1}:=c_{1}+1; \mathrm{goto}~q_{\fin}) \in \delta$.


\section{Game of repeating values}%
\label{games}
The game of repeating values is played between two players, called \env and
\sys. The set $\bvars$ is partitioned as $\bvars^{e},
\bvars^{s}$, owned by \env and \sys respectively.
The set $\dvars$ is partitioned similarly. Let $\bvalse$
(resp.~$\dvalse$, $\bvalss$, $\dvalss$) be the set of all mappings
$\bvale:\bvars^{e} \to \set{\top, \bot}$ (resp.,~$\dvale:\dvars^{e}
\to \Domain$, $\bvals: \bvars^{s} \to \set{\top, \bot}$, $\dvals:
\dvars^{s} \to \Domain$).  Given two mappings $\val_{1}: V_{1} \to
\Domain \cup \set{\top, \bot}$, $\val_{2}: V_{2} \to \Domain \cup
\set{\top, \bot}$ for disjoint sets of variables $V_{1}, V_{2}$, we
denote by $\val = \val_{1} \oplus v_{2}$ the mapping defined as
$\val(x_{1}) = \val_{1}(x_{1})$ for all $x_{1} \in V_{1}$ and
$\val(x_{2}) = \val_{2}(x_{2})$ for all $x_{2} \in V_{2}$. Let
$\vals^{e}$ (resp.,~$\vals^{s}$) be the set of mappings $\set{\bvale
\oplus \dvale \mid \bvale \in \bvalse, \dvale \in \dvalse}$
(resp.~$\set{\bvals \oplus \dvals \mid \bvals \in \bvalss, \dvals \in
\dvalss}$). The first round of a game of repeating values is begun by
\env choosing a mapping $\val^{e}_{1} \in
\vals^{e}$, to which \sys responds by choosing a mapping
$\val^{s}_{1} \in \vals^{s}$. Then \env continues
with the next round by choosing a mapping from $\vals^{e}$ and so on.
The game continues forever and results in an infinite model $\sigma =
(\val^{e}_{1} \oplus \val^{s}_{1}) (\val^{e}_{2} \oplus \val^{s}_{2})
\cdots$. The winning condition is given by a \pmainlogic{} formula
$\phi$ --- \sys wins iff $\sigma, 1 \models \phi$.

Let $\vals$ be the set of all valuations. For any model $\sigma$ and  $i>0$,  let $\sigma\pref{i}$ denote
the valuation sequence $\sigma(1) \cdots \sigma(i)$, and $\sigma \pref{0}$ denote the empty sequence. A strategy for
\env is a mapping $\te: \vals^{*} \to \vals^{e}$. A
strategy for \sys is a mapping $\ts: \vals^{*}\cdot
\vals^{e} \to \vals^{s}$. We say that \env plays
according to a strategy $\te$ if the resulting model $(\val^{e}_{1}
\oplus \val^{s}_{1}) (\val^{e}_{2} \oplus \val^{s}_{2}) \cdots$ is
such that $\val^{e}_{i} = \te(\sigma\pref{(i-1)})$ for all positions
$i \in \PNat$. \Sys plays according to a
strategy $\ts$ if the resulting model is such that
$\val^{s}_{i} = \ts(\sigma\pref{(i-1)} \cdot \val^{e}_{i})$ for all
positions $i \in \PNat$. A strategy $\ts$ for
\sys is winning if \sys wins all games
that she plays according to $\ts$, irrespective of the strategy used
by \env. Given a formula $\phi$ in (some fragment
of) \pmainlogic{}, we are interested in the decidability of checking
whether \sys has a winning strategy in the game of
repeating values whose winning condition is $\phi$.

We illustrate the utility of this game with an example. Consider a
scenario in which the system is trying to schedule tasks on
processors. The number of tasks can be unbounded and task identifiers
can be data values. Assume that a system variable $\textsf{init}$ carries
identifiers of tasks that are initialized. \review{If a task is initialized at
a certain moment of time, then the variable $\textsf{init}$ carries
the identifier of that task at that moment; at moments when no tasks
are initialized, $\textsf{init}$ is blank. We assume that at most one
task can be initialized at a time, so $\textsf{init}$ is either blank
or carries one task identifier.} Additionally, another
system variable $\textsf{proc}$ carries
identifiers of tasks that are processed. \review{If a task is
processed at a certain moment of time, then the variable
$\textsf{proc}$ carries the identifier of that task at that
moment. We assume for simplicity that processing a task takes only one
unit of time and at most one task can be processed in one unit of time.} The formula
$G ~ (\toblieqp{\textsf{proc}}{\textsf{init}})$ specifies that all tasks
that are processed must have been initialized beforehand. Assume the
system
variable $\textsf{log}$ carries identifiers of tasks that have been 
processed and are being logged into an audit table. \review{If a task is
logged at a certain moment of time, then the variable
$\textsf{log}$ carries the identifier of that task at that 
moment.} The formula
$G~(\textsf{proc} \oblieqlocal \mynext~\textsf{log})$ specifies that 
all processed tasks are logged into the audit table in the next step.
Suppose there is a Boolean variable $\textsf{lf}$ belonging to the
environment. The formula $G~( \lnot \textsf{lf} ~ \Rightarrow ~ \lnot
(\textsf{log} \oblieqlocal \mynext^{-1} ~ \textsf{proc}))$ specifies 
that if $\textsf{lf}$ is false (denoting that the logger is not
working), then the logger can not put the task that was processed in
the previous step into the audit table in this step. The combination
of the last two specifications is not realizable by any system since
as soon as the system processes a task, the environment can make the
logger non-functional in the next step. This can be algorithmically
determined by the fact that for the conjunction of the last two
formulas, there is no winning strategy for \sys in the game of
repeating values.


\section{Undecidability of \texorpdfstring{\pmainlogic[$\top$, $\oblieqlocal$, $\leftarrow$]}{LRV[⟙,≈,←]} games}%
\label{undec-plrvgame}
Here we establish that determining if \sys has a winning strategy in
the \pmainlogic[$\top, \oblieqlocal, \leftarrow]$ game is undecidable.
This uses a fragment of \pmainlogic{} in which there are no future
demands, no disequality 
demands $\oblineqlocal$, and every sub-formula
$\oblieqp{x}{y}{\phi}$ is such that $\phi = \top$. 
\review{Further, this
undecidability result holds even for the case where the \pmainlogic{} formula
contains only one data variable of \env and one of \sys
and moreover}, the distance of local
demands is bounded by $3$, that is, all local demands of the form $x
\oblieqlocal X^i y$ are so that $-3 \le i \le 3$. Simply put, the
result shows that bounding the distance of local demands and the number of data variables does not help in obtaining decidability.
\newcommand{\thmUndecPast}{%
The winning strategy existence problem for the \pmainlogic$[\top, \oblieqlocal, \leftarrow]$ game is undecidable, \review{even when the \pmainlogic{} formula
contains one data variable of \env and one of \sys}, and the distance of local demands is bounded by $3$.}
\begin{thm}\label{thm:UndecPast}
\thmUndecPast
\end{thm}
As we shall see in the next section, if we further restrict the game so that 
\review{the \pmainlogic{} formula does not
contain any \env data variable}, we obtain decidability.

Undecidability is shown by reduction from the reachability problem for counter machines. The reduction will be first shown for the case 
\review{where the formula
consists of a \sys data variable $y$, an \env data variable $x$ and some Boolean variables of \env, }
 encoding \review{\emph{instructions} of a 2-counter machine}. In a second part we show how to eliminate these Boolean variables.

\subsection{Reduction with Boolean variables}
\newcommand{\countx}{c_x}%
\newcommand{\county}{c_y}%
\begin{lem}\label{lem:undec-plrvgame-withbool}
The winning strategy existence problem for the \pmainlogic$[\top,
\oblieqlocal, \leftarrow]$ game is undecidable when
\review{the formula
consists of a \sys data variable, an \env data variable and unboundedly many Boolean variables of \env.}
\end{lem}
\begin{proof}
    We first give a short description of the ideas used. For
    convenience, we name the counters of the 2-counter machines
    $\countx$ and $\county$ instead of $c_{1}$ and $c_{2}$. To
    simulate counters $\countx$ and $\county$, we use \env's
    variable $x$ and \sys's variable $y$. There are a few more Boolean
    variables that \env uses for the simulation. We define a
    \pmainlogic[$\top$, $\oblieqlocal$, $\leftarrow$] formula to force
    \env and \sys to simulate runs of 2-counter machines as follows.
    Suppose $\sigma$ is the concrete model built during a game. The
    value of counter $\countx$ (resp.~$\county$) before the
    $i$\textsuperscript{th} transition is the cardinality of the set
    $\set{d \in \Domain \mid \exists j \in \set{1, \ldots, i}:
    \sigma(j)(x)=d, \forall j' \in \set{1, \ldots, i}: \sigma(j')(y)
\ne d}$ (resp.~$\set{d \in \Domain \mid \exists j \in \set{1, \ldots,
i}, \sigma(j)(y)=d, \forall j' \in \set{1, \ldots, i}, \sigma(j')(x)
\ne d}$). Intuitively, the value of counter $\countx$ is the number of
data values that have appeared under variable $x$ but not under $y$.
In each round, \env chooses the transition of the 2-counter machine to
be simulated and sets values for its variables accordingly. If
everything is in order, \sys cooperates and sets the value of the
variable $y$ to complete the simulation.  Otherwise, \sys can win
immediately by setting the value of $y$ to a value that certifies that
the actions of \env violate the semantics of the 2-counter machine. If
any player deviates from this behavior at any step, the other player
wins immediately. The only other way \sys can win is by reaching the
halting state and the only other way \env can win is by properly
simulating the 2-counter machine for ever and never reaching the
halting state.

Now we give the details. To increment $\countx$, a fresh new
data value is assigned to $x$ and the data value assigned to
$y$ should be one that has already appeared before \review{under
$y$.} To decrement $\countx$, the same data value should be assigned to
$x$ and $y$ and it should have appeared before under $x$ but not under
$y$. In order to test that $\countx$ has the value zero, the same data
value should be assigned to $x$ and $y$. In addition, every increment
for $\countx$ should have been matched by a subsequent decrement for
$\countx$. The operations for $\county$ should follow similar rules,
with $x$ and $y$ interchanged.

For every \review{instruction} $t$ of the 2-counter machine, there is a Boolean
variable $p_t$ owned by \env. The $i$\textsuperscript{th} \review{instruction}
chosen by \env is in the $(i+1)$\textsuperscript{st} valuation.

\review{We will build the winning condition formula $\phi$ from two sets of formulas $\Phi^e$ and $\Phi^s$ as $\phi = \bigvee \Phi^{e} \lor \bigwedge \phi^{s}$.  Hence, if any of the formulas from $\Phi^e$ is true, then  $\phi$ is true and \sys wins the game.}
The set $\Phi^e$ consists of the following formulas, each of which denotes
a mistake made by \env.
\begin{itemize}
\item \env chooses some \review{instruction} in the first position.
\begin{align*}
\bigvee_{t \text{ is any \review{instruction}}} p_t
\end{align*}
\item The first \review{instruction} is not an initial \review{instruction}.
\begin{align*}
\mynext (\bigvee_{t \text{ is not an initial \review{instruction}}} p_t)
\end{align*}
\item \env chooses more or less than one \review{instruction}.
\begin{align*}
  \mynext F( \bigvee_{t \ne t'} (p_t \land p_{t'}) \lor \bigwedge_{t\text{ is any
\review{instruction}}}(\lnot p_t))
\end{align*}
\item Consecutive \review{instructions} are not compatible.
\begin{align*}
\mynext F( \bigvee_{t' \text{ cannot come after } t} (p_t \land \mynext
p_{t'}))
\end{align*}
\item An \review{instruction} increments \review{$c_x$} but the data value for $x$ is old.
\begin{align*}
\mynext F (\bigvee_{t \text{ increments } c_x} p_t \land (\toblieqp{x}{x}
\lor \toblieqp{x}{y}))
\end{align*}
\item An \review{instruction} decrements \review{$c_x$} but the data value for $x$ has not
appeared under $x$ or it has appeared under $y$.
\begin{align*}
\mynext F (\bigvee_{t \text{ decrements } c_x} p_t \land (\lnot
\toblieqp{x}{x} \lor \toblieqp{x}{y}) )
\end{align*}
\item An \review{instruction} increments $c_y$ but the data value for $x$ is
new.
\begin{align*}
\mynext F (\bigvee_{t \text{ increments } c_y} p_t \land
(\lnot \toblieqp{x}{x} \lor \lnot \toblieqp{x}{y}))
\end{align*}
\item An \review{instruction} decrements $c_y$ but the data value for $x$ has not
    appeared before in $y$ or it has appeared before in $x$.
\begin{align*}
\mynext F (\bigwedge_{t \text{ decrements } c_y} p_t \land (\lnot
(\toblieqp{x}{y}) \lor \toblieqp{x}{x}))
\end{align*}
\item An \review{instruction} tests that the value in the counter $c_x$ is zero,
but there is a data value that has appeared under $x$ but not under
$y$. In such a case, \sys can map that value to $y$, make the
following formula true and win immediately.
\begin{align*}
\mynext F ( \bigvee_{t \text{ tests } c_x=0} p_t \land
\toblieqp{y}{x} \land \lnot \toblieqp{y}{y})
\end{align*}
\item An \review{instruction} tests that the value in the counter $c_y$ is zero,
but there is a data value that has appeared under $y$ but not under
$x$. In such a case, \sys can map that value to $y$, make the
following formula true and win immediately.
\begin{align*}
\mynext F ( \bigvee_{t \text{ tests } c_y=0} p_t \land
\toblieqp{y}{y} \land \lnot \toblieqp{y}{x})
\end{align*}
\end{itemize}

\noindent
The set $\Phi^s$ consists of the following formulas, each of which
denotes constraints that \sys has to satisfy after \env makes a move. \review{Remember that, assuming \env has done none of the  mistakes above, if any of the formulas below is false, then the final formula $\phi$ is false, and \env wins the game.}
\begin{itemize}
\item The first position contains the same data value under $x$ and $y$.
\begin{align*}
	x \oblieqlocal y
\end{align*}
This will ensure that the initial value of the counters is $0$.
\color{black}
\item If an \review{instruction} increments $c_x$, then the data value of $y$ must
already have appeared in the past \review{under $y$}.
\begin{align*}
\mynext G ( \bigwedge_{t \text{ increments } c_x} ( p_t \Rightarrow  \toblieqp{y}{y}))
\end{align*}\color{black}
\item If an \review{instruction} increments $c_y$, then the data value of $y$ must
be a fresh one.
\begin{align*}
\mynext G( \bigwedge_{t \text { increments } c_y} p_t \Rightarrow \lnot
(y \oblieqlocal x) \land \lnot (\toblieqp{y}{x}) \land \lnot
(\toblieqp{y}{y}))
\end{align*}
\item If an \review{instruction} decrements $c_x$ or $c_y$ or tests one of them
    for zero, then the data value of $y$ must
be equal to that of $x$.
\begin{align*}
\mynext G( \bigwedge_{t \text { decrements/zero tests some counter}} p_t \Rightarrow y
\oblieqlocal x)
\end{align*}
\item The halting state is reached.
    \begin{align*}
        \mynext F \bigvee_{t \text{ is an \review{instruction} whose target state
        is halting}} p_t
    \end{align*}
\end{itemize}
The winning condition of the \pmainlogic[$\top$, $\oblieqlocal$,
$\leftarrow$] game is given by the formula $\phi = \bigvee \Phi^{e} \lor
\bigwedge \phi^{s}$. For \sys to win, one of the formulas in
$\Phi^e$ must be true or all the formulas in $\Phi^s$ must be true.
Hence, for \sys to win, \env should make a mistake during simulation
or no one makes any mistake and the halting state is reached. Hence,
\sys has a winning strategy iff the 2-counter machine reaches the
halting state.
\end{proof}

\subsection{Getting rid of Boolean variables}%
\label{sec:nobooleanvars}

The reduction above makes use of some Boolean variables to encode
\review{instructions} of the 2-counter machine. However, one can modify the
reduction above to do the encoding  inside equivalence classes of the variable $x$. Suppose there are $m-1$ labels that we want to encode. A data word prefix of the form
\[
\arraycolsep=5pt\def\arraystretch{.8}
   \begin{array}{rllcl}
     \textit{label}:&l_1&l_2&&l_n\\
     x:&x_1&x_2&\cdots&x_n\\
     y:&y_1&y_2&&y_n
   \end{array}
\]
where $l_i$, $x_i$, $y_i$ are, respectively, the label, value of $x$, and value of $y$ at position $i$, is now encoded as
\[\arraycolsep=3pt\def\arraystretch{0}
  \begin{array}{rlllllllllllllllll}
    x:&d&d&&x_1&d&d&&x_2&d&d&
&d&d&&x_n&d&d\\
    &&&w_1&&&&w_2&&&&
\cdots&&&w_n&&&\\
y:&d&d&&y_1&d&d&&y_2&d&d&&d&d&&y_n&d&d
  \end{array}\tag{$\dag$}
\]
where each $w_i$ is a data word of the form $(d_1,d_1) \dotsb
(d_m,d_m)$; further the data values of $w_i$ are so that $d \not\in
\set{d_1, \dotsc, d_m}$, and so that every pair of $w_i, w_j$ with $i
\neq j$ has disjoint sets of data values. The purpose of $w_i$ is to
encode the label $l_i$; the purpose of the repeated data value $(d,d)$
is to delimit the boundaries of each encoding of a label, which we will call a `block'; the purpose of repeating $(d,d)$ at each occurrence is to avoid confusing this position with the encoding position $(x_i,y_i)$ ---\ie, a boundary position is one whose data value is repeated at distance m+3 \emph{and} at distance 1.

This encoding can be enforced using a \pmainlogic{} formula. Further, the encoding of values
of counters in the reduction before is not broken since the additional positions have the property of having the same data value under $x$ as under $y$, and in this the encoding of counter $\countx$ ---\ie, the number of data values that have appeared under $x$ but not under $y$--- is not modified; similarly for counter $\county$.


\begin{lem}%
\label{lem:undec-plrvgame-nobool}
The winning strategy existence problem for the \pmainlogic$[\top, \oblieqlocal, \leftarrow]$ game is undecidable
\review{when the formula
consists of a \sys data variable and an \env data variable.}
\end{lem}
\begin{proof}
Indeed, note that assuming the above encoding, we can make sure that we are standing at the left boundary of a block using the \pmainlogic{} formula
\[
\phi_{\textit{block}(0)} = (x \oblieqlocal \mynext^{m+3}x) \land x \oblieqlocal \mynext x;
\]
and we can then test that we are in position $i\in \set{1, \dotsc, m+2}$ of a block through the formula
\[
\phi_{\textit{block}(i)}=\mynext^{-i} \phi_{\textit{block}(0)}.
\]

For any fixed linear order on the set of labels $\lambda_1 < \dotsb < \lambda_{m-1}$, we will encode that the current block has the $i$-th label $\lambda_i$ as
\[
\phi_{\lambda_i} = \phi_{\textit{block}(0)} \land \mynext^2(x \oblieqlocal \mynext^i x).
\]
Notice that in this coding of labels, a block could have several labels, but of course this is not a problem, if need be one can ensure that exactly one label holds at each block.
\[
\phi_{\textit{1-label}} = \bigvee_i \phi_{\lambda_i} \land \lnot \bigwedge_{i \neq j} \phi_{\lambda_i} \land \phi_{\lambda_j}
\]
Now the question is: How do we enforce this shape of data words?

Firstly, the structure of a block on variable $x$ can be enforced through the following formula
\begin{align*}
\phi_{\textit{block-str}} =~ &\mynext^2(\lnot (\toblieqp{x}{x})) ~\land\\
&\bigwedge_{1< i \leq m+1} \mynext^i ( (x \oblieqlocal \mynext^{1-i}x) \lor \lnot (\toblieqp{x}{x}))  ~\land\\
& \phi_{\textit{1-label}} \land  \mynext^{m+1}( x \oblieqlocal \mynext x).
\end{align*}
The first two lines ensure that the data values of each $w_i$ are `fresh' (\ie, they have not appeared before the current block); while the last line ensures that the two last positions repeat the data value and that each blocks encodes exactly one label.
Further, a formula can inductively enforce that this structure is repeated on variable $x$:
\begin{enumerate}[align=left]
\item[(8)] The first position verifies $\phi_{\textit{block}(0)}$; and for every position we have $\phi_{\textit{block}(0)} \Rightarrow \phi_{\textit{block-str}} \land \mynext^{m+2}\phi_{\textit{block}(0)}$
\end{enumerate}

And secondly, we can make sure that the $y$ variable must have the same data value as the $x$ variable in all positions ---except, of course, the $(m+3)$-rd positions of blocks. This can be enforced by making false the formula as soon as the following property holds.
\begin{enumerate}[align=left]
\item[(F)] There is some $i \in \set{0,\dotsc, m+1}$ and some position verifying
  \[
    \phi_{\textit{block}(i)} \land \lnot (x \oblieqlocal y).
  \]
\end{enumerate}

\noindent
In each of the formulas $\phi$ described in the previous section, consider now guarding all positions with $\phi_{\textit{block}(m+2)}$\footnote{That is, where it said ``there exists a position where $\psi$ holds'', now it should say ``there exists a position where $\phi_{\textit{block}(m+2)}\land \psi$ holds'', where it said ``for every position $\psi$ holds'' it should now say ``for every position $\phi_{\textit{block}(m+2)} \Rightarrow \psi$ holds''.}; replacing  each test for a label $\lambda_i$ with $\mynext^{-(m+2)}\phi_{\lambda_i}$; and replacing each $\mynext^i$ with $\mynext^{(m+3)i}$, obtaining a new formula $\hat\phi$ that works over the block structure encoding we have just described.

Then, for the resulting formula $\bigvee \hat\Phi^{e} ~ \lor ~
\bigwedge \hat\Phi^{s}$
there is a winning strategy for \sys if, and only if, there is an accepting run of the 2-counter machine.
\end{proof}

Observe that, in the reduction above, through a binary encoding one could encode the labels in blocks of logarithmic length, and it is therefore easy produce a formula whose $\mynext$-distance is logarithmic in the size of the 2-counter machine. However, the $\mynext$-distance would remain unbounded. One obvious question would then be: is the problem decidable when the $\mynext$-distance is bounded? Unfortunately, in the next section we will see that in fact even when the $\mynext$-distance is bounded by $3$ the problem is still undecidable.
\color{black}

\subsection{Unbounded local tests}
\color{black}
The previous undecidability
\review{results} use either an unbounded number of variables or a bounded
number of variables but an unbounded $\mynext$-distance of local
demands.
However, through a more clever encoding one can
avoid testing whether two  positions at distance $n$ have the same
data value by a chained series of tests. This is a standard coding
which does not break the 2-counter machine reduction. This proves the theorem:

\smallskip

\noindent
\review{\textbf{Theorem~\ref{thm:UndecPast}.} \thmUndecPast}
\begin{proof}
\newcommand{\gr}[1]{\color{teal}\textit{\textbf{d}}\color{black}}
Remember that in the reduction of Lemma~\ref{lem:undec-plrvgame-nobool}, we enforce the encoding $(\dag)$ of the shape
\[\arraycolsep=3pt\def\arraystretch{0}
  \begin{array}{rlllllllllllllllll}
    x:&\gr d&\gr d&&x_1&\gr d&\gr d&&x_2&\gr d&\gr d&
&&&&&&\\
    &&&w_1&&&&w_2&&&&
\cdots&&&&&&\\
y:&\gr d&\gr d&&y_1&\gr d&\gr d&&y_2&\gr d&\gr d&&&&&&&
  \end{array}\tag{$\dag$}
\]
where each $w_i$ has length $m$, where $m$ is the number of
\review{instructions} of the machine (plus one), and hence unbounded. We can, instead, enforce a slightly more involved encoding of the shape
\[\arraycolsep=1.5pt\def\arraystretch{0}
  \begin{array}{rllllllllllll}
    x:&\gr d&\gr d&&\gr d&&\gr d&&\gr d&&&\gr d&x_1
\\
    &&&w_1[1] \, w_1[2]&&w_1[1] \, w_1[3]&&w_1[1] \, w_1[4]&&\cdots&w_1[1] \, w_1[m]&&
\hphantom{x_1}\lcurvearrowsw
\\
y:&\gr d&\gr d&&\gr d&&\gr d&&\gr d&&&\gr d&y_1\\
\vphantom{x}&&&&&&&&&&&&\\
\vphantom{x}&&&&&&&&&&&&\\
&\gr d&\gr d&&\gr d&&\gr d&&&&&&\\
\rcurvearrowse&&&w_2[1]\, w_2[2]&&w_2[1]\,w_2[2]&&\cdots&&&&&\\
&\gr d&\gr d&&\gr d&&\gr d&&&&&&
  \end{array}\tag{$\ddag$}
\]
where $w_i[j]$ stands for the $j$-th pair of data values of $w_i$. That is, the $i$-th block in the encoding
\[
  \begin{array}{rlllllllllll}
    x:&\gr d&\gr d&d_1&d_2&d_3&d_4& \dotsb &d_m&x_i&\gr d&\gr d\\
    y:&\gr d&\gr d&d_1&d_2&d_3&d_4& \dotsb &d_m&y_i&\gr d&\gr d
  \end{array},
\]
where $(d_j,d_j) = w_i[j]$ for every $j\leq m$, will now look like
\[\arraycolsep=3.5pt\def\arraystretch{1}
  \begin{array}{rlllllllllllllllll}
    x:&\gr d&\gr d&d_1&d_2&\gr d&d_1&d_3&\gr d&d_1&d_4& \dotsb &\gr d&d_1&d_m&x_i&\gr d&\gr d\\
    y:&\gr d&\gr d&d_1&d_2&\gr d&d_1&d_3&\gr d&d_1&d_4& \dotsb &\gr d&d_1&d_m&y_i&\gr d&\gr d
  \end{array}.
\]
Note that $d,d_1$ repeats along the whole block, there is hence a lot of redundancy of information; a block has now length $3(m-1)+1$ instead of $m+1$.

The idea is that this redundant encoding is done in such a way that testing if $d_i = d_1$ and ensuring that the first two data values are equal to the last two data values can be done using data tests of bounded $\mynext$-distance. This encoding, albeit being more cumbersome, can still be enforced by a \pmainlogic{} formula in such a way that it has a bounded $\mynext$-distance. To see this, let us review the changes that need to be applied to the formulas described in Lemma~\ref{lem:undec-plrvgame-nobool}.
\begin{align*}
	\phi_{\textit{block}(0)} =~& x \oblieqlocal \mynext x \land \mynext(x \oblieqlocal \mynext^{3}x);\\
	\phi_{\lambda_i} =~& \phi_{\textit{block}(0)} \land \mynext^{2+3(i-1)}(x \oblieqlocal \mynext x).\\
\phi_{\textit{block-str}} =~ &\mynext^2(\lnot (\toblieqp{x}{x})) ~\land
\underbrace{\bigwedge_{0 \leq i \leq m-1} \mynext^{2+3i} \left( (x \oblieqlocal \mynext x) \lor \mynext(\lnot (\toblieqp{x}{x}))\right)}_{A}  ~\land\\
& \phi_{\textit{1-label}} \land  \underbrace{\bigwedge_{0 \leq i \leq m-2} \left(\mynext^{2+ 3i}(x \oblieqlocal \mynext^3 x) \land \mynext^{1 + 3i}(x \oblieqlocal \mynext^3 x)\right)}_{B} \land \mynext^{3(m-1)+1}( x \oblieqlocal \mynext x).
\end{align*}
Observe that in $\phi_{\textit{block-str}}$ the subformula $A$ ensures that each data value of $w_i$ is either fresh or equal to the first data value, and subformula $B$ enforces that $d$ and $d_1$ are repeated every third position, all along the block.
Also, conditions 8 and F need to be modified accordingly, as follows.
\begin{enumerate}[align=left]
    \item[($8'$)] The first position verifies $\phi_{\textit{block}(0)}$; and for every position we have $\phi_{\textit{block}(0)} \Rightarrow \phi_{\textit{block-str}} \land \mynext^{3(m-1)+2}\phi_{\textit{block}(0)}$
    \item[(F${}'$)] There is some $i \in \set{0,\dotsc, 3(m-1)+1}$ and some position verifying
  \[
    \phi_{\textit{block}(i)} \land \lnot (x \oblieqlocal y).
  \]
\end{enumerate}
Observe that the encoding of counter values in the reduction before is not broken. This is because the new positions of the encoding have the property of having the same data value under $x$ and $y$, and thus the encoding of counter $\countx$ ---\ie, the number of data values that have appeared under $x$ but not under $y$--- is not modified; and similarly for counter $\county$. Notice that the above encoding has a $\mynext$-distance of $3$. Therefore, determining the winner of a \pmainlogic{} game is still undecidable if both the variables and the $\mynext$-distance is bounded.
\end{proof}
\color{black}



\section{Decidability of single-sided \pmainlogic[\texorpdfstring{$\top, \leftarrow$}{⟙,←}]}%
\label{single-past-decidable}
In this section we show that the single-sided \pmainlogic$[\top,\leftarrow]$-game
is decidable. We first observe that we do not need to consider
$\oblineqlocal$ formulas for our decidability argument, since there is
a reduction of the winning strategy existence problem that removes
all sub-formulas of the from $\toblineqp{x}{y}$.
\begin{prop}%
\label{prop:disequalityRemoval}
There is a polynomial-time reduction from the winning
strategy existence problem for \pmainlogic$[\top,\leftarrow]$ into the problem
on \pmainlogic$[\top, \oblieqlocal,\leftarrow]$.
\end{prop}
This is done as it was done for the satisfiability problem~\cite[Proposition~4]{DFP2016}. The key observation is that
\begin{itemize}
\item $\lnot (\toblineqp{x}{y})$ is equivalent to
  $\lnot \mynext^{-1} \top \lor (x \oblieqlocal \mynext^{-1} y \land
  \always^{-1}(\lnot \mynext^{-1} \top \lor y \oblieqlocal \mynext^{-1}
  y))$;
\item $\toblineqp{x}{y}$ can be translated into $\lnot(x \oblieqlocal
    x_{\oblieqlocal \Diamond^{-1} y}) \land \toblieqp{x_{\oblieqlocal
        \Diamond^{-1} y}}{y}$  for a new variable $x_{\oblieqlocal \Diamond^{-1} y}$ belonging to the same player as $x$.
\end{itemize}
Given a formula $\phi$ in negation normal form (\ie, negation is only applied to boolean variables and data tests), consider the formula
$\phi'$ resulting from the replacements listed above. It follows that $\phi'$ does not make use of $\not\oblieqlocal$. It is easy to see that there is a winning strategy for \sys in the game with winning condition $\phi$ if and only if she has a winning strategy for the game with condition $\phi'$.

\smallskip

\review{We consider games where the formula specifying the winning condition
only uses Boolean
variables belonging to \env while it can use data variables belonging
to \sys. Boolean variables can be simulated by data variables ---
for every Boolean variable $q$, we can have two data variables
$x_{q},y_{q}$. The Boolean variable $q$ will be true at a position if
$x_{q}$ and $y_{q}$ are assigned the same value at that position.
Otherwise, $q$ will be false. Hence, the formula specifying the
winning condition can also use Boolean variables belonging to \sys
without loss of generality. We
call this the single-sided \pmainlogic[$\top$, $\leftarrow$] games and
show that winning strategy existence problem is
decidable. We should remark that the decidability result of this section is subsumed by the
one in Section~\ref{dec-singlesided-nestedpast}. However, we prefer to retain this
section since Section~\ref{dec-singlesided-nestedpast} is technically
more tedious. The underlying intuitions used in both sections can be
more easily explained here without getting buried in technical
details.}

 The main concept we use for decidability is a symbolic
representation of models, introduced in~\cite{DDG2012}. The building
blocks of the symbolic representation are \emph{frames}, which we
adapt here. We finally show effective reductions between single-sided
\pmainlogic[$\top$, $\leftarrow$] games and
single-sided VASS games. This implies decidability of
single-sided \pmainlogic[$\top$, $\leftarrow$] games.
From Proposition~\ref{prop:disequalityRemoval}, it suffices to show
effective reductions between single-sided \pmainlogic[$\top$,
$\oblieqlocal$, $\leftarrow$] games and single-sided VASS games.

Given a formula in \pmainlogic[$\top$, $\oblieqlocal$, $\leftarrow$],
we replace sub-formulas of the form $x \oblieqlocal \mynext^{-j} y$
with $\mynext^{-j}(y \oblieqlocal \mynext^{j}x)$ if $j > 0$. For a
formula $\phi$ obtained after such replacements, let $l$ be the
maximum $i$ such that a term of the form $\mynext^{i}x$ appears in
$\phi$. We call $l$ the $\mynext$-length of $\phi$. Let
$\bvars^{\phi} \subseteq \bvars$ and $\dvars^{\phi} \subseteq \dvars$
be the set of Boolean and data variables used in $\phi$. Let $\Omega_{l}^{\phi}$
be the set of constraints of the form $\mynext^{i} q$, $\mynext^{i}x
\oblieqlocal \mynext^{j}y$ or $\mynext^{i}(\toblieqp{x}{y})$, where $q
\in \bvars^{\phi}$, $x,y\in \dvars^{\phi}$ and $i,j \in \{0, \ldots,
l\}$. For $e \in \set{0, \ldots, l}$, an $(e, \phi)$-frame is a set of
constraints $\symbval \subseteq \Omega_{l}^{\phi}$ that satisfies the
following conditions:
\begin{enumerate}[align=left]
    \item[(F0)] For all constraints $\mynext^{i} q, \mynext^{i}x
\oblieqlocal \mynext^{j}y, \mynext^{i}(\toblieqp{x}{y}) \in \symbval$,
$i,j \in \set{0, \ldots, e}$.
    \item[(F1)] For all $i \in \{0,\ldots,e\}$ and $x \in
        \dvars^{\phi}$, $\mynext^{i}x\oblieqlocal \mynext^{i}x \in \symbval$.
    \item[(F2)] For all $i,j \in \set{0,\ldots, e}$ and $x,y \in
        \dvars^{\phi}$, $\mynext^{i}x \oblieqlocal \mynext^{j}y \in \symbval$
        iff $\mynext^{j}y \oblieqlocal \mynext^{i}x \in \symbval$.
    \item[(F3)] For all $i,j,j' \in \set{0,\ldots,e}$ and $x,y,z\in
        \dvars^{\phi}$, if $\set{\mynext^{i}x \oblieqlocal \mynext^{j}y,
        \mynext^{j}y\oblieqlocal \mynext^{j'}z} \subseteq \symbval$,
        then $\mynext^{i}x \oblieqlocal \mynext^{j'}z \in \symbval$.
    \item[(F4)] For all $i,j\in \set{0,\ldots, e}$ and $x,y \in
      \dvars^{\phi}$ such that
      $\mynext^{i}x \oblieqlocal \mynext^{j}y \in \symbval$:
      \begin{itemize}
        \item if $i=j$, then for every $z \in \dvars^{\phi}$ we have
          $\mynext^{i}(\toblieqp{x}{z}) \in \symbval$ iff
          $\mynext^{j}(\toblieqp{y}{z}) \in \symbval$.
        \item if $i < j$, then $\mynext^{j}(\toblieqp{y}{x}) \in
          \symbval$ and for any $z \in \dvars^{\phi}$,
          $\mynext^{j}(\toblieqp{y}{z}) \in \symbval$ iff either
          $\mynext^{i}(\toblieqp{x}{z}) \in \symbval$ or there exists
          $i \le j' < j$ with $\mynext^{j}y \oblieqlocal
          \mynext^{j'}z \in \symbval$.
      \end{itemize}
\end{enumerate}
\noindent
The condition (F0) ensures that a frame can constrain at most
$(e+1)$ contiguous valuations. The next three conditions ensure that equality
constraints in a frame form an equivalence relation. The last
condition ensures that obligations for repeating values in the past
are consistent among various variables.

\review{Intuitively, an $(e,\phi)$-frame captures equalities among data values
within $(e+1)$ contiguous valuations of a model. If there are more
than $(e+1)$ valuations in a model, the first $(e+1)$ will be
considered by the first frame and valuations in positions $2$ to
$(e+2)$ by another frame. The valuations in positions $2$ to
$(e+1)$ are considered by both the frames, so two adjacent frames
should be consistent about what they say about overlapping positions.
This is formalized in the following definition.}

A pair of $(l, \phi)$-frames $(\symbval, \symbval')$ is said to be
one-step consistent \review{if}
\begin{enumerate}[align=left]
  \item[(O1)] for all $\mynext^{i}x \oblieqlocal \mynext^{j}y \in
    \Omega_{l}^{\phi}$ with $i,j>0$, we have $\mynext^{i}x \oblieqlocal
    \mynext^{j}y \in \symbval$ iff $\mynext^{i-1}x \oblieqlocal
    \mynext^{j-1}y \in \symbval'$,
  \item[(O2)] for all $\mynext^{i}(\toblieqp{x}{y}) \in
    \Omega_{l}^{\phi}$ with $i > 0$, we have $\mynext^{i}(\toblieqp{x}{y})
    \in \symbval$ iff $\mynext^{i-1}(\toblieqp{x}{y}) \in \symbval'$
    and
\item[(O3)] for all $\mynext^{i} q \in \Omega_{l}^{\phi}$ with $i > 0$, we
    have $\mynext^{i}q \in \symbval$ iff $\mynext^{i-1}q \in
    \symbval'$.
\end{enumerate}
For $e \in \{0, \ldots, l-1\}$, an $(e, \phi)$ frame $\symbval$ and an
$(e+1, \phi)$ frame $\symbval'$, the pair $(\symbval,\symbval')$ is said
to be one step consistent iff $\symbval \subseteq \symbval'$ and
for every constraint in $\symbval'$ of the form $\mynext^{i}
x \oblieqlocal \mynext^{j} y$, $\mynext^{i} q$ or
$\mynext^{i}( \toblieqp{x}{y} )$ with $i,j \in \set{0, \ldots,
e}$, the same constraint also belongs to $\symbval$.

An (infinite) $(l, \phi)$-symbolic model $\rho$ is an infinite sequence of
$(l, \phi)$-frames such that for all $i \in \Nat$, the pair $(\rho(i),
\rho(i+1))$ is one-step consistent. Let us define the symbolic
satisfaction relation $\rho,i \symbmodels \phi'$ where $\phi'$ is a
sub-formula of $\phi$. The relation $\symbmodels$ is defined in the
same way as $\models$ for $\pmainlogic$, except that for every
element $\phi'$ of $\Omega_{l}^{\phi}$, we have $\rho,i \symbmodels
\phi'$ whenever $\phi' \in \rho(i)$. We say that a concrete model $\sigma$
realizes a symbolic model $\rho$ if for every $i \in \PNat$,
$\rho(i) = \set{\phi' \in \Omega_{l}^{\phi} \mid \sigma, i
\models \phi'}$. The next result follows easily from definitions.

\begin{lem}[symbolic vs.~concrete models]%
    \label{lem:symbolicToConcreteSat}
    Suppose $\phi$ is a $\pmainlogic[\top, \oblieqlocal, \leftarrow]$
    formula of $\mynext$-length $l$, $\rho$ is a $(l,\phi)$-symbolic model and
    $\sigma$ is a concrete model realizing $\rho$. Then $\rho$
    symbolically satisfies $\phi$ iff $\sigma$ satisfies $\phi$.
\end{lem}

\review{The main idea behind the symbolic model approach is that we temporarily
    forget that the semantics of constraints like $\toblieqp{x}{y}$
require looking at past positions and not just the current position.
We forget the special semantics of $\toblieqp{x}{y}$ and treat it to
be true in a symbolic model at some position if the frame at that
position contains $\toblieqp{x}{y}$; in other words, we symbolically
assume $\toblieqp{x}{y}$ to be true by looking only at the current
position. This way, a $\pmainlogic[\top, \oblieqlocal, \leftarrow]$
formula can be treated as if it is a propositional LTL formula and the
existence of winning strategies can be solved using games on
deterministic parity automata corresponding to the propositional LTL
fromula. However, this comes at a price --- we may assume too many
constraints of the form $\toblieqp{x}{y}$ to be true in a symbolic
model and not all of them may be simultaneously satisfiable in any
concrete model. Suppose a symbolic model assumes, at the second
position, both
$\toblieqp{x}{y}$ and $\toblieqp{z}{y}$ to be true and $x \oblieqlocal
z$ to be false. The three constraints cannot be satisfied by any concrete
model since there is only one
past postion where the value assigned to $y$ could either be the value
assigned to $x$ in the second position or the value assigned to
$z$ in the second position, but not both. In order to detect which
symbolic models can be realized by concrete models, we keep count of
how many distinct data values can be repeated in the past, using
counters. We explain this in more detail in the following paragraphs.}


We fix a \pmainlogic[$\top$, $\oblieqlocal$, $\leftarrow$] formula
$\phi$ of $\mynext$-length $l$. For $e \in \{0,\ldots, l\}$, an $(e,
\phi)$-frame $\symbval$, $i \in \{0, \ldots, e\}$ and a variable $x$,
the set of past obligations of the variable $x$ at level $i$ in
$\symbval$ is defined to be the set $\pobli_{\symbval}(x,i) = \{y \in
\dvars^{\phi} \mid \mynext^{i}(\toblieqp{x}{y}) \in \symbval\}$.
The equivalence class of $x$ at level $i$ in $\symbval$ is defined
to be ${[(x,i)]}_{\symbval} = \{y \in \dvars^{\phi} \mid \mynext^{i}
x \oblieqlocal \mynext^{i}y \in \symbval\}$.

\begin{figure}[!htp]
\begin{center}
\vspace{-5mm}
\includegraphics[width=.55\textwidth]{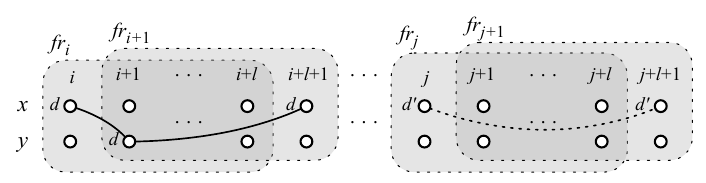}
\vspace{-2mm}
\end{center}
    \caption{Intuition for points of increment and decrement}%
    \label{fig:pointOfIncDec}
\end{figure}

Consider a concrete model $\sigma$ restricted to two variables $x,y$
as shown in Fig.~\ref{fig:pointOfIncDec}. The top row indicates the positions $i, (i+1), \ldots,
(i+l), (i+l+1), j, (j+1), \ldots, (j+l), (j+l+1)$.
The left column indicates the two variables $x,y$ and the remaining
columns indicate valuations. E.g., $\sigma(i+1)(y)=d$ and
$\sigma(j+l+1)(x) = d'$. Let $\symbval_{i} = \set{\phi' \in
\Omega^{\phi}_{l} \mid \sigma,i\models \phi'}$. We have indicated this
pictorially by highlighting the valuations that determine the
contents of $\symbval_{i}$. The data values for $x$ at positions
$i$ and $(i+l+1)$ are equal, but the
positions are too far apart to be captured by any one constraint of
the form $\mynext^{\alpha} x \oblieqlocal \mynext^{\beta} x$ in
$\Omega^{\phi}_{l}$. However, the
intermediate position $(i+1)$ has the same data value and is less than
$l$ positions apart from both positions. One constraint from
$\Omega^{\phi}_{l}$ can capture the data repetition between positions
$i$ and $(i+1)$ while another one captures the repetition between
positions $(i+1)$ and $(i+l+1)$, thus indirectly capturing the
repetition between positions $i$ and $(i+l+1)$.
For $e \in \{0,\ldots, l\}$, an $(e,
\phi)$-frame $\symbval$, $i \in \{0, \ldots, e\}$ and a variable $x$,
we say that there is a forward
(resp.~backward) reference from $(x,i)$ in $\symbval$ if $\mynext^{i}
x \oblieqlocal \mynext^{i+j}y \in \symbval$ (resp.~$\mynext^{i} x
\oblieqlocal \mynext^{i-j}y \in \symbval$) for some $j > 0$ and $y \in
\dvars^{\phi}$. The constraint $x \oblieqlocal \mynext y$ in
$\symbval_{i}$ above is a forward reference from $(x,0)$ in
$\symbval_{i}$, while the constraint $\mynext^{l}x \oblieqlocal y$ is
a backward reference from $(x,l)$ in $\symbval_{i+1}$.

In Figure~\ref{fig:pointOfIncDec}, the data values of $x$ at positions $j$ and $(j+l+1)$
are equal, but the two positions are too far apart to be captured by any
constraint of the form $\mynext^{\alpha} z \oblieqlocal
\mynext^{\beta}w$ in $\Omega^{\phi}_{l}$. Neither are there any
intermediate positions with the same data value to capture the
repetition indirectly.
We maintain a counter to keep track of the number of such remote data
repetitions.
Let $X \subseteq \dvars^{\phi}$ be a set of
variables. A \emph{point of decrement} for counter $X$ in an $(e, \phi)$-frame
$\symbval$ is an equivalence class of the form ${[(x,e)]}_{\symbval}$
such that there is no backward reference from $(x,e)$ in $\symbval$
and $\pobli_{\symbval}(x,e)=X$. In the above picture, the equivalence class
${[(x,l)]}_{\symbval_{j+1}}$ in the frame $\symbval_{j+1}$ is a point of
decrement for $\set{x}$.
A \emph{point of increment for
$X$ in an $(l, \phi)$-frame $\symbval$} is an equivalence class of the
form ${[(x,0)]}_{\symbval}$ such that there is no forward reference from
$(x,0)$ in $\symbval$ and ${[(x,0)]}_{\symbval} \cup
\pobli_{\symbval}(x,0)= X$. In the above picture, the equivalence
class ${[(x,0)]}_{\symbval_{j}}$ in the frame $\symbval_{j}$ is a point
of increment for $\set{x}$. Points of increment are not present in
$(e, \phi)$-frames for $e < l$ since such frames do not contain complete
information about constraints in the next $l$ positions. We denote by
$\inc(\symbval)$ the vector indexed by non-empty subsets of
$\dvars^{\phi}$, where each coordinate contains the number of points
of increment in $\symbval$ for the corresponding subset of variables.
Similarly, we have the vector $\dec(\symbval)$ for points of
decrement.

\review{Intuitively, points of increment are positions where there is
an opportunity to assign a value to variable $y$ in order to satisfy a data
repetition constraint like $\toblieqp{x}{y}$ that may occur later in
a symbolic model. On the other hand, points of decrements are those
positions of the symbolic models where we are obliged to ensure that
some data value repeats in the past. So if there are lots of points of
decrement, we have lots of obligations to repeat lots of data values
in the past. If one needs to be able to do this, there should be lots of
opportunities (points of increment) that have occured in the past. We
can ensure that there are sufficient points of increment in the past
by using counters --- every time we see a point of increment along a
symbolic model, we increment the counter. Every time we see a point of
decrement, we decrement the counter. There will be sufficiently many
points of increment to satisfy all the data repetition constraints if
the value of the counter always stays above zero. This is exactly the
constraint imposed on counters in energy games (the counter value is
intuitively the ``energy'' stored in a system and it should never be
below zero) and that's why energy games are useful to solve
\pmainlogic[$\top$, $\oblieqlocal$, $\leftarrow$] games.  Energy games
are effectively equivalent to single-sided VASS games and we use the
later since, technically, it is easier to adapt to our context. The
value of a counter at a position maintains the number of points of
increment before that position that are free to be used to satisfy
constraints that may occur later. We give the formal construction
below. The resulting single-sided VASS game is basically a product of two
components. The first one is a deterministic parity automaton which checks
whether a symbolic model symbolically satisfies the given
\pmainlogic[$\top$, $\oblieqlocal$, $\leftarrow$] formula. The second
component is a VASS which keeps track of the number of points of
increment and decrement. By playing two player games on these two
components in parallel, we can determine whether \sys has a strategy to
build a symbolic model that symbolically satisfies the given
\pmainlogic[$\top$, $\oblieqlocal$, $\leftarrow$] formula while, at the same time,
ensuring that the symbolic model is realizable.}

Given a \pmainlogic[$\top$, $\oblieqlocal$, $\leftarrow$] formula
$\phi$ in which \review{$(\dvars^{e} \cap \dvars^{\phi}) = \emptyset =
(\bvars^{s} \cap \bvars^{\phi})$}, we construct a
single-sided VASS game as follows. Let $l$ be the $\mynext$-length of
$\phi$ and $\symbvals$ be the set of all $(e, \phi)$-frames for all $e \in
\set{0, \ldots, l}$. Let $A^{\phi}$ be a deterministic parity automaton that accepts
a symbolic model iff it symbolically satisfies $\phi$, with set of
states $Q^{\phi}$ and initial state $q^{\phi}_{\init}$. The
single-sided VASS game will have set of counters
$\nepwrset(\dvars^{\phi})$,
set of environment states $\set{-1, 0, \ldots, l} \times Q^{\phi}
 \times (\symbvals \cup \set{\bot})$ and set of system states
$\set{-1, 0, \ldots, l} \times Q^{\phi} \times (\symbvals
\cup \set{\bot}) \times \pwrset(\bvars^{\phi})$. Every state will inherit the
colour of its $Q^{\phi}$ component. For convenience, we let
$\bot$ to be the only $(-1, \phi)$-frame and $(\bot, \symbval')$ be one-step
consistent for every $0$-frame $\symbval'$. The initial state is
$(-1, q^{\phi}_{\init}, \bot)$, the initial counter values are all $0$
and the transitions are as follows ($\lceil \cdot
\rceil l$ denotes the mapping that is identity on $\set{-1, 0,
\ldots, l-1}$ and maps all others to $l$).
\begin{itemize}
    \setlength{\itemsep}{0.7em}
    \item $(e,q,\symbval) \act{\vec{0}} (e,q, \symbval, V)$ for
every $e \in \set{-1, 0, \ldots, l}$, $q \in Q^{\phi}$,
$\symbval \in \symbvals \cup \set{\bot}$ and $V \subseteq
\bvars^{\phi}$.
\item $(e, q^{\phi}_{\init}, \symbval, V)
\act{\inc(\symbval)-\dec(\symbval')}
(e+1,q^{\phi}_{\init}, \symbval')$ for every $V
\subseteq \bvars^{\phi}$, $e \in \set{-1, 0,
\ldots, l-2}$, $(e, \phi)$-frame $\symbval$ and $(e+1, \phi)$-frame $\symbval'$,
where the pair $(\symbval, \symbval')$ is one-step consistent and
$\set{p \in \bvars^{\phi} \mid \mynext^{e+1}p \in \symbval'} = V$.
\item $(e, q, \symbval, V) \act{\inc(\symbval) -
\dec(\symbval')} (\lceil e+1
\rceil l, q', \symbval')$
for every $e \in \set{l-1,l}$, $(e, \phi)$-frame $\symbval$, $V \subseteq
\bvars^{\phi}$, $q, q' \in Q^{\phi}$ and $(\lceil e+1
\rceil l, \phi)$-frame $\symbval'$, where the pair $(\symbval, \symbval')$
is one-step consistent, $\set{p \in \bvars^{\phi} \mid \mynext^{\lceil e+1
    \rceil l}p \in
\symbval'} = V$ and $q \act{\symbval'} q'$ is a
transition in ${A^{\phi}}$.
\end{itemize}

\noindent
Transitions of the form $(e,q,\symbval) \act{\vec{0}} (e,q, \symbval, V)$ let
the environment choose any subset $V$ of $\bvars^{\phi}$ to be true in the
next round. In transitions of the form $(e, q,
\symbval, V) \act{\inc(\symbval) - \dec(\symbval')} (\lceil e+1 \rceil
l, q', \symbval')$, the condition $\set{p \in \bvars^{\phi} \mid \mynext^{\lceil e+1
\rceil l}p \in \symbval'} = V$  ensures that the frame $\symbval'$ chosen by the
system is compatible with the subset $V$ of $\bvars^{\phi}$ chosen by the
environment in the preceding step. By insisting that the pair
$(\symbval,\symbval')$ is one-step consistent, we ensure that the
sequence of frames built during a game is a symbolic model. The
condition $q \act{\symbval'} q'$ ensures that the symbolic model is
accepted by $A^{\phi}$ and hence symbolically satisfies $\phi$. The
update vector $\inc(\symbval) - \dec(\symbval')$ ensures that symbolic
models are realizable, as explained in the proof of the following
result.

\begin{lem}[repeating values to VASS]%
    \label{lem:repValuesToVASS}
    Let $\phi$ be a \pmainlogic$[\top$, $\oblieqlocal$,
    $\leftarrow]$ formula with \review{$(\dvars^{e} \cap \dvars^{\phi}) =
    \emptyset$} and \review{$(\bvars^{s} \cap \bvars^\phi) = \emptyset$}.
    Then \sys has a winning strategy in the corresponding
    single-sided \pmainlogic$[\top$, $\oblieqlocal$, $\leftarrow]$
    game iff she has a winning strategy in the
    single-sided VASS game constructed above.
\end{lem}
\begin{proof}
    We begin with a brief description of the ideas used. A game on the
    single-sided VASS game results in a sequence of frames. The
    single-sided VASS game embeds automata which check that these
    sequences are symbolic models that symbolically satisfy $\phi$.
    This in conjunction with Lemma~\ref{lem:symbolicToConcreteSat}
    (symbolic vs.~concrete models) will prove the result, provided the
    symbolic models are also realizable. Some symbolic models are not
    realizable since frames contain too many constraints about data
    values repeating in the past and no concrete model can satisfy all
    those constraints.  To avoid this, the single-sided VASS game
    maintains counters for keeping track of the number of such
    constraints. Whenever a frame contains such a past repetition
    constraint that is not satisfied locally within the frame itself,
    there is an absence of backward references in the frame and it
    results in a point of decrement.  Then the $-\dec(\symbval')$ part
    of transitions of the form $(e, q, \symbval, V)
    \act{\inc(\symbval) - \dec(\symbval')} (\lceil e+1 \rceil l, q',
    \symbval')$ will decrement the corresponding counter. In order for
    this counter to have a value of at least $0$, the counter should
    have been incremented earlier by $\inc(\symbval)$ part of earlier
    transitions. This ensures that symbolic models resulting from the
    single-sided VASS games are realizable.

    Now we give the details for the forward direction.
    Suppose the system player has a strategy $\ts: \vals^{*}\cdot
    \vals^{e} \to \vals^{s}$ in the single-sided \pmainlogic$[\top$,
    $\oblieqlocal$, $\leftarrow]$ game. We will show that the system
    player has a strategy $\ssys: {(Q \times \Nat^{C})}^{*} \cdot (Q^{s}
    \times \Nat^{C}) \to T$ in the single-sided VASS game. It is
routine to construct such a strategy from the mapping $\mu:
{(\pwrset(\bvars^{\phi}))}^* \to \symbvals \cup \set{\bot}$ that we
define now. For every sequence $\chi \in {(\pwrset(\bvars^{\phi}))}^*$,
we will define $\mu(\chi)$ and a concrete
model of length $|\chi|$, by induction on $|\chi|$. For the base case
$|\chi|=0$, the concrete model is the empty sequence and the frame is
$\bot$.

    For the induction step, suppose $\chi$ is of
    the form $\chi' \cdot V$ and $\sigma$ is the concrete
    model defined for $\chi'$ by induction hypothesis. Let $\upsilon^{e}:\bvars^{e} \to
    \set{\top, \bot}$ be the mapping defined as $\upsilon^{e}(p) =
    \top$ iff $p \in V$. The system player's strategy $\ts$ in the
    single-sided \pmainlogic$[\top$, $\oblieqlocal$, $\leftarrow]$
    game will give a valuation $\ts(\sigma \cdot \upsilon^{e}) = \upsilon^{s}:
    \dvars^{s} \to \Domain$. We define the finite concrete model
    to be $\sigma \cdot (\upsilon^{e} \oplus \upsilon^{s})$ and
$\mu(\chi)$ to be the
    frame
    $\symbval' = \set{\phi' \in \Omega^{\phi}_{l} \mid \sigma \cdot
    (\upsilon^{e} \oplus \upsilon^{s}), |\sigma|+1-\lceil |\sigma| \rceil l
    \models \phi'}$.

    Next we will prove that the strategy $\ssys$ defined above is
    winning for the system player. Suppose the system player plays
    according to $\ssys$ in the single-sided VASS game, resulting in
    the sequence of states
    \begin{align*}
        (-1, q^{\phi}_{\init}, \bot) (-1,
        q^{\phi}_{\init}, \bot, V_{1}) (0, q^{\phi}_{\init},
        \symbval_{1}) (0, q^{\phi}_{\init}, \symbval_{1}, V_{2})\\
        (1,q^{\phi}_{\init}, \symbval_{2}) \cdots (l, q, \symbval_{l+1})
        (l,q,\symbval_{l+1}, V_{l+2}) (l,q',\symbval_{l+2}) \cdots
    \end{align*}
    The sequence $\symbval_{l+1} \symbval_{l+2} \cdots$ is an infinite
    $(l,\phi)$-symbolic model; call it $\rho$. It is clear from the construction
    that $\rho$ is realized by a concrete model $\sigma$, which is the
    result of the system player playing according to the winning
    strategy $\ts$ in the \pmainlogic[$\top$, $\oblieqlocal$,
    $\leftarrow$] game. So $\sigma,1 \models \phi$ and by
    Lemma~\ref{lem:symbolicToConcreteSat} (symbolic vs.~concrete
    models), $\rho$ symbolically satisfies $\phi$. By definition of
    $A^{\phi}$, the unique run of $A^{\phi}$ on $\rho$ satisfies the
    parity condition and hence the play satisfies the parity condition in the
    single-sided VASS game. It remains to prove that if a transition
    given by $\ssys$ decrements some counter, that counter will have
    sufficiently high value. Any play starts with all counters
    having zero and a counter is decremented by a transition if the
    frame chosen by that transition has points of decrement for the
    counter. For $e \in \set{1, \ldots, l+1}$ and
    $x \in \dvars^{\phi}$, ${[(x,e)]}_{\symbval_{e}}$ cannot be a point
    of decrement in $\symbval_{e}$ --- if it were, the data value
    $\sigma(e)(x)$ would have appeared in some position in
    $\set{1, \ldots, e-1}$, creating a backward reference from
    $(x,e)$ in $\symbval_{e}$.

    For $i > l+1$, $x \in \dvars^{\phi}$ and $X \in
    \nepwrset(\dvars^{\phi})$, suppose
    ${[(x,l)]}_{\symbval_{i}}$ is a point of decrement for $X$ in
    $\symbval_{i}$. Before decrementing the counter $X$, it is
    incremented for every point of increment for $X$ in every frame
    $\symbval_{j}$ for all $j < i$. Hence, it suffices to associate with
    this point of decrement a point of
    increment for $X$ in a frame earlier than $\symbval_{i}$ that is
    not associated to any other point of decrement. Since
    ${[(x,l)]}_{\symbval_{i}}$ is a point of decrement for $X$ in
    $\symbval_{i}$, the data value $\sigma(i)(x)$ appears
    in some of the positions $\set{1, \ldots, i-l-1}$. Let $i' =
    \max\set{j \in \set{1, \ldots, i-l-1} \mid \exists y \in
    \dvars^{\phi}, \sigma(j)(y) = \sigma(i)(x)}$. Let $x' \in
    X$ be such that $\sigma(i')(x') = \sigma(i)(x)$ and associate with
    ${[(x,l)]}_{\symbval_{i}}$ the class ${[(x',0)]}_{\symbval_{i'+l}}$,
    which is a point of increment for $X$ in $\symbval_{i'+l}$. The
    class ${[(x',0)]}_{\symbval_{i'+l}}$ cannot be associated with any
    other point of decrement for $X$ --- suppose it were associated with
    ${[(y,l)]}_{\symbval_{j}}$, which is a point of decrement for
    $X$ in $\symbval_{j}$. Then $\sigma(j)(y) = \sigma(i)(x)$. If
    $j = i$, then ${[(x,l)]}_{\symbval_{i}} = {[(y,l)]}_{\symbval_{j}}$
    and the two points of decrement are the same. So $j < i$ or
    $j>i$. We compute $j'$ for ${[(y,l)]}_{\symbval_{j}}$ with
    $j' < j$ just like we computed $i'$ for ${[(x,l)]}_{\symbval_{i}}$.
    If $j < i$, then $j$ would be one of the positions in $\set{1,
    \ldots, i-l-1}$
    where the data value
    $\sigma(i)(x)$ appears ($j$ cannot be in the interval $[i-l, i-1]$
    since those positions do not contain the data value
    $\sigma(i)(x)$; if they did, there would have been a backward
    reference from $(x,l)$ in $\symbval_{i}$ and
    ${[(x,l)]}_{\symbval_{i}}$ would not have been a point of decrement), so
    $j \le i'$ (and hence $j' < i'$). If $j > i$, then $i$ is one of
    the positions in $\set{1, \ldots, j-l-1}$ where the data value
    $\sigma(j)(y)$ appears ($i$ cannot be in the interval $[j-l, j-1]$
    since those positions do not contain the data value
    $\sigma(j)(y)$; if they did, there would have been a backward
    reference from $(y,l)$ in $\symbval_{j}$ and
    ${[(y,l)]}_{\symbval_{j}}$ would not have been a point of decrement), so $i \le j'$ (and hence $i' < j'$). In
    both cases, $j' \ne i'$ and hence, the class
    ${[y',0]}_{\symbval_{j'+l}}$ we associate with
    ${[(y,l)]}_{\symbval_{j}}$ would be different from
    ${[(x',0)]}_{\symbval_{i'+l}}$.

    Next we give the details for the reverse direction.
    Suppose the system player has a strategy $\ssys: {(Q \times
    \Nat^{C})}^{*} \cdot (Q^{s} \times \Nat^{C}) \to T$ in the
    single-sided VASS game. We will show that the system player has a
    strategy $\ts: \vals^{*}\cdot \vals^{e} \to \vals^{s}$ in the
    single-sided \pmainlogic$[\top$, $\oblieqlocal$, $\leftarrow]$
    game. For every $\sigma \in \vals^{*}$ and every $\val^{e} \in
    \vals^{e}$, we will define $\ts(\sigma \cdot \val^{e}):
    \dvars^{\phi} \to \Domain$ and a sequence of configurations
    $\chi \cdot ((e,q,\symbval),\vec{n}_{\inc} - \vec{n}_{\dec})$ in
    ${(Q \times \Nat^{C})}^{*}\cdot (Q^{e}\times\Nat^{C})$ of length
    $2|\sigma|+3$ such that for every counter $X \in
    \nepwrset(\dvars^{\phi})$, $\vec{n}_{\inc}(X)$ is the sum of the
    number of points of increment for $X$ in all the frames occurring
    in $\chi$ and $\vec{n}_{\dec}(X)$ is the sum of the number of points
    of decrement for $X$ in all the frames occurring in $\chi$ and in
    $\symbval$. We will do this by induction on
    $|\sigma|$ and prove that the resulting strategy is winning for the
    system player. By \emph{frames occurring in $\chi$}, we refer to
    frames $\symbval$ such that there are consecutive configurations
    $( (e, q, \symbval), \vec{n}) ( (e,q,\symbval,V), \vec{n})$ in
    $\chi$. By $\Pi_{\symbvals}(\chi)(i)$, we refer to
    $i$\textsuperscript{th} such occurrence of a frame in $\chi$. Let
    $\set{d_{0}, d_{1}, \ldots} \subseteq \Domain$ be a countably
    infinite set of data values.

    For the base case $|\sigma|=0$, let $V \subseteq \bvars^{e}$ be
    defined as $p \in V$ iff $\val^{e}(p)=\top$. Let $\ssys((
    (-1, q^{\phi}_{\init},\bot),\vec{0})\cdot((-1,
    q^{\phi}_{\init},\bot,V),\vec{0}))$ be the transition $(-1,
    q^{\phi}_{\init},\bot,V) \act{\vec{0}-\dec(\symbval_{1})} (0, q,
    \symbval_{1})$. Since $\ssys$ is a winning strategy for \sys in
    the single-sided VASS game, $\dec(\symbval_{1})$ is necessarily
    equal to $\vec{0}$. The
    set of variables $\dvars^{\phi}$ is partitioned into equivalence
    classes by the $(0,\phi)$-frame $\symbval_{1}$. We define
    $\ts(\val^{e})$ to be the valuation that assigns to each such
    equivalence class a data value $d_{j}$, where $j$ is the
    smallest number such that $d_{j}$ is not assigned to any variable
    yet. We let the sequence of configurations be $(
    (-1,q^{\phi}_{\init}, \bot),\vec{0}) \cdot
    ( (-1,q^{\phi}_{\init}, \bot, V), \vec{0}) \cdot (
    (0,q,\symbval_{1}), -\dec(\symbval_{1}))$.

    For the induction step, suppose $\sigma\cdot \val^{e} = \sigma'
    \cdot (\val^{e}_{1}\oplus \val^{s}_{1} ) \cdot \val^{e}$ and
    $\chi' \cdot ((e,q,\symbval), \vec{n})$ is the sequence of
    configurations given by the induction hypothesis for $\sigma'
    \cdot \val^{e}_{1}$. If $\set{\phi' \in \Omega^{\phi}_{l} \mid
    \sigma' \cdot (\val^{e}_{1} \oplus \val^{s}_{1}),|\sigma'|+1-e
    \models \phi'} \ne \symbval$, it corresponds to the case where the
    system player in the \pmainlogic$[\top$, $\oblieqlocal$,
    $\leftarrow]$ game has already deviated from the strategy we have
    defined so far. So in this case, we define $\ts(\sigma\cdot
    \val^{e})$ and the sequence of configurations to be arbitrary.
    Otherwise, we have $\set{\phi' \in \Omega^{\phi}_{l} \mid
    \sigma' \cdot (\val^{e}_{1} \oplus \val^{s}_{1}),|\sigma'|+1-e
    \models \phi'} = \symbval$. Let $V \subseteq \bvars^{e}$ be
    defined as $p \in V$ iff $\val^{e}(p)=\top$ and let $\ssys( \chi' \cdot (
    (e,q,\symbval),\vec{n}) \cdot  ((e,q,\symbval,V),\vec{n}))$ be the
    transition $(e,q,\symbval,V) \act{\inc(\symbval) - \dec(\symbval')} (\lceil e+1 \rceil l, q',
    \symbval')$. We define the sequence of configurations as
    $\chi'\cdot ((e,q,\symbval), \vec{n}) \cdot ( (e,q,\symbval,V)
    \vec{n}) \cdot ((\lceil e+1 \rceil l, q',\symbval'),
    \vec{n} + \inc(\symbval) - \dec(\symbval'))$. Since $\ssys$ is a
    winning strategy for the system player in the single-sided VASS
    game, $ \vec{n} + \inc(\symbval) - \dec(\symbval') \ge
    \vec{0}$. The valuation $\ts(\sigma \cdot
    \val^{e}): \dvars^{\phi} \to \Domain$ is defined as follows. The
    set $\dvars^{\phi}$ is partitioned by the
    equivalence classes at level $\lceil e+1 \rceil l$ in
    $\symbval'$. For every such equivalence class ${[(x,\lceil e+1
    \rceil l)]}_{\symbval'}$, assign the data value $d'$ as defined
    below.
    \begin{enumerate}
        \item If there is a backward reference $\mynext^{\lceil e+1
            \rceil l}x \oblieqlocal \mynext^{\lceil e+1 \rceil l -j}y$ in
            $\symbval'$, let $d' = \sigma'\cdot (\val^{e}_{1} \oplus
            \val^{s}_{1})(|\sigma'|+2-j)(y)$.
        \item If there are no backward references from $(x,\lceil e+1
            \rceil l)$ in $\symbval'$ and the set
            $\pobli_{\symbval}(x,\lceil e+1 \rceil l)$ of past
            obligations of $x$ at level $\lceil e+1 \rceil l$ in
            $\symbval'$ is empty, let $d'$ be $d_{j}$, where $j$ is
            the smallest number such that $d_{j}$ is not assigned to
            any variable yet.
        \item If there are no backward references from $(x,\lceil e+1
            \rceil l)$ in $\symbval'$ and the set
            $\pobli_{\symbval}(x,\lceil e+1 \rceil l)$ of past
            obligations of $x$ at level $\lceil e+1 \rceil l$ in
            $\symbval'$ is the non-empty set $X$, then ${[(x,\lceil e+1
            \rceil l)]}_{\symbval'}$ is a point of decrement for $X$ in
            $\symbval'$. Pair off this with a point of increment for
            $X$ in a frame that occurs in $\chi' \cdot (
            (e,q,\symbval),\vec{n}) \cdot  ((e,q,\symbval,V),\vec{n})$
            that has not been paired off before. It is possible to do
            this for every point of decrement for $X$ in $\symbval'$,
            since $(\vec{n} + \inc(\symbval))(X)$ is the number of
            points of increment for $X$  occurring in $\chi' \cdot (
            (e,q,\symbval),\vec{n}) \cdot  ((e,q,\symbval,V),\vec{n})$
            that have not yet been paired off and $(\vec{n} +
            \inc(\symbval))(X) \ge \dec(\symbval')(X)$. Suppose we
            pair off ${[(x,\lceil e+1 \rceil l)]}_{\symbval'}$ with a
            point of increment ${[(y,0)]}_{\symbval_{i}}$ in the frame
            $\symbval_{i} = \Pi_{\symbvals}(\chi' \cdot (
            (e,q,\symbval),\vec{n}) \cdot
            ((e,q,\symbval,V),\vec{n}))(i)$, then let $d'$ be
            $\sigma'\cdot (\val^{e}_{1} \oplus \val^{s}_{1})(i)(y)$.
    \end{enumerate}
    Suppose the system player plays according to the strategy
    $\ts$ defined above, resulting in the model
    $\sigma = (\val^{e}_{1} \oplus \val^{s}_{1}) \cdot (\val^{e}_{2} \oplus
    \val^{s}_{2}) \cdots$. It is clear from the construction that
    there is a sequence of configurations
    \begin{align*}
        ((-1, q^{\phi}_{\init}, \bot), \vec{0}) ((-1,
        q^{\phi}_{\init}, \bot, V_{1}), \vec{0})\\ ((0, q^{\phi}_{\init},
        \symbval_{1}), \vec{n}_{1}) ((0, q^{\phi}_{\init},
        \symbval_{1}, V_{2}), \vec{n}_{1})\\
        ((1,q^{\phi}_{\init}, \symbval_{2}), \vec{n}_{2}) \cdots ((l,
        q, \symbval_{l+1}), \vec{n}_{l+1})\\
        ((l,q,\symbval_{l+1}, V_{l+2}), \vec{n}_{l+1})
        ((l,q',\symbval_{l+2}), \vec{n}_{l+2}) \cdots
    \end{align*}
    that is the result of the system player playing according to the
    strategy $\ssys$ in the single-sided VASS game such that the
    concrete model $\sigma$ realizes the symbolic model
    $\symbval_{l+1} \symbval_{l+2} \cdots$. Since $\ssys$ is a winning
    strategy for the system player, the sequence of configurations
    above satisfy the parity condition of the single-sided VASS game,
    so $\symbval_{l+1} \symbval_{l+2} \cdots$ symbolically satisfies
    $\phi$. From Lemma~\ref{lem:symbolicToConcreteSat} (symbolic
    vs.~concrete models), we conclude that $\sigma$ satisfies
    $\phi$.
\end{proof}
\begin{cor}%
    \label{cor:upperBound}
    The winning strategy existence problem for single-sided
    \pmainlogic$[\top$, $\oblieqlocal$, $\leftarrow]$ game of
    repeating values (without past-time temporal modalities) is in
    \textnormal{\textsc{3ExpTime}}.
\end{cor}
\begin{proof}
We recall from~\cite[Corollary 5.7]{CJLS2017} that the winning
strategy existence problem for energy games (and hence
single-sided VASS games) can be solved in time ${(|V|\cdot
\parallel E \parallel)}^{2^{O(d \cdot \log(d+p))}}+O(d \cdot c)$,
where $V$ is the set of vertices, $\parallel E \parallel$ is the
maximal absolute value of counter updates in the edges, $d$ is the
number of counters, $p$ is the number of even priorities and $c$
is the maximal value of the initial counter values. \review{For a
\pmainlogic$[\top$, $\oblieqlocal$, $\leftarrow]$ formula $\phi$ with
$\dvars^{e} \cap \dvars^{\phi}= \bvars^{s} \cap \bvars^{\phi} = \emptyset$ and
no past-time temporal modalities, a deterministic parity automaton for
symbolic models can be constructed in \textsc{2ExpTime}, having doubly exponentially many states. A frame is a subset of atomic
constraints, so there are exponentially many frames. Hence, the number
of vertices $|V|$ in the constructed single-sided VASS game is doubly
exponential.
The value of ${\parallel} E {\parallel}$ is polynomial, since it depends
on the number of points of increment and the number of points of
decrement in frames.
The value of $p$ is bounded by the number of priorities used in the
parity automaton and hence, it is at most doubly exponential. The
value of $c$ is zero. The number of counters $d$ is exponential, since
there is one counter for every subset of data variables used in
$\phi$.} Hence, the upper bound for
energy games translates to \textsc{3ExpTime} for single-sided
\pmainlogic$[\top$, $\oblieqlocal$, $\leftarrow]$ games.
\end{proof}


Our decidability proof thus depends ultimately on energy games, as
hinted in the title of this paper. Next we show that single-sided VASS
games can be effectively reduced to single-sided \pmainlogic[$\top$,
$\oblieqlocal$, $\leftarrow$] games.

\begin{thm}%
    \label{thm:ssVASSGamesToGRV}
    Given a single-sided VASS game, a single-sided \pmainlogic[$\top$,
$\oblieqlocal$, $\leftarrow$] game can be
    constructed in polynomial time so that \sys has a
    winning strategy in the first game iff \sys has a
    winning strategy in the second one.
\end{thm}
\begin{proof}
We begin with a brief description of the ideas used. We will simulate runs of
single-sided VASS games with models of formulas in \pmainlogic{}.
The formulas satisfied at position $i$ of the concrete model will
contain
information about counter values before the
$i$\textsuperscript{th} transition and the identity of the
$i$\textsuperscript{th} transition chosen by the environment and
the system players in the run of the single-sided
VASS game.
For simulating a counter $x$, we use two \sys variables $x$ and
$\overline{x}$. The data values assigned to these variables from
positions $1$ to $i$ in a concrete model $\sigma$ will represent the
counter value that is equal to the cardinality of the set $\set{d \in
\Domain \mid \exists j \in \set{1, \ldots, i}, \sigma(j)(x)=d,
\forall j' \in \set{1, \ldots, i}, \sigma(j')(\overline{x}) \ne
d}$. Using formulas in \pmainlogic[$\top$, $\oblieqlocal$,
$\leftarrow$], the two players can be enforced to correctly update the
concrete model to faithfully reflect the moves in the single-sided
VASS game. A formula can also be written to ensure that \sys wins the
single-sided \pmainlogic[$\top$, $\oblieqlocal$,$\leftarrow$] game iff
the single-sided VASS game being simulated satisfies the parity
condition.

Now we give the details.
Given a single-sided VASS game, we
will make the following assumptions about it without loss of
generality.
\begin{itemize}
    \item The initial state belongs to the environment player (if it
        doesn't, we can add an extra state and a transition to achieve
        this).
    \item The environment and system players strictly alternate (if
        there are transitions between states belonging to the same
        player, we can add a dummy state belonging to the other player
        in between).
    \item The initial counter values are zero (if they aren't, we can
        add extra transitions before the initial state and force the
        system player to get the counter values from zero to the
        required values).
\end{itemize}

\noindent
The formula giving the winning condition of the single-sided \pmainlogic[$\top$,
$\oblieqlocal$, $\leftarrow$] game is made up of the following variables. Suppose $T^e$ and $T^s$
are the sets of environment and systems transitions respectively. For every
transition $t \in T^e$, there is an environment variable $p_t$. We
indicate that the environment player chooses a transition $t$ by
setting $p_t$ to true.
For every transition $t \in T^s$ of the single-sided VASS game, there
is a system variable $t$. There is a
system variable $\curtrs$ to indicate the moves made by the
system
player. We indicate that the system player chooses a
transition $t$ by mapping $t$ and $\curtrs$ to the same data value. For every counter $x$ of the single-sided VASS game, there are
system variables $x$ and $\overline{x}$.

The formula $\phi_{e}$ indicates that the environment player makes
some wrong move and it is the disjunction of the following formulas.
\begin{itemize}
    \item The environment does not choose any transition in some
        round.
        \begin{align*}
            F(~ \bigwedge_{t \in T^{e}} \lnot p_t~)
        \end{align*}
    \item The environment chooses more than one transition in some
        round.
        \begin{align*}
            F (~ \bigvee_{t\ne t' \in T^{e}} ( p_t
            \land p_{t'}) ~)
        \end{align*}
    \item The environment does not start with a transition originating
        from the designated initial state.
        \begin{align*}
            \bigvee_{t \in T^{e}, \text{ origin of } t \text{ is not
            the initial state}} p_t
        \end{align*}
    \item The environment takes some transition that cannot be
        taken after the previous transition by the system player.
        \begin{align*}
            \bigvee_{t \in T^{s}} F(~ t \oblieqlocal \curtrs ~ \land
            \bigwedge_{t' \in T_{s}\setminus \set{t}} \lnot (t'
            \oblieqlocal \curtrs) ~ \land \\ ~ \bigvee_{t' \in
                T^{e},~ t' \text{ can not come after } t}
                \mynext( p_{t'}) ~)
        \end{align*}
\end{itemize}

\noindent
For simulating a counter $x$, we use two variables $x$ and
    $\overline{x}$. The data
    values assigned to these variables from positions $1$ to
    $i$ in a concrete model $\sigma$ will represent the counter value
that is
    equal to the cardinality of the set $\set{d \in \Domain \mid \exists j
    \in \set{1, \ldots, i}, \sigma(j)(x)=d, \forall j' \in \set{1, \ldots, i},
    \sigma(j')(\overline{x}) \ne d}$.  We will use special formulas for
    incrementing, decrementing and retaining previous values of
    counters.
\begin{itemize}
        \item To increment a counter represented by $x, \overline{x}$,
            we force the next data values of $x$ and $\overline{x}$ to
            be new ones that have never appeared before in $x$ or
            $\overline{x}$.
            \begin{align*}
                \phi_{\inc}(x, \overline{x}) = \mynext \lnot
(&~(\toblieqp{x}{x}) ~ \lor~
                (\toblieqp{x}{\overline{x}})~\lor ~
(\toblieqp{\overline{x}}{x})
                ~ \lor~\\ & ~
                (\toblieqp{\overline{x}}{\overline{x}}) ~ \lor ~ (x
\oblieqlocal
                \overline{x})~)
            \end{align*}
        \item To decrement a counter represented by $x,
            \overline{x}$, we force the next position to have a data
            value for $x$ and $\overline{x}$ such that it has appeared
            in the past for $x$ but not for $\overline{x}$.
            \begin{align*}
                \phi_{\dec}(x, \overline{x}) =& \mynext (~ x
\oblieqlocal
                \overline{x} ~ \land~ \toblieqp{x}{x} ~\land ~ \lnot
                (\toblieqp{x}{\overline{x}}) ~)
            \end{align*}
        \item To ensure that a counter represented by $x,
            \overline{x}$ is not changed, we force the next position
            to have a data value for $x$ that has already appeared in
            the past for $x$ and we force the next position to have a
            data value for $\overline{x}$ that has never appeared in
            the past for $x$ or $\overline{x}$.
            \begin{align*}
                \phi_{\nci}(x, \overline{x}) = \mynext(~
                \toblieqp{x}{x} ~ \land ~
                \lnot(\toblieqp{\overline{x}}{x}) ~ \land ~
                \lnot(\toblieqp{\overline{x}}{\overline{x}}) ~)
            \end{align*}
\end{itemize}

\noindent
The formula $\phi_{s}$ indicates that the system player makes all the
right moves and it is the conjunction of the following formulas.
\begin{itemize}
    \item The system player always chooses at least one move.
        \begin{align*}
            G(~ \bigvee_{t \in T^{s}} t \oblieqlocal \curtrs ~)
        \end{align*}
    \item The system player always chooses at most one move.
        \begin{align*}
            G(~ \bigwedge_{t\ne t' \in T^{s}} \lnot(t \oblieqlocal
            \curtrs \land t' \oblieqlocal \curtrs) ~)
        \end{align*}
    \item The system player always chooses a transition that can come
        after the previous transition chosen by the environment.
        \begin{align*}
            \bigwedge_{t \in T^{e}}G(~ p_t ~
            \Rightarrow ~ \bigvee_{t'\in T_{s},~t' \text{ can come
            after } t}t' \oblieqlocal \curtrs  ~)
        \end{align*}
    \item The system player sets the initial counter values to zero.
        \begin{align*}
            \bigwedge_{x \text{ is a counter}}x \oblieqlocal
            \overline{x}
        \end{align*}
    \item The system player updates the counters properly.
        \begin{align*}
            G(~ & \bigwedge_{(q, x++,q')=t \in T^{s}} (~ t
\oblieqlocal
            \curtrs ~\Rightarrow ~ \phi_{\inc}(x, \overline{x})~ \land
            \bigwedge_{x' \ne x} \phi_{\nci}(x',\overline{x'}))\\
            &\bigwedge_{(q,\nop,q')=t\in T^{s}} (~ t\oblieqlocal
            \curtrs ~ \Rightarrow ~ \bigwedge_{x \text{ is a counter}}
            \phi_{\nci}(x,\overline{x}) ~)\\
            & \bigwedge_{(q, x--,q')=t \in T^{s}} (~ t \oblieqlocal
            \curtrs ~\Rightarrow ~ \phi_{\dec}(x, \overline{x})~ \land
            \bigwedge_{x' \ne x} \phi_{\nci}(x',\overline{x'}))~)
        \end{align*}
    \item The maximum colour occurring infinitely often is even.
        \begin{align*}
            \bigvee_{j \text{ is an even colour}} GF(\bigvee_{t\in T^e,
            \text{origin of } t \text{ has colour } j}(p_t) \\ \lor  \bigvee_{t\in T^s,
            \text{origin of } t \text{ has colour } j}(t
\oblieqlocal \curtrs))~ \land\\
             FG(\bigwedge_{t \in T, \text{ origin of } t \text{ has
colour greater than }
        j}\lnot( p_t \lor t \oblieqlocal \curtrs))
        \end{align*}
\end{itemize}

The system player wins if the environment player makes any mistake or
the system player makes all the moves correctly and satisfies the
parity condition. We set the winning condition for the system player
in the single-sided \pmainlogic[$\top$, $\oblieqlocal$, $\leftarrow$] game to be $\phi_{e} \lor
\phi_{s}$. If the system player has a winning strategy in the
single-sided VASS game, the system player simply makes choices in the
single-sided \pmainlogic[$\top$, $\oblieqlocal$, $\leftarrow$] game to
imitate the moves in the single-sided VASS game. Since the resulting
concrete model satisfies $\phi_{e} \lor \phi_{s}$, the system player
wins. Conversely, suppose the system player has a winning strategy in
the single-sided \pmainlogic[$\top$, $\oblieqlocal$, $\leftarrow$] game. In
the case where the environment does not make any mistake, the system
player has to choose data values such that the simulated sequence of
states of the VASS satisfy the parity condition.  Hence, the system
player in the single-sided VASS game can follow the strategy of the
system player in the single-sided \pmainlogic[$\top$, $\oblieqlocal$,
$\leftarrow$] game and irrespective of how the environment player plays, the
system player wins.
\end{proof}


\section{Single-sided \pmainlogic[\texorpdfstring{$\top$, $\oblieqlocal$, $\rightarrow$}{⟙,≈,→}] is undecidable}%
\label{undec-singlesided-plrvgame}

In this section we show that the positive decidability result for the single-sided \pmainlogic[$\top, \leftarrow$] game cannot be replicated for the future demands fragment, even in a restricted setting.
\begin{thm}\label{thm:undec-singlesided-plrvgame}
The winning strategy existence problem for single-sided
\pmainlogic$[\top, \oblieqlocal, \rightarrow]$  games is undecidable,
even when the formula giving the winning condition uses one Boolean
variable belonging to \env and three data variables belonging to \sys.
\end{thm}
We don't know the decidability status for the case where the formula
uses less than three data variables belonging to \sys.

\review{First we explain why the technique used for proving decidability in
Section~\ref{single-past-decidable} cannot be applied here. In
Section~\ref{single-past-decidable}, atomic constraints can only test
if a current data value appeared in the past. At any point of a game,
\sys can satisfy such an atomic constraint by looking at data values
that have appeared in the past and assigning such a data value to some
variable in the current position of the game. However, this cannot be
done when atomic constraints can refer to repetitions in the future
--- if \sys decides to satisfy such an atomic constraint at some point
in the game, then \sys will be obligated to repeat a data value at
some point in the future. The opponent \env can prevent this, if the
formula specifying the winning condition for \sys is cleverly set up
so that as soon as \sys commits itself to repeating some value in the
future, it will \emph{not} be able to make the repetition. Indeed, in
this section, we use such formulas to force \sys to faithfully simulate a
2-counter machine --- the formulas are set up so that if \sys makes a
mistake in the simulation, he will have to commit to repeating some
value in the future, and \env will not let the repetition happen, thus
defeating \sys.}

As in the previous undecidability results in
Section~\ref{undec-plrvgame}, \review{Theorem~\ref{thm:undec-singlesided-plrvgame}} is proven by a reduction
from the reachability problem for 2-counter machines. \Sys
makes use of \emph{labels} to encode the sequence of transitions of a
witnessing run of the counter machine. This time, \sys uses 3 data
variables $x, y, z$ (in addition to a number of Boolean variables
which encode the labels); and \env uses just one Boolean variable $b$.
Variables $x$, $y$ are used to encode the counters $\countx$ and
$\county$ as before, and variables $z$, $b$ are used to ensure that there are no `illegal' transitions --- namely, no decrements of a zero-valued counter, and no tests for zero for a non-zero-valued counter.

Each transition in the run of the 2-counter machine will be
encoded using \emph{two} consecutive positions of the game. Concretely, while in the previous coding of Section~\ref{undec-plrvgame} a witnessing reachability run $t_1 \, t_2 \dotsb t_n \in \delta^*$ was encoded with the label sequence $\textit{begin}\, t_1  \, t_2  \dotsb t_n  \hat t^\omega$,  in this encoding  transitions are interspersed with a special $\textit{bis}$ label, and thus the run is encoded as $t_1 \textit{bis} \,t_2 \,\textit{bis} \dotsb t_n \, \textit{bis} \, {(\hat t \, \textit{bis})}^\omega \in {(\delta \cup \set{\textit{bis}})}^\omega$.

Suppose a position has the label of a $\countx++$ \review{instruction} and the
variable $x$ has the data value $d$. Our encoding will ensure that if
the data value $d$ repeats in the future, it will be only once and at
a position that has the label of a $\countx--$ \review{instruction}. A
symmetrical property holds for $\county$ and variable $y$.
The value of counter $\countx$ (resp.\ $\county$) before the $i^{\text{th}}$ transition (encoded in the $2i^{\textit{th}}$ and ${(2i+1)}^{\textit{st}}$ positions) is
the number of positions $j < 2i$ satisfying the following two
conditions: (i) the position $j$ should have the label of a $\countx++$
\review{instruction} and (ii) $\sigma(j)(x) \not\in\set{\sigma(j')(x) \mid
j+1<j'<2i}$. Intuitively, if $2i$ is the current position, the value
of $\countx$ (resp.\ $\county$) is the number of previous positions
that have the label of a $\countx++$ \review{instruction} whose data value is
not yet matched by a position with the label of a $\countx--$
\review{instruction}. In this reduction we
assume that \sys plays first and \env plays next at each round, since it
is easier to understand (the reduction also holds for the game where turns are inverted by shifting
\env behavior by one position
). At each
round, \sys will play a label \emph{bis} if the last label played was
an \review{instruction}. Otherwise, she will choose the next transition of the
2-counter machine to simulate and she will chose the values for
variables $x$, $y$, $z$ in such a way that the aforementioned encoding for counters $\countx$ and $\county$ is preserved. To this end, \sys is bound by the following set of rules, described here pictorially:
\begin{center}
  \includegraphics[width=.7\textwidth]{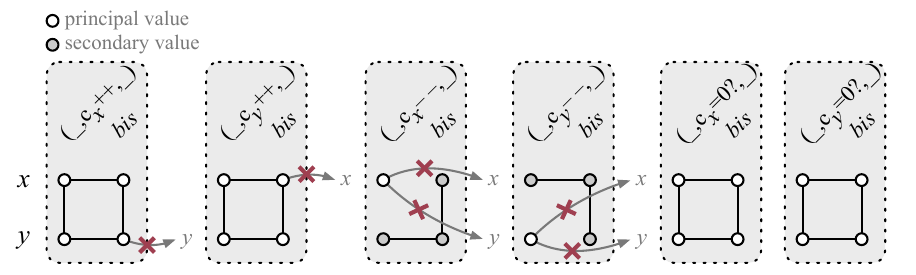}
\end{center}
The first (leftmost) rule, for example, reads that whenever there is a
$\countx++$ transition label, then all four values  for $x$ and $y$ in
both positions (\ie, the \review{instruction} position and the next \textit{bis}
position) must have the same data value $d$ (which
we call `principal'), which  does not occur in the future
under variable $y$. The third rule says that $\countx--$ is encoded by having $x$ on the first
position to carry the `principal' data value $d$ of the \review{instruction},
which is \emph{final} (that is, it is not repeated in the future under
$x$ or $y$), and all three remaining positions have the same data
value $d'$ different from $d$.  In this way, \sys can make sure that
the value of $\countx$ is decremented, by playing a data value $d$
that has occurred in a $\countx++$ position that is not yet matched.
(While \sys could also play some data value which does not match any
previous $\countx++$ position, this `illegal' move leads to a
losing play for \sys, as we will show.)
\review{In this rule, the fact that one transition of the $2$-counter
    machine is encoded using two positions of the game is used to ensure that the
data value $d'$ of $y$ (for which $d' \neq d$) appears in the future
both in $x$ and $y$.} Thus, the presence of $d'$ doesn't affect the
value of $\county$ or $\countx$ ---to affect either, the data value
should repeat in only one variable.
\review{If we do not force $d'$ to repeat in both variables in the
future, this position can potentially be treated as an increment for
$c_y$. Using two positions per transitions is a simple way of
preventing this.}

From these rules, it follows that every $c_k++$ can be matched to at
most one future $c_k--$. However, there can be two ways in which this
coding can fail: (a) there could be invalid tests $c_k=0?$, that is, a 
situation in which the preceding positions of the test contain a
$c_k++$ \review{instruction} which is not matched with a $c_k--$ \review{instruction}; and
(b) there could  be some $c_k--$ with no previous matching $c_k++$. As
we will see next, variables $z$ and $b$ play a crucial role in the
game whenever any of these two cases occurs. In all the rounds, \env
always plays $\top$, except if he detects that one of these two
situations, (a) or (b), have arisen, in which case he plays $\bot$. In
the following rounds \sys plays a value in $z$ that will enable to test, with an \pmainlogic{} formula, if there was indeed an (a) or (b) situation, in which case \sys will lose, or if \env was just `bluffing', in which case \sys will win. Since this is the most delicate point in the reduction, we dedicate the remaining of this section to the explanation of how these two situations (a) and (b) are treated. 

Remember that \env has just \emph{one bit} of information to play with. The \pmainlogic{} property we build ensures that the sequence of $b$-values must be from the set $\top^*\bot^*\top^\omega$.

\begin{figure}
  \centering
  \includegraphics[width=.57\textwidth]{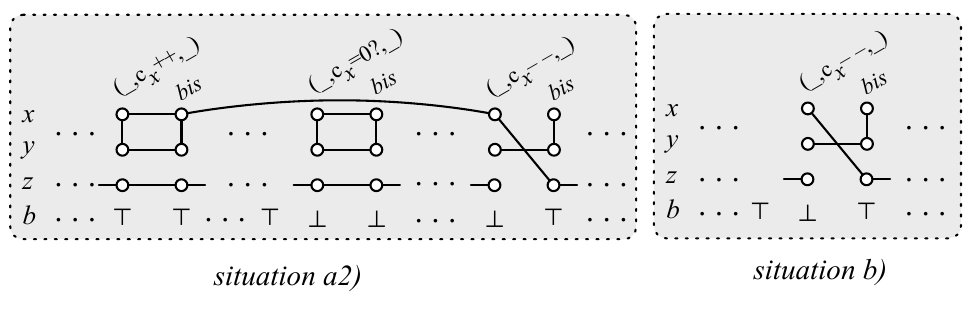}
  \caption{Depiction of best strategies in both situations.}%
\label{fig:situation-ab}
\end{figure}

\paragraph{(a) Avoiding illegal tests for zero.}
Suppose that at some point of the $2$-counter machine simulation, \sys
decides to play a $c_k=0?$ \review{instruction}. Suppose there is some preceding 
$c_k++$ \review{instruction} for which either: (a1) there is no matching $c_k--$
\review{instruction}; or (a2) there is a matching $c_{k}--$ \review{instruction} but it
occurs after the $c_{k}=0?$ \review{instruction}. Situation a1 can be easily 
avoided by ensuring that any winning play must satisfy the formula
$
\mu = \always (\tau_{(c_k++)} \land \future \tau_{(c_k=0?)} \Rightarrow \toblieq{k}{k})
$
for every $k \in \set{x,y}$. Here, $\tau_{inst}$ tests if the current position is labelled with an instruction of type $inst$.
On the other hand, Situation a2 requires \env to play a certain
strategy (represented in Figure~\ref{fig:situation-ab}-a2).
This means that $c_k$ is non-zero at the position of the $c_k=0?$ 
\review{instruction}, and that this is an illegal transition; thus, \env must
respond accordingly. Further, suppose this is the \emph{first} illegal
transition that has occurred so far. \Env, who so far has been playing
only $\top$, decides to play $\bot$ to mark the cheating point.
Further, he will continue playing $\bot$ until the matching $c_k--$
\review{instruction} is reached (if it is never reached, it is situation a1
and \sys loses as explained before), after which he will play $\top$
forever afterwards. In some sense, \env provides a \emph{link} between the illegal transition and the proof of its illegality through a $\bot^*$-path. The following characterizes \env{}'s denouncement of an illegal test for zero:

\vspace{1mm}

\noindent
\textit{Property 1:} $b$ becomes $\bot$ at a $c_k=0?$ position and stops being $\bot$ at a $c_k--$ position thereafter. 

\vspace{1mm}
Note that Property 1 is clearly definable by a formula $\pi_1$ of
\pmainlogic$[\top,\oblieqlocal,\rightarrow]$.  If Property 1 holds, a
formula $\phi_1$ can constrain \sys to play $z$ according to the
following: $z$ always carries the same data value, distinct from the values of all other variables, but  as soon as the last $\bot$ value is played, which has to be on a $c_k--$ position, the value of $z$ changes and holds the principal value of that $c_k--$ \review{instruction},\footnote{To make sure it is the \emph{last} $\bot$ element, \sys has to wait for $\top$ to appear, hence variable $z$ changes its value at the next position after the last $\bot$.} and it continues to hold that value forever after (\cf~Figure\ \ref{fig:situation-ab}-$a2$).
%
Further, if \env cheated in his denouncement by linking a $c_k=0?$ \review{instruction} with a future $c_k--$ with a matching $c_k++$ that falls in-between the test for zero and the decrement, then a property $\pi'_1$ can catch this: there exists a $c_k++$ with  $\bot$ whose principal value matches that of a future $z$-value. 

Finally, assuming \env correctly denounced an illegal test for zero and system played accordingly on variable $z$, a property $\phi'_1$ can test that \env exposed an illegal transition, by testing that there exists a $c_k++$ \review{instruction} whose principal value corresponds to the $z$-value of some future position. Thus, the encoding for this situation is expressed with the formula $\psi_1 = \mu \land ((\pi_1 \land \lnot \pi'_1) \Rightarrow (\phi_1 \land \lnot \phi'_1))$.

\paragraph{(b) Avoiding illegal decrements.}
Suppose now that at some point of the $2$-counter machine simulation, \sys decides to play a $c_k--$ \review{instruction} for which there is no preceding $c_k++$ \review{instruction} matching its final data value. This is a form of cheating, and thus \env should respond accordingly. Further, suppose this is the first cheating that has occurred so far. \Env, who so far has been playing only $\top$, decides then to mark this position with $\bot$; and for the remaining of the play \env plays only $\top$ (even if more illegal transitions are performed in the sequel). Summing up, for this situation \env{}'s best strategy has a value sequence from $\top^*\bot\top^\omega$, and this property characterizes \env{}'s denouncement of an illegal decrement (\cf~Figure~\ref{fig:situation-ab}-$b$).

\vspace{1mm}

\noindent
\textit{Property 2:} $b$ becomes $\bot$ at a $c_k--$ position and stops being $\bot$ immediately after.

\vspace{1mm}

A formula $\pi_2$ can test Property 2; and a formula $\phi_2$ can
constrain variable $z$ to always have the same data value ---distinct
from all other data values played on variables $x,y$--- while $b$
contains $\top$ values; and as soon as $b$ turns to $\bot$ on a
$c_k--$ position, then $z$ at the next position takes the value of the current variable $k$, and maintains that value (\cf~Figure~\ref{fig:situation-ab}-$b$).
Further, a formula $\phi'_2$ tests that in this case there must be some $c_k++$ position with a data value equal to variable $z$ of a future position. The formula corresponding to this case is then $\psi_2 = \pi_2 \Rightarrow \phi_2 \land \phi'_2$.

\smallskip

The final formula to test is then of the form $\phi = \phi_{\textit{lab}} \land \phi_{x,y} \land  \psi_1 \land \psi_2$, where $\phi_{\textit{lab}}$ ensures the finite-automata behavior of labels, and in particular that a final state can be reached, and $\phi_{x,y}$ asserts the correct behavior of the variables $x,y$ relative to the labels. It follows that \sys has a winning strategy for the game with input $\phi$ if, and only if, there is a positive answer to the reachability problem for the 2-counter machine.
 Finally, labels can be eliminated by means of special data values
 encoding \emph{blocks} exactly as done in Section~\ref{sec:nobooleanvars}, and in  this way Theorem~\ref{thm:undec-singlesided-plrvgame} follows.

 \begin{proof}[{Proof of Theorem~\ref{thm:undec-singlesided-plrvgame}}]

We briefly discuss why the properties $\phi_{\textit{lab}}$,
$\phi_{x,y}$, $\psi_1$ and $\psi_2$ can be described in
\pmainlogic$[\top, \oblieqlocal, \review{\rightarrow}]$.

Encoding $\phi_{\textit{lab}}$ using some Boolean variables belonging to \sys is easy since it does not involve the use of data values.

The formula $\phi_{x,y}$ can be encoded as $\always(\bigwedge_{a \in A} \tau_{a} \Rightarrow \zeta_{a})$ for $A = \set{\countx++, \countx--, \countx = 0?, \county++, \county--, \county = 0?}$ and
\[\tau_a = \bigvee_{q,q' \in Q, (q,a,q') \in \delta} \lambda_{(q,a,q')},\] where $\lambda_{(q,a,q')}$ tests that we are standing on a position labelled with \review{instruction} $(q,a,q')$ (in particular not a \emph{bis} position). Finally, $\zeta_a$ encodes the rules as already described. That is,
\begin{align*}
  \zeta_{\countx++} = {}&x \oblieqlocal y \land x \oblieqlocal \mynext x \land y
  \oblieqlocal \mynext y \land \lnot \mynext(\toblieq{x}{y} ),\\
  \zeta_{\countx--} = {}&\lnot x \oblieqlocal y \land \lnot \toblieq{x}{x} \land \lnot \toblieq{x}{y} \land  y \oblieqlocal \mynext y \land y \oblieqlocal \mynext x,\\
  \zeta_{\countx=0?} = {}&x \oblieqlocal y \land x \oblieqlocal \mynext x \land y \oblieqlocal \mynext y,
\end{align*}
and similarly for the rules on $\county$.

The formula $\psi_1 = \mu \land ( (\pi_1 \land \lnot \pi'_1) \Rightarrow (\phi_1 \land \lnot \phi'_1))$ is actually composed of two conjunctions $\psi_1 = \psi_1^x \land \psi_1^y$, one for $k=x$ and another for $k=y$, let us first suppose that $k=x$. Then,
\begin{itemize}
\item $\pi_1$, which checks Property 1, which is simply
\[
\pi_1 = b \until (\lnot b \land \tau_{\countx = 0?} \land \mynext(\lnot b \until \tau_{\countx--}))
\]
\item $\pi'_1$, expresses that there exists a $\countx ++$ with $\lnot b$ with value matching that of a future $z$-value:
\[
\pi'_1 = \future(\tau_{\countx++} \land \lnot b \land \toblieq{x}{z})
\]
\item $\phi_1$, on the other hand, checks that $z$ carries always the same data value, disjoint from the values of all other variables, but  as soon as the last $\bot$ value is played the value of $z$ in the next position changes and holds now the $x$ value of that position, and it continues to hold it forever:
  \begin{align*}
\phi_1 = {} &\lnot (\toblieq{z}{x} \lor \toblieq{z}{y}) \land
(z \oblieqlocal \mynext z )\until \big(\lnot(z \oblieqlocal \mynext z) \land\lnot b \land \mynext b  \land  x \oblieqlocal \mynext z \land \mynext\always(z \oblieqlocal \mynext z)\big)
  \end{align*}
\item finally, $\phi'_1$  tests there exists a $\countx ++$ \review{instruction} whose principal value corresponds to the $z$-value of some future position:
  \begin{align*}
    \phi'_1 &= \future (\tau_{\countx++} \land \toblieq{x}{z}).
  \end{align*}
\end{itemize}

\noindent
The formula $\psi_2 = \pi_2 \Rightarrow \phi_2 \land \phi'_2$ is also composed of two conjuncts, one for $k=x$ and one for $k=y$, let us only show the case $k=x$. Then,
\begin{itemize}
\item $\pi_2$ checks Property 2:
\[
\pi_2 = b \until (\lnot b \land \tau_{\countx--} \land \mynext(\always b))
\]
\item $\phi_2$ checks  that as soon as $b$ turns to $\bot$ then $z$ at the next position takes the value as current variable $x$, and maintains that value:
  \begin{align*}
\phi_2 = {} &\lnot (\toblieq{z}{x} \lor \toblieq{z}{y}) \land (z \oblieqlocal \mynext z )\until \big(\lnot(z \oblieqlocal \mynext z) \land \lnot b  \land  x \oblieqlocal \mynext z \land \mynext\always(z \oblieqlocal \mynext z)\big)
  \end{align*}
\item finally, $\phi'_2$  tests that in this situation there must be some $\countx++$ position with a data value equal to variable $z$ of a future position:
  \begin{align*}
    \phi'_2 &= \future (\tau_{\countx++} \land \toblieq{x}{z}).
  \end{align*}
\end{itemize}

\subsection*{Correctness.}
Suppose first that the $2$-counter machine has an accepting run of the form $(q_0, I_1,q_1) \dotsb (q_{n-1},I_n,q_n)$ with $q_n=q_f$. \Sys{}'s strategy is then to play (the encoding of) the labels
\[(q_0, I_1,q_1) \, \textit{bis}\, \dotsb (q_{n-1},I_n,q_n) \, \textit{bis}\, {(\hat t \, \textit{bis})}^\omega.\]
In this way, the formula $\phi_{\textit{lab}}$ holds.

With respect to the data values on $x,y$, \sys will respect the rules depicted in Section~\ref{undec-singlesided-plrvgame}, making $\phi_{x,y}$ true.

Finally, \sys will play a data value in $z$ that at the beginning will be some data value which is not used on variables $x$ nor $y$. She will keep this data value all the time, but keeping an eye on the value of $b$ that is being played by \env. If \env plays a first $\bot$ on a $c_k--$ \review{instruction}, \sys will then play on $z$, at the next round, the data value of variable $k$ at this round. If \env plays a first $\bot$ at a $c_k=0?$ \review{instruction} and a last $\bot$ at a $c_k--$ \review{instruction}, again \sys will change the value of $z$ to have the principal value of the $c_k--$ \review{instruction}. In this way, \sys is sure to make true the formula $\phi_1 \land \lnot \phi_1'$ in one case, and formula $\phi_2 \land \phi'_2$ in the other case. All other cases for $b$ are going to be winning situations for \sys due to the preconditions $\pi_1 \land \lnot \pi'_1$ and $\pi_2$ in the formulas $\psi_1$ and $\psi_2$. 

\smallskip

On the other hand, if there is no accepting run for the $2$-counter
machine, then each play of \sys on variables $x,y$ and the variables
verifying both $\phi_{\textit{lab}}$ and $\phi_{x,y}$ must have an
illegal transition of type (a) or (b). At the first illegal transition
\env will play $\bot$. If it is an  illegal transition with the
\review{instruction} $c_k--$, then \env will continue playing $\top$
in subsequent positions; if it is an illegal transition with the
instruction $c_k=0?$, then \env will keep playing $\bot$ until the corresponding $c_k--$ matching to a witnessing $c_k++$ played before the $c_k=0?$ \review{instruction} is reached. In either of these situations the antecedent of $\psi_1$ or $\psi_2$ will be true while the consequent will be false; and thus the final formula will not hold, making \sys incapable of finding a winning strategy. 

\smallskip

Finally, let us explain further how this reduction can be turned into a reduction for the game in which \env plays first and \sys plays second at each round. For the final formula $\phi$ of the reduction, let $\phi'$ be the formula in which \env conditions are shifted one step to the right. This is simply done by replacing every sub-expression of the form $\mynext^i b$ with $\mynext^{i+1}b$. It follows that if \env starts playing $\top$ and then continues the play reacting to \sys strategy in the same way as before, \sys will have no winning strategy if, and only if, \sys had no winning strategy in the game where the turns are inverted.
 \end{proof}

 \review{Now we explain why the technique used to prove undecidability
 in this section cannot be used in
 Section~\ref{single-past-decidable}. The crucial dependency on atomic
 constraints checking for repetition in the future occurs in avoiding
 illegal tests for zero and avoiding illegal decrements. For avoiding
 either of the errors, we let \env win by using atomic constraints
 that specify that the data value in the error position repeats in the
 \emph{future}. Without the ability to test for repetition of values
 in the future, such a strategy for catching errors in simulation will
 not work. Indeed, the decidability result of
 Section~\ref{single-past-decidable} implies that with atomic
 constraints that can only test repetition of values in the past in
 \sys variables, counter machines cannot be simulated.}


\section{Single-sided \pmainlogic[\texorpdfstring{$\langle \previous, \since\rangle$, $\oblieqlocal$, $\leftarrow$}{(X⁻¹,S),≈,←}] is decidable}%
\label{dec-singlesided-nestedpast}
In this section, we prove that if we restrict nested formulas to use
only past temporal operators and only allow past obligations, then
single sided games are decidable. We enrich symbolic models used in
Section~\ref{single-past-decidable} with information related to nested
formulas and reduce the winning strategy existence problem to the
same problem in single-sided VASS games.

\review{The main idea is same as the one used in
Section~\ref{single-past-decidable}. We use symbolic models to ignore
the special semantics of constraints like $\toblieqp{x}{y}$ and treat
such constraints like Boolean atomic propositions, whose truth value
doesn't depend on past positions. However, now we can have constraints
of the form $\oblieqp{x}{y}{\psi}$, which says that the current data
value assigned to $x$ should have been assigned to $y$ in some past
position, and that past position should satisfy the formula $\psi$.
Instead of tracking the number of points of decrement for the set
$\set{y}$, we now track the number of points of decrement for the set
of pairs $\set{(y,\psi)}$. We formalize this in the next two
paragraphs.}

For the rest of this section, we fix a \pmainlogic[$\langle \previous,
\since\rangle$, $\oblieqlocal$, $\leftarrow$] formula $\phi$. Let
$\bvars^{\phi} \subseteq \bvars$ and $\dvars^{\phi} \subseteq \dvars$
be the set of Boolean and data variables used in $\phi$. Let $\Phi$ be
the set $\set{\psi \mid \exists x,y \in \dvars^{\phi} \text{ s.t.~}
\oblieqp{x}{y}{\psi} \text{ is a sub-formula in } \phi} \cup
\set{\top}$. Intuitively, $\Phi$ is the set of formulas used as nested
formulas inside $\phi$. We will need to keep track of positions in the
past where a data value repeats and the formulas in $\Phi$ that are
satisfied at those positions. Formally, this is done by a
\emph{repetition history}, which is a subset of
$\pwrset(\dvars^{\phi}) \times \pwrset(\Phi)$. Intuitively, every
element of a repetition history corresponds to one position in the
past where a data value repeats.  For example, the repetition history
$\set{(\set{x,y}, \set{\psi_{1}, \psi_{2}}), (\set{y,z},
\set{\psi_{2}, \psi_{3}})}$ indicates that a data value
repeats at two positions in the past. The first position satisfies
$\psi_{1}$ and $\psi_{2}$ and at this position, the data value is
assigned to variables $x$ and $y$. The second position satisfies
$\psi_{2}$ and $\psi_{3}$ and at this position, the data
value is assigned to variables $y$ and $z$. We denote the set of
all repetition histories by $\RH$.

We will also need to keep track of the variables in which a data value
is required to be repeated, and the nested formulas that need to be
satisfied at the positions where the repetitions happen. Formally, this is
done by a \emph{past obligation}, which is a subset of
$\dvars^{\phi} \times \Phi$. For example, the past obligation
$\set{(y,\psi_{1}), (x,\psi_{2}), (x,\psi_{1}), (y, \psi_{2}), (y,
\psi_{3}), (z,\psi_{3}), (z,\psi_{2})}$
indicates that a data value needs to be repeated in the past (i) in
 variable $y$ at a position that satisfies $\psi_{1}$, (ii) in
variable $x$ at a position that satisfies $\psi_{2}$, (iii) in
variable $x$ at a position that satisfies $\psi_{1}$, (iv) in
variable $y$ at a position that satisfies $\psi_{2}$, (v) in
variable $y$ at a position that satisfies $\psi_{3}$, (vi) in
variable $z$ at a position that satisfies $\psi_{3}$ and (vii) in
variable $z$ at a position that satisfies $\psi_{2}$.

\review{A repetition history keeps track of past postions where a data
value appeared and the nested formulas that are satisfied in those
past positions. A past obligation keeps track of the variables where a
data value needs to be repeated in the past, and the nested formulas
that need to be satisfied in those past positions. If we want to make
use of a repetition history to satisfy the requirements contained in a
past obligation, we need to check that all the variables that are
specified by the past obligation are covered by the repetition history
and all the nested formulas are satisfied in the past positions. This
is formalized in the next paragraph.}

A repetition history $\rh$ \emph{matches} a past obligation $\po$ if
there is a function $m:\po \to \rh$ satisfying the following
conditions:
\begin{itemize}
    \item For any $x\in \dvars^{\phi}$ and $\psi\in \Phi$, if
        $(x,\psi) \in \po$ and $m((x,\psi ))=(V',\Phi') \in \rh$, then
        $x \in V'$ and $\psi \in \Phi'$.
    \item\label{matching-second-condition} For every
        $(V',\Phi')\in\rh$, for every $x \in V'$ and $\psi \in \Phi'$,
        $(x,\psi) \in \po$.
\end{itemize}
Intuitively, in the first condition above, the element $(V',\Phi')$ of
the repetition history $\rh$ denotes a position in the past where a
data value is assigned to all the variables in $V'$ and that the
position satisfies all the formulas in $\Phi'$. The conditions $x \in
V'$ and $\psi \in \Phi'$ then ensure that the data value indeed
repeats in the past in variable $x$ at a position that satisfies the
formula $\psi$. The second condition above ensures that all variables
that appear in the repetition history $\rh$ are utilized and none of
them are wasted. For example, the repetition history $\set{(\set{x,y},
\set{\psi_{1}, \psi_{2}}), (\set{y,z}, \set{\psi_{2}, \psi_{3}
})}$ matches the past obligation $\set{(y,\psi_{1}),
(x,\psi_{2}), (x,\psi_{1}), (y,\psi_{2}), (y, \psi_{3}), (z,\psi_{3}), (z,\psi_{2})}$ by setting
$m:(y,\psi_{1}) \mapsto (\set{x,y}, \set{\psi_{1}, \psi_{2}})$,
$m:(x,\psi_{2}) \mapsto (\set{x,y}, \set{\psi_{1}, \psi_{2}})$,
$m:(x,\psi_{1}) \mapsto (\set{x,y}, \set{\psi_{1}, \psi_{2}})$,
$m:(y,\psi_{2}) \mapsto (\set{x,y}, \set{\psi_{1}, \psi_{2}})$,
$m:(y,\psi_{3}) \mapsto (\set{y,z}, \set{\psi_{2}, \psi_{3}
})$, $m:(z,\psi_{3})\mapsto (\set{y,z}, \set{\psi_{2}, \psi_{3}
})$ and $m: (z,\psi_{2}) \mapsto (\set{y,z}, \set{\psi_{2}, \psi_{3}
})$.

Let $\closure(\Phi)$ be the smallest set that contains $\Phi$ and
satisfies the following conditions:
\begin{itemize}
    \item If $\psi_{1} \land \psi_{2} \in \closure(\Phi)$, then
        $\psi_{1}, \psi_{2} \in \closure(\Phi)$,
    \item if $\lnot \psi \in \closure(\Phi)$, then $\psi \in
        \closure(\Phi)$,
    \item if $\previous \psi \in \closure(\Phi)$, then $\psi \in
        \closure(\phi)$,
    \item if $\psi_{1} \since \psi_{2} \in \closure(\Phi)$, then
        $\psi_{1}, \psi_{2} \in \closure(\Phi)$ and
    \item if $\psi \in \closure(\Phi)$, then $\lnot \psi \in
        \closure(\Phi)$, where we identify $\lnot \lnot \psi$ with
        $\psi$.
\end{itemize}

\noindent
Let $l$ be the $\mynext$-length of $\phi$ and $\Omega_{l}^{\phi}$ be
the set of constraints of the form $\mynext^{i} \top$, $\mynext^{i}q$, $\mynext^{i}x
\oblieqlocal \mynext^{j}y$ or $\mynext^{i}(\oblieqp{x}{y}{\psi})$,
where $q \in \bvars^{\phi}$, $x,y \in \dvars^{\phi}$, $i,j \in
\set{0,\ldots, l}$ and $\psi \in \Phi$. Intuitively,
$\Omega_{l}^{\phi}$ contains atomic constraints that are potentially
satisfied at positions of a model, while $\Phi$ contains formulas that
use Boolean and/or temporal operators. \review{We will use concepts from the
classical B\"{u}chi automaton construction from propositional LTL
formulas. The conditions in the next paragraph are analogous to
conditions on atoms in the classical B\"{u}chi automaton construction,
adapted for the past fragment of LTL.}

A set $\Phi_{1} \subseteq \closure(\Phi)$ is said to be
\emph{Boolean consistent} if the following conditions are satisfied:
\begin{itemize}
    \item $\top \in \Phi_{1}$.
    \item For every $\psi_{1} \land \psi_{2} \in \closure(\Phi)$,
        $\psi_{1} \land \psi_{2} \in \Phi_{1}$ iff $\psi_{1}, \psi_{2} \in
        \Phi_{1}$.
    \item For every $\lnot \psi_{1} \in \closure(\Phi)$,
        $\lnot \psi \in \Phi_{1}$ iff $\psi_{1} \notin \Phi_{1}$.
\end{itemize}

\noindent
\review{The following condition is analogous to the conditions used in the
classical B\"{u}chi automaton construction to determine when there is
a transition between two atoms.} Two sets $\Phi_{1}, \Phi_{2} \subseteq \closure(\Phi)$ are said to
be \emph{one step consistent} if the following conditions are
satisfied:
\begin{itemize}
    \item For every $\previous \psi_{1} \in \closure(\Phi)$,
        $\previous \psi_{1} \in \Phi_{2}$ iff $\psi_{1} \in \Phi_{1}$.
    \item For every $\psi_{1} \since \psi_{2} \in \closure(\Phi)$,
        $\psi_{1} \since \psi_{2} \in \Phi_{2}$ iff either $\psi_{2} \in \Phi_{2}$ or
        ($\psi_{1} \in \Phi_{2}$ and $\psi_{1} \since \psi_{2} \in
        \Phi_{1}$).
\end{itemize}

\noindent
\review{We will later use sets such as $\Phi_{1},\Phi_{2}$ above for
the same purpose atoms are used in the classical B\"{u}chi automaton
construction.} A set $\Phi_{1} \subseteq
\closure(\Phi)$ is said to
be \emph{initially consistent} if the following conditions are
satisfied:
\begin{itemize}
    \item The set $\Phi_{1}$ is Boolean consistent.
    \item For every $\previous \psi_{1} \in \closure(\Phi)$,
        $\previous \psi_{1} \notin \Phi_{1}$.
    \item For every $\psi_{1} \since \psi_{2} \in \closure(\Phi)$,
        $\psi_{1} \since \psi_{2} \in \Phi_{1}$ iff $\psi_{2} \in \Phi_{1}$.
\end{itemize}

For any numbers $n_{1}, n_{2} \in \Int$, let $[n_{1}, n_{2}]$ denote the
set $\set{n_{1}, \ldots, n_{2}}$. We now extend the definition of
frames to include information about nested formulas and consistency
among them. Recall that $\RH$ is the set of all repetition histories.
\review{As before, a frame will contain a subset of
$\Omega_{l}^{\phi}$. Additionally, the frame will specify, for every
position of the frame, the formulas in $\Phi$ that are satisfied at
that position. This will help us identify which nested formulas
($\Phi$ is the set of nested formulas) are satisfied in each
position of a symbolic model. In addition, the frame will specify, for
every variable $x$ in $\dvars^{\phi}$ and every position $i$ of the frame, a
repetition history. This will be the repetition history that should be
used to match the past obligation of the variable $x$ in the position
$i$.}

For $e \in [0,l]$, an $(e,\phi)$-frame $\symbval$ is a triple
$(\Omega_{\symbval}, \Phi_{\symbval}, H_{\symbval})$ where
$\Omega_{\symbval} \subseteq \Omega_{l}^{\phi}$, $\Phi_{\symbval} :
[0,e] \to \pwrset(\closure(\Phi))$ and $H_{\symbval}:\dvars^{\phi}
\times [0,e] \to \RH$ satisfying the following conditions:
\begin{enumerate}[align=left]
    \item[(F0)] For all constraints $\nxt^{i} q, \nxt^{i}x
\oblieqlocal \nxt^{j}y, \nxt^{i}(\oblieqp{x}{y}{\psi}) \in \Omega_{\symbval}$,
$i,j \in [0,e]$.
    \item[(F1)] For all $i \in [0,e]$ and $x \in
        \dvars^{\phi}$, $\nxt^{i}x\oblieqlocal \nxt^{i}x \in \Omega_{\symbval}$.
    \item[(F2)] For all $i,j \in [0,e]$ and $x,y \in
        \dvars^{\phi}$, $\nxt^{i}x \oblieqlocal \nxt^{j}y \in \Omega_{\symbval}$
        iff $\nxt^{j}y \oblieqlocal \nxt^{i}x \in \Omega_{\symbval}$.
    \item[(F3)] For all $i,j,j' \in [0,e]$ and $x,y,z\in
        \dvars^{\phi}$, if $\set{\nxt^{i}x \oblieqlocal \nxt^{j}y,
        \nxt^{j}y\oblieqlocal \nxt^{j'}z} \subseteq \Omega_{\symbval}$,
        then $\nxt^{i}x \oblieqlocal \nxt^{j'}z \in \Omega_{\symbval}$.
    \item[(F4)] For all $i,j\in [0,e]$ and $x,y \in
      \dvars^{\phi}$ such that
      $\nxt^{i}x \oblieqlocal \nxt^{j}y \in \Omega_{\symbval}$:
      \begin{itemize}
        \item If $i=j$, then for every $z \in \dvars^{\phi}$ and every $\psi \in \Phi$, we have
          $\nxt^{i}(\oblieqp{x}{z}{\psi}) \in \Omega_{\symbval}$ iff
          $\nxt^{j}(\oblieqp{y}{z}{\psi}) \in \Omega_{\symbval}$.
        \item If $i < j$, then $\nxt^{j}(\oblieqp{y}{x}{\psi}) \in
            \Omega_{\symbval} \ \forall \psi \in
            \Phi_{\symbval}(i) \cap \Phi$ and
          for every $z \in \dvars^{\phi}$ and every $\psi' \in \Phi$,
          $\nxt^{j}(\oblieqp{y}{z}{\psi'}) \in \Omega_{\symbval}$ iff either
          $\nxt^{i}(\oblieqp{x}{z}{\psi'}) \in \Omega_{\symbval}$ or there exists
          $i \le j' < j$ with $\nxt^{j}y \oblieqlocal
          \nxt^{j'}z \in \Omega_{\symbval}$ and $\psi' \in
          \Phi_{\symbval}(j')\cap \Phi$.
      \end{itemize}
  \item[(F5)] For all $i,j\in [0,e]$ and $x,y \in
        \dvars^{\phi}$ such that
        $\nxt^{i}x \oblieqlocal \nxt^{j}y \in \Omega_{\symbval}$,
        \begin{itemize}
            \item If $i=j$, then $H_{\symbval}(x,i) = H_{\symbval}(y,j)$.
          \item If $i < j$, and there is no $j'$ such that $i < j' < j$ satisfying
            $\nxt^{i}x \oblieqlocal \nxt^{j'}z \in \Omega_{\symbval}$ for any $z$, then
            $H_{\symbval}(y,j) = H_{\symbval}(x,i) \cup
            \{({[(x,i)]}_{\symbval}, \Phi_{\symbval}(i)\cap\Phi)\}$,
            where ${[(x,i)]}_{\symbval} = \set{z \in \dvars^{\phi} \mid
            \nxt^{i}x \oblieqlocal \nxt^{i}z \in \Omega_{\symbval}}$
            is the \emph{equivalence class of $x$ at level $i$ in
            $\symbval$}.
        \end{itemize}
    \item[(F6)] For all $i \in [0,e]$ and for all $x \in
        \dvars^{\phi}$, the repetition history $H_{\symbval}(x,i)$
        should match the past obligation $\pobli_{\symbval}(x,i) =
        \set{(y,\psi) \in \dvars^{\phi} \times \Phi \mid
        \nxt^{i}(\oblieqp{x}{y}{\psi}) \in \Omega_{\symbval}}$.
    \item[(F7)] For every $\nxt^{-j}q \in \closure(\Phi)$
        and every $i \in [0,e]$ with $i-j \ge 0$, we have
        $\nxt^{-j}q \in
        \Phi_{\symbval}(i)$ iff $\nxt^{i-j}q \in \Omega_{\symbval}$.
    \item[(F8)] For every $\nxt^{-j}(\oblieqp{x}{y}{\psi}) \in \closure(\Phi)$
        and every $i \in [0,e]$ with $i -j \ge 0$, we have
        $\nxt^{-j}(\oblieqp{x}{y}{\psi}) \in
        \Phi_{\symbval}(i)$ iff $\nxt^{i-j}(\oblieqp{x}{y}{\psi}) \in \Omega_{\symbval}$.
    \item[(F9)] For every $\nxt^{-j}(x \oblieqlocal \nxt^{-j'} y) \in
        \closure(\Phi)$ and for every $i \in [0,e]$ with
        $i-j - j' \ge 0$, we have $\nxt^{-j}(x \oblieqlocal \nxt^{-j'}
        y) \in
        \Phi_{\symbval}(i)$ iff $\nxt^{i-j} x \oblieqlocal
        \nxt^{i-j-j'} y
        \in \Omega_{\symbval}$.
    \item[(F10)] For every $i \in [0,e]$, $\Phi_{\symbval}(i)$ is Boolean
        consistent and $\Phi_{\symbval}(i), \Phi_{\symbval}(i+1)$ are
        one step consistent whenever $i < e$.
\end{enumerate}
Intuitively, an $(e,\phi)$-frame captures information about
$(e+1)$ consecutive positions of a model of $\phi$. The set
$\Omega_{\symbval}$ contains all the atomic constraints satisfied at a
position. \review{The function $\Phi_{\symbval}$ is the one which specifies, for every
position of the frame, the formulas in $\Phi$ that are satisfied at
that position.} The set $\Phi_{\symbval}(i)$ contains all the formulas in
$\Phi$ that are satisfied at the $i$\textsuperscript{th} position
under consideration. \review{The function $H_{\symbval}$ is the one
which specifies, for
every variable $x$ in $\dvars^{\phi}$ and every position $i$ of the frame, a
repetition history.} The repetition history
$H_{\symbval}(x,i)$ is one that should be used to satisfy the past
obligation arising from constraints of the form
$\nxt^{i}(\oblieqp{x}{y}{\psi})$ contained in $\Omega_{\symbval}$.
\review{The
condition (F5) above ensures consistency among repetition histories
assigned to different variables at different positions. If
$\Omega_{\symbval}$ contains the formula $\nxt^{i}x\oblieqlocal
\nxt^{i}y$, it means the same data value will be assigned to $x$ and
$y$ at position $i$. Hence, the same repetition histories should be
used for these two variables. If $\Omega_{\symbval}$ contains the
formula $\nxt^{i}x \oblieqlocal \nxt^{j}y$ and $i< j$, then the data
value assigned to $x$ at position $i$ should be taken into account in the repetition
history assigned to $y$ at position $j$, which is ensured by the
second point of condition (F5).} The conditions (F7)--(F9) above ensure that the 
constraints contained in $\Phi_{\symbval}(i)$ are consistent with the
atomic constraints in $\Omega_{\symbval}$. \review{Suppose
$\Phi_{\symbval}(i)$ contains the formula $\nxt^{-j}q$. It means
that at the position $i$ steps to the right from the current one, the
formula $\nxt^{-j}q$ should be satisfied. In turn, this means that
$\nxt^{i-j}q$ should be contained in $\Omega_{\symbval}$, since the set
$\Omega_{\symbval}$ should contain all the atomic constraints
satisfied at the current position. This is what  condition (F7)
above ensures. The conditions (F8) and (F9) ensure similar consistency
for other types of atomic constraints.}

Next we extend the definition of one-step consistency of frames, to
include the extra information about nested formulas. A pair of
$(l,\phi)$-frames $(\symbval, \symbval')$ is said to be one-step
consistent iff the following conditions are satisfied.
\begin{enumerate}[align=left]
  \item[(O1)] For all $\nxt^{i}x \oblieqlocal \nxt^{j}y \in
    \Omega_{l}^{\phi}$ with $i,j>0$, we have $\nxt^{i}x \oblieqlocal
    \nxt^{j}y \in \Omega_{\symbval}$ iff $\nxt^{i-1}x \oblieqlocal
    \nxt^{j-1}y \in \Omega_{\symbval'}$,
  \item[(O2)] For all $\nxt^{i}(\oblieqp{x}{y}{\psi}) \in
    \Omega_{l}^{\phi}$ with $i > 0$, we have $\nxt^{i}(\oblieqp{x}{y}{\psi})
    \in \Omega_{\symbval}$ iff $\nxt^{i-1}(\oblieqp{x}{y}{\psi}) \in
    \Omega_{\symbval'}$,
  \item[(O3)] For all $\nxt^{i} q \in \Omega_{l}^{\phi}$ with $i > 0$, we
    have $\nxt^{i}q \in \Omega_{\symbval}$ iff $\nxt^{i-1}q \in
    \Omega_{\symbval'}$,
\item[(O4)] For all $x \in \dvars^{\phi}$ and $i \in [1,l]$,
        $H_{\symbval}(x,i) = H_{\symbval'}(x,i-1)$.
\item[(O5)] For every $i \in [1,l]$, $\Phi_{\symbval}(i) =
    \Phi_{\symbval'}(i-1)$.
\item[(O6)] The sets $\Phi_{\symbval}(0),\Phi_{\symbval'}(0)$
    are one-step consistent, as well as the sets
    $\Phi_{\symbval}(l),\Phi_{\symbval'}(l)$.
\end{enumerate}

\noindent
For $e \in [0,l-1]$, an $(e,\phi)$-frame $\symbval$ and an
$(e+1,\phi)$-frame $\symbval'$, the pair $(\symbval,\symbval')$ is
one-step consistent iff the following conditions are satisfied.
\begin{enumerate}
    \item $\Omega_{\symbval} \subseteq \Omega_{\symbval'}$ and for
        every constraint in $\Omega_{\symbval'}$ of the form
        $\nxt^{i}x\oblieqlocal \nxt^{j}y$, $\nxt^{i}q$ or
        $\nxt^{i}(\oblieqp{x}{y}{\psi})$ with $i,j \in [0,e]$, the
        same constraint also belong to $\Omega_{\symbval}$.
    \item For every $i \in [0,e]$ and every $x \in \dvars^{\phi}$,
        $\Phi_{\symbval}(i) = \Phi_{\symbval'}(i)$ and
        $H_{\symbval}(x,i) = H_{\symbval'}(x,i)$.
    \item The sets $\Phi_{\symbval}(e),\Phi_{\symbval'}(e+1)$ are one-step
        consistent.
\end{enumerate}
For $e \in [0,l]$, an $(e,\phi)$-frame $\symbval$ is \emph{initially consistent} if the
set $\Phi_{\symbval}(0)$ is initially consistent.

An (infinite) $(l, \phi)$-symbolic model $\rho$ is an infinite sequence of
$(l, \phi)$-frames such that for all $i \in
\PNat$, the pair $(\rho(i),
\rho(i+1))$ is one-step consistent and the first frame $\rho(1)$ is
initially consistent. Let us extend the definition of the symbolic
satisfaction relation $\rho,i \symbmodels \phi'$ where $\phi'$ is a
sub-formula of $\phi$. The relation $\symbmodels$ is defined in the
same way as $\models$ for $\pmainlogic$, except that for every
element $\phi'$ of $\Omega_{l}^{\phi}$, we have $\rho,i \symbmodels
\phi'$ whenever $\phi' \in \Omega_{\rho(i)}$. We say that a concrete model $\sigma$
realizes a symbolic model $\rho$ if for every $i \in \PNat$,
$\Omega_{\rho(i)} = \set{\phi' \in \Omega_{l}^{\phi} \mid \sigma, i
\models \phi'}$. The second part of the following lemma is not used in
the rest of the paper. The conditions (F7) to (F9) in the definition
of frames ensure that the information contained in
$\Phi_{\rho(i)}$ can be obtained from $\Omega_{\rho(i)}$ itself. We
have still included the second part to give some intuition about the
role of $\Phi_{\rho(i)}$ --- the sequence of sets of sub-formulas given
by ${(\Phi_{\rho(i)})}_{i \in \PNat}$ forms a
deterministic automaton that tells us which nested formulas are true
in which positions. This is a convenience compared to refering to
$\Omega_{\rho(i)}$ --- if we want the nested formula $\psi$,
$\Omega_{\rho(i)}$ may contain $\mynext^{l}\psi$.
\begin{lem}[symbolic vs.~concrete models]%
    \label{lem:symbolicToConcreteSatNested}
    Suppose $\phi$ is a $\pmainlogic[\langle \previous, \since\rangle,
    \oblieqlocal, \leftarrow]$ formula of $\mynext$-length $l$, $\rho$
    is a $(l,\phi)$-symbolic model and $\sigma$ is a concrete model
    realizing $\rho$. Then the following are true.
    \begin{enumerate}
        \item $\rho$ symbolically satisfies $\phi$ iff $\sigma$
            satisfies $\phi$.
        \item For every formula $\psi \in \closure(\Phi)$ and every $i \in
            \PNat$, $\sigma,i \models \psi$ iff
            $\psi \in \Phi_{\rho(i)}(0)$.
    \end{enumerate}
\end{lem}
\begin{proof}
    For proving (1), we prove by induction on structure that for every
    position $i$ and for every sub-formula $\phi'$ of $\phi$,
    $\rho,i \symbmodels \phi'$ iff $\sigma,i \models \phi'$. The base
    cases of this induction on structure comprise of $\phi'$ being an
    atomic constraint in $\Omega_{l}^{\phi}$. The result follows from
    the definition of the concrete model $\sigma$ realizing the
    symbolic model $\rho$. The induction steps follow directly since
    in these cases, symbolic satisfaction coincides with the semantics of
    \pmainlogic{} by definition.

    We prove (2) by induction on the lexicographic order of the pair
    $(i,\psi)$ where the structural order is used on $\psi$. In the
    base case, $i=1$ and $\psi$ is of the form either $q$ or
    $\oblieqp{x}{y}{\psi'}$ or $x \oblieqlocal \nxt^{-j}y$. If
    $\psi$ is of the form $q$, then from condition (F7) we have that
    $q \in \Phi_{\rho(1)}(0)$ iff $q \in \Omega_{\rho(1)}$. The result
    then follows from the proof of part (1). If $\psi$ is of the form
    $\oblieqp{x}{y}{\psi'}$, then we conclude from the semantics that
    $\sigma,1 \not\models \oblieqp{x}{y}{\psi'}$ and from part (1) and
    condition (F8) that $\oblieqp{x}{y}{\psi'} \notin
    \Phi_{\rho(1)}(0)$. If $\psi$ is of the form $x \oblieqlocal
    \nxt^{-j}y$ with $j \ge 1$, then we conclude from the semantics that
    $\sigma,1 \not\models x \oblieqlocal \nxt^{-j}y$ and from part (1) and
    condition (F9) that $x \oblieqlocal \nxt^{-j}y \notin
    \Phi_{\rho(1)}(0)$. If $\psi$ is of the form $x \oblieqlocal y$,
    then we have $\sigma, 1 \models x \oblieqlocal y$ iff $x
    \oblieqlocal y \in \Omega_{\rho(1)}$ iff $x \oblieqlocal y \in
    \Phi_{\rho(1)}(0)$ (the first equality follows from part (1) and
    the second one follows from condition (F9)).

    For the induction step, either $i=1$ and $\psi$ is of the form
    $\psi_{1} \land \psi_{2}$, $\lnot \psi_{1}$, $\nxt^{-1}
    \psi_{1}$ or $\psi_{1} \since \psi_{2}$ or $i > 1$. Suppose
    $i=1$ and $\psi$ is of the form $\psi_{1} \land \psi_{2}$ or
    $\lnot \psi_{1}$. The result follows from induction hypothesis and
    Boolean consistency of $\Phi_{\rho(1)}(0)$. If $i=1$ and
    $\psi$ is of the form $\nxt^{-1}\psi_{1}$, we have from semantics
    that $\sigma, 1 \not\models \nxt^{-1}\psi_{1}$ and from initial
    consistency of $\Phi_{\rho(1)}(0)$, we have that
    $\nxt^{-1}\psi_{1} \notin \Phi_{\rho(1)}(0)$. If $i=1$ and
    $\psi$ is of the form $\psi_{1} \since \psi_{2}$, we have from
    semantics that $\sigma,1 \models \psi_{1} \since \psi_{2}$ iff
    $\sigma,1 \models \psi_{2}$ and from initial consistency of
    $\Phi_{\rho(1)}(0)$, we have that $\psi_{1} \since \psi_{2} \in
    \Phi_{\rho(1)}(0)$ iff $\psi_{2} \in \Phi_{\rho(1)}(0)$. The
    result then follows from induction hypothesis.

    Finally for the induction step when $i > 1$, we do an induction on
    structure of $\psi$. If $\psi$ is of the form $q$, then from
    condition (F7) we have that $q \in \Phi_{\rho(i)}(0)$ iff $q \in
    \Omega_{\rho(i)}$. The result then follows from the proof of
    part (1). If $\psi$ is of the form $\oblieqp{x}{y}{\psi'}$, we
    have from condition (F8) that $\oblieqp{x}{y}{\psi'} \in
    \Phi_{\rho(i)}(0)$ iff $\oblieqp{x}{y}{\psi'} \in
    \Omega_{\rho(i)}$. The result then follows from the proof of
    part (1). Suppose $\psi$ is of the form $x \oblieqlocal
    \nxt^{-j}y$. We have from semantics that $\sigma, i \models x \oblieqlocal
    \nxt^{-j}y$ iff $i \ge j$ and $\sigma(i)(x) \oblieqlocal \sigma(i-j)(y)$.
    We infer from proof of part (1) that $\sigma(i)(x) \oblieqlocal \sigma(i-j)(y)$
    iff $\nxt^{j}x \oblieqlocal \nxt^{0}y \in
    \Omega_{\rho(i-j)}$. We infer from condition (F9) that $\nxt^{j}x \oblieqlocal \nxt^{0}y \in
    \Omega_{\rho(i-j)}$ iff $x \oblieqlocal \nxt^{-j}y \in
    \Phi_{\rho(i-j)}(j)$. By applying the condition (O5) $j$ times, we
    infer that $x \oblieqlocal \nxt^{-j}y \in
    \Phi_{\rho(i-j)}(j)$ iff $x \oblieqlocal \nxt^{-j}y \in
    \Phi_{\rho(i)}(0)$. Hence $\sigma, i \models x \oblieqlocal
    \nxt^{-j}y$ iff $x \oblieqlocal \nxt^{-j}y \in
    \Phi_{\rho(i)}(0)$. If $\psi$ is of the form $\psi_{1} \land
    \psi_{2}$ or $\lnot \psi_{1}$, the result follows from induction
    hypothesis and Boolean consistency of $\Phi_{\rho(i)}(0)$. If
    $\psi$ is of the form $\nxt^{-1} \psi_{1}$, we have from condition
    (O6) that $\nxt^{-1}\psi_{1} \in \Phi_{\rho(i)}(0)$ iff
    $\psi_{1} \in \Phi_{\rho(i-1)}(0)$. The result then follows by
    induction hypothesis and the semantics of $\nxt^{-1}\psi_{1}$. If
    $\psi$ is of the form $\psi_{1} \since \psi_{2}$, we have from
    condition (O6) that $\psi_{1} \since \psi_{2} \in
    \Phi_{\rho(i)}(0)$ iff either $\psi_{2} \in \Phi_{\rho(i)}(0)$ or
    $\psi_{1} \in \Phi_{\rho(i)}(0)$ and $\psi_{1} \since
    \psi_{2} \in \Phi_{\rho(i-1)}(0)$. The result then follows by
    induction hypothesis and the semantics of $\psi_{1} \since
    \psi_{2}$.
\end{proof}

Similar to Section~\ref{single-past-decidable}, we say that there is a forward
(resp.~backward) reference from $(x,i)$ in $\symbval$ if $\mynext^{i}
x \oblieqlocal \mynext^{i+j}y \in \Omega_{\symbval}$ (resp.~$\mynext^{i} x
\oblieqlocal \mynext^{i-j}y \in \Omega_{\symbval}$) for some $j > 0$ and $y \in
\dvars^{\phi}$. Now we extend the definitions of points of increments
and decrements to take into account extra information about nested
formulas.
\begin{itemize}
    \item In a $(l,\phi)$-frame $\symbval$, if there are no forward
        references from $(x,0)$, then ${[(x,0)]}_{\symbval}$ is a point
        of increment for the repetition history
        $H_{\symbval}(x,0) \cup ({[(x,0)]}_{\symbval},
        \Phi_{\symbval}(0)\cap \Phi)$.
    \item In an $(e,\phi)$-frame $\symbval$ for some $e \in [0,l]$, if
        there is no backward reference from $(x,e)$, then
        ${[(x,e)]}_{\symbval}$ is a point of decrement for the
        repetition history $H_{\symbval}(x,e)$.
\end{itemize}
We denote by $\inc(\symbval)$ the vector indexed by non-empty
repetition histories, where each coordinate contains the number of
points of increments in $\symbval$ for the corresponding repetition
history. Similarly we have the vector $\dec(\symbval)$ for points of
decrement.

Given a \pmainlogic[$\langle \previous,
\since\rangle$, $\oblieqlocal$, $\leftarrow$] formula
$\phi$ in which \review{$\dvars^{e}\cap \dvars^{\phi} = \emptyset = \bvars^{s}
\cap \bvars^{\phi}$}, we construct a
single-sided VASS game as follows. Let $l$ be the $\mynext$-length of
$\phi$ and $\symbvals$ be the set of all $(e, \phi)$-frames for all $e \in
[0,l]$. Let $A^{\phi}$ be a deterministic parity automaton that accepts
a symbolic model iff it symbolically satisfies $\phi$, with set of
states $Q^{\phi}$ and initial state $q^{\phi}_{\init}$. The
single-sided VASS game will have one counter corresponding to every
non-empty repetition history in $\RH$,
set of environment states $[-1,l] \times Q^{\phi}
 \times (\symbvals \cup \set{\bot})$ and set of system states
 $[-1,l] \times Q^{\phi} \times (\symbvals
\cup \set{\bot}) \times \pwrset(\bvars^{\phi})$. Every state will inherit the
colour of its $Q^{\phi}$ component. For convenience, we let
$\bot$ to be the only $(-1, \phi)$-frame and $(\bot, \symbval')$ be one-step
consistent for every initially consistent $0$-frame $\symbval'$. The initial state is
$(-1, q^{\phi}_{\init}, \bot)$, the initial counter values are all $0$
and the transitions are as follows ($\lceil \cdot
\rceil l$ denotes the mapping that is identity on $[-1,l-1]$ and maps
all others to $l$).
\begin{itemize}
    \item $(e,q,\symbval) \act{\vec{0}} (e,q, \symbval, V)$ for every
        $e \in \set{-1, 0, \ldots, l}$, $q \in Q^{\phi}$, $\symbval
        \in \symbvals \cup \set{\bot}$ and $V \subseteq
        \bvars^{\phi}$.
    \item $(e, q^{\phi}_{\init}, \symbval, V)
        \act{\inc(\symbval)-\dec(\symbval')} (e+1,q^{\phi}_{\init},
        \symbval')$ for every $V \subseteq \bvars^{\phi}$, $e \in
        \set{-1, 0, \ldots, l-2}$, $(e, \phi)$-frame $\symbval$ and
        $(e+1, \phi)$-frame $\symbval'$, where the pair $(\symbval,
        \symbval')$ is one-step consistent and $\set{p \in
            \bvars^{\phi} \mid \mynext^{e+1}p \in \Omega_{\symbval'}} = V$.
    \item $(e, q, \symbval, V) \act{\inc(\symbval) - \dec(\symbval')}
        (\lceil e+1 \rceil l, q', \symbval')$ for every $e \in
        \set{l-1,l}$, $(e, \phi)$-frame $\symbval$, $V \subseteq
        \bvars^{\phi}$, $q, q' \in Q^{\phi}$ and $(\lceil e+1 \rceil
        l, \phi)$-frame $\symbval'$, where the pair $(\symbval,
        \symbval')$ is one-step consistent, $\set{p \in \bvars^{\phi}
        \mid \mynext^{\lceil e+1 \rceil l}p \in \Omega_{\symbval'}} = V$ and $q
        \act{\symbval'} q'$ is a transition in ${A^{\phi}}$.
\end{itemize}

\noindent
Transitions of the form $(e,q,\symbval) \act{\vec{0}} (e,q, \symbval, V)$ let
the environment choose any subset $V$ of $\bvars^{\phi}$ to be true in the
next round. In transitions of the form $(e, q,
\symbval, V) \act{\inc(\symbval) - \dec(\symbval')} (\lceil e+1 \rceil
l, q', \symbval')$, the condition $\set{p \in \bvars^{\phi} \mid \mynext^{\lceil e+1
\rceil l}p \in \Omega_{\symbval'}} = V$  ensures that the frame $\symbval'$ chosen by the
system is compatible with the subset $V$ of $\bvars^{\phi}$ chosen by the
environment in the preceding step. By insisting that the pair
$(\symbval,\symbval')$ is one-step consistent, we ensure that the
sequence of frames built during a game is a symbolic model. The fact
that $(\bot,\symbval')$ is one-step consistent only when $\symbval'$
is an initially consistent $(0,\phi)$-frame ensures that the first
frame in the sequence of frames built during a game is initially
consistent. The condition $q \act{\symbval'} q'$ ensures that the
symbolic model is accepted by $A^{\phi}$ and hence symbolically
satisfies $\phi$. The update vector $\inc(\symbval) - \dec(\symbval')$
ensures that symbolic models are realizable, as explained in the proof
of the following result.

\begin{thm}[repeating values to VASS]%
    \label{thm:repValuesToVASSNested}
    Let $\phi$ be a \pmainlogic$[\langle \previous, \since\rangle,
    \oblieqlocal, \leftarrow]$ formula with $\dvars^{e} \cap
    \dvars^{\phi}= \bvars^{s} \cap \bvars^{\phi}=
    \emptyset$.  Then \sys has a winning strategy in the corresponding
    single-sided \pmainlogic$[\langle \previous, \since\rangle,
    \oblieqlocal, \leftarrow]$ game iff she has a winning strategy in
    the single-sided VASS game constructed above.
\end{thm}
\begin{proof}
    First we prove the forward direction.
    Suppose that \sys has a strategy $\ts: \vals^{*}\cdot
    \vals^{e} \to \vals^{s}$ in the single-sided \pmainlogic$[\langle \previous, \since\rangle,
    \oblieqlocal, \leftarrow]$ game. We will show that \sys
    has a strategy $\ssys: {(Q \times \Nat^{C})}^{*} \cdot (Q^{s}
    \times \Nat^{C}) \to T$ in the single-sided VASS game. It is
routine to construct such a strategy from the mapping $\mu:
{(\pwrset(\bvars^{\phi}))}^* \to \symbvals \cup \set{\bot}$ that we
define now. For every sequence $\chi \in {(\pwrset(\bvars^{\phi}))}^*$,
we will define $\mu(\chi)$ and a concrete
model of length $|\chi|$, by induction on $|\chi|$. For the base case
$|\chi|=0$, the concrete model is the empty sequence and
$\mu(\chi)$ is $\bot$.

    For the induction step, suppose $\chi$ is of
    the form $\chi' \cdot V$ and $\sigma$ is the concrete
    model defined for $\chi'$ by induction hypothesis. Let $\upsilon^{e}:\bvars^{e} \to
    \set{\top, \bot}$ be the mapping defined as $\upsilon^{e}(p) =
    \top$ iff $p \in V$. The \sys's strategy $\ts$ in the
    single-sided \pmainlogic$[\top$, $\oblieqlocal$, $\leftarrow]$
    game will give a valuation $\ts(\sigma \cdot \upsilon^{e}) = \upsilon^{s}:
    \dvars^{s} \to \Domain$. We define the finite concrete model
    to be $\sigma \cdot (\upsilon^{e} \oplus \upsilon^{s})$ and
    $\mu(\chi)$ to be the $(\lceil |\sigma| \rceil l, \phi)$-frame $\symbval'$ such that
    $\Omega_{\symbval'} = \set{\phi' \in \Omega^{\phi}_{l} \mid \sigma
    \cdot (\upsilon^{e} \oplus \upsilon^{s}), |\sigma|+1-\lceil
    |\sigma| \rceil l \models \phi'}$. Suppose
    $\mu(\chi')=\symbval$. Then we define
    $H_{\symbval'}(x,e-1)=H_{\symbval}(x,e)$ and
    $\Phi_{\symbval'}(e-1)=\Phi_{\symbval}(e)$ for every $e \in
    [1,\lceil |\sigma| \rceil l]$ and every $x \in \dvars^{\phi}$. We
    define $\Phi_{\symbval'}(\lceil |\sigma| \rceil l)$ to be the set
    $\set{\psi \in \closure(\Phi) \mid \sigma
    \cdot (\upsilon^{e} \oplus \upsilon^{s}), |\sigma|+1 \models
    \psi}$. Let $d \in \Domain$ and $\mathit{pos}(\sigma,d)$ be the set
    $\set{i \in [1,|\sigma|] \mid \sigma(i)(x)=d \text{ for some } x \in
    \dvars^{\phi}}$ of positions of $\sigma$ in which at least one
    variable is assigned to $d$. For every $x \in \dvars^{\phi}$, we
    define $H_{\symbval'}(x,\lceil |\sigma| \rceil l)$ to be the
    repetition history $\cup_{i \in
        \mathit{pos}(\sigma,\upsilon^{s}(x))}\set{(  \set{y \in
            \dvars^{\phi} \mid \sigma(i)(y)=\upsilon^{s}(x)},
            \set{ \psi \in \Phi \mid \sigma,i \models \psi} )}$.

    It is routine to verify that the frame $\symbval'$ defined above
    indeed satisfies all the conditions (F1)--(F10). Intuitively, 
    $\Omega_{\symbval'}$ is the set of atomic constraints in
    $\Omega_{l}^{\phi}$ that are satisfied at the position $|\sigma|+1-\lceil
    |\sigma| \rceil l$ of the concrete model $\sigma
    \cdot (\upsilon^{e} \oplus \upsilon^{s})$. The definition of
    $H_{\symbval'}(x,e-1)$ and $\Phi_{\symbval'}(e-1)$ are borrowed
    from the previous frame. We have defined $\Phi_{\symbval'}(\lceil
    |\sigma| \rceil l)$ to be the set of all formulas in
    $\closure(\Phi)$ that are true in the last position of the
    concrete model $\sigma \cdot (\upsilon^{e} \oplus \upsilon^{s})$.
    The repetition history $H_{\symbval'}(x,\lceil |\sigma| \rceil l)$
    is obtained by looking at all the positions in $\sigma$ that
    assigns at least one variable to the data value
    $\upsilon^{s}(x)$, which are all the positions where the data
    value of $x$ at position $|\sigma|+1$ repeats in the past. This
    step crucially uses the fact that nested formulas do not refer to
    future positions --- if they did, we couldn't have constructed the
    frame $\symbval'$ by looking only at the past positions of the
    concrete model.

    Next we will prove that the strategy $\ssys$ defined above is
    winning for \sys. Suppose \sys{} plays
    according to $\ssys$ in the single-sided VASS game, resulting in
    the sequence of states
    \begin{align*}
        (-1, q^{\phi}_{\init}, \bot) (-1,
        q^{\phi}_{\init}, \bot, V_{1}) (0, q^{\phi}_{\init},
        \symbval_{1}) (0, q^{\phi}_{\init}, \symbval_{1}, V_{2})\\
        (1,q^{\phi}_{\init}, \symbval_{2}) \cdots (l, q, \symbval_{l+1})
        (l,q,\symbval_{l+1}, V_{l+2}) (l,q',\symbval_{l+2}) \cdots
    \end{align*}
    The sequence $\symbval_{l+1} \symbval_{l+2} \cdots$ is an infinite
    $(l,\phi)$-symbolic model; call it $\rho$. It is clear from the construction
    that $\rho$ is realized by a concrete model $\sigma$, which is the
    result of \sys{} playing according to the winning
    strategy $\ts$ in the \pmainlogic$[\langle \previous, \since\rangle,
    \oblieqlocal, \leftarrow]$ game. So $\sigma,1 \models \phi$ and by
    Lemma~\ref{lem:symbolicToConcreteSatNested} (symbolic vs.~concrete
    models), $\rho$ symbolically satisfies $\phi$. By definition of
    $A^{\phi}$, the unique run of $A^{\phi}$ on $\rho$ satisfies the
    parity condition and hence the play satisfies the parity condition in the
    single-sided VASS game. It remains to prove that if a transition
    given by $\ssys$ decrements some counter, that counter will have
    sufficiently high value. Any play starts with all counters
    having zero and a counter is decremented by a transition if the
    frame chosen by that transition has points of decrement for the
    counter. For $e \in \set{1, \ldots, l+1}$ and
    $x \in \dvars^{\phi}$, ${[(x,e)]}_{\symbval_{e}}$ cannot be a point
    of decrement in $\symbval_{e}$ --- if it were, the data value
    $\sigma(e)(x)$ would have appeared in some position in
    $\set{1, \ldots, e-1}$, creating a backward reference from
    $(x,e)$ in $\symbval_{e}$.

    For $i > l+1$, $x \in \dvars^{\phi}$ and $H \in \RH$, suppose
    ${[(x,l)]}_{\symbval_{i}}$ is a point of decrement for $H$ in
    $\symbval_{i}$. Before decrementing the counter $H$, it is
    incremented for every point of increment for $H$ in every frame
    $\symbval_{j}$ for all $j < i$. Hence, it suffices to associate with
    this point of decrement a point of
    increment for $H$ in a frame earlier than $\symbval_{i}$ that is
    not associated to any other point of decrement. Since
    ${[(x,l)]}_{\symbval_{i}}$ is a point of decrement for $H$ in
    $\symbval_{i}$, the data value $\sigma(i)(x)$ appears
    in some of the positions $\set{1, \ldots, i-l-1}$. Let $i' =
    \max\set{j \in \set{1, \ldots, i-l-1} \mid \exists y \in
    \dvars^{\phi}, \sigma(j)(y) = \sigma(i)(x)}$. Let $x' \in
    \dvars^{\phi}$ be such that $\sigma(i')(x') = \sigma(i)(x)$ and associate with
    ${[(x,l)]}_{\symbval_{i}}$ the class ${[(x',0)]}_{\symbval_{i'+l}}$,
    which is a point of increment for $H$ in $\symbval_{i'+l}$. The
    class ${[(x',0)]}_{\symbval_{i'+l}}$ cannot be associated with any
    other point of decrement for $H$ --- suppose it were associated with
    ${[(y,l)]}_{\symbval_{j}}$, which is a point of decrement for
    $H$ in $\symbval_{j}$. Then $\sigma(j)(y) = \sigma(i)(x)$. If
    $j = i$, then ${[(x,l)]}_{\symbval_{i}} = {[(y,l)]}_{\symbval_{j}}$
    and the two points of decrement are the same. So $j < i$ or
    $j>i$. We compute $j'$ for ${[(y,l)]}_{\symbval_{j}}$ with
    $j' < j$ just like we computed $i'$ for ${[(x,l)]}_{\symbval_{i}}$.
    If $j < i$, then $j$ would be one of the positions in $\set{1,
    \ldots, i-l-1}$
    where the data value
    $\sigma(i)(x)$ appears ($j$ cannot be in the interval $[i-l, i-1]$
    since those positions do not contain the data value
    $\sigma(i)(x)$; if they did, there would have been a backward
    reference from $(x,l)$ in $\symbval_{i}$ and
    ${[(x,l)]}_{\symbval_{i}}$ would not have been a point of decrement), so
    $j \le i'$ (and hence $j' < i'$). If $j > i$, then $i$ is one of
    the positions in $\set{1, \ldots, j-l-1}$ where the data value
    $\sigma(j)(y)$ appears ($i$ cannot be in the interval $[j-l, j-1]$
    since those positions do not contain the data value
    $\sigma(j)(y)$; if they did, there would have been a backward
    reference from $(y,l)$ in $\symbval_{j}$ and
    ${[(y,l)]}_{\symbval_{j}}$ would not have been a point of decrement), so $i \le j'$ (and hence $i' < j'$). In
    both cases, $j' \ne i'$ and hence, the class
    ${[y',0]}_{\symbval_{j'+l}}$ we associate with
    ${[(y,l)]}_{\symbval_{j}}$ would be different from
    ${[(x',0)]}_{\symbval_{i'+l}}$.

    Next we prove the reverse direction.
    Suppose \sys{} has a strategy $\ssys: {(Q \times
    \Nat^{C})}^{*} \cdot (Q^{s} \times \Nat^{C}) \to T$ in the
    single-sided VASS game. We will show that \sys{} has a
    strategy $\ts: \vals^{*}\cdot \vals^{e} \to \vals^{s}$ in the
    single-sided \pmainlogic$[\langle \previous, \since\rangle, \oblieqlocal,
    \leftarrow]$
    game. For every $\sigma \in \vals^{*}$ and every $\val^{e} \in
    \vals^{e}$, we will define $\ts(\sigma \cdot \val^{e}):
    \dvars^{\phi} \to \Domain$ and a sequence of configurations
    $\chi \cdot ((e,q,\symbval),\vec{n}_{\inc} - \vec{n}_{\dec})$ in
    ${(Q \times \Nat^{C})}^{*}\cdot (Q^{e}\times\Nat^{C})$ of length
    $2|\sigma|+3$ such that for every repetition history $H \in
    \RH$, $\vec{n}_{\inc}(H)$ is the sum of the
    number of points of increment for $H$ in all the frames occurring
    in $\chi$ and $\vec{n}_{\dec}(H)$ is the sum of the number of points
    of decrement for $H$ in all the frames occurring in $\chi$ and in
    $\symbval$. We will do this by induction on
    $|\sigma|$ and prove that the resulting strategy is winning for
    \sys. By \emph{frames occurring in $\chi$}, we refer to
    frames $\symbval$ such that there are consecutive configurations
    $( (e, q, \symbval), \vec{n}) ( (e,q,\symbval,V), \vec{n})$ in
    $\chi$. By $\Pi_{\symbvals}(\chi)(i)$, we refer to
    $i$\textsuperscript{th} such occurrence of a frame in $\chi$. Let
    $\set{d_{0}, d_{1}, \ldots} \subseteq \Domain$ be a countably
    infinite set of data values.

    For the base case $|\sigma|=0$, let $V \subseteq \bvars^{e}$ be
    defined as $p \in V$ iff $\val^{e}(p)=\top$. Let $\ssys((
    (-1, q^{\phi}_{\init},\bot),\vec{0})\cdot((-1,
    q^{\phi}_{\init},\bot,V),\vec{0}))$ be the transition $(-1,
    q^{\phi}_{\init},\bot,V) \act{\vec{0}-\dec(\symbval_{1})} (0, q,
    \symbval_{1})$. Since $\ssys$ is a winning strategy for \sys in
    the single-sided VASS game, $\dec(\symbval_{1})$ is necessarily
    equal to $\vec{0}$. The
    set of variables $\dvars^{\phi}$ is partitioned into equivalence
    classes by the $(0,\phi)$-frame $\symbval_{1}$. We define
    $\ts(\val^{e})$ to be the valuation that assigns to each such
    equivalence class a data value $d_{j}$, where $j$ is the
    smallest number such that $d_{j}$ is not assigned to any variable
    yet. We let the sequence of configurations be $(
    (-1,q^{\phi}_{\init}, \bot),\vec{0}) \cdot
    ( (-1,q^{\phi}_{\init}, \bot, V), \vec{0}) \cdot (
    (0,q,\symbval_{1}), -\dec(\symbval_{1}))$.

    For the induction step, suppose $\sigma\cdot \val^{e} = \sigma'
    \cdot (\val^{e}_{1}\oplus \val^{s}_{1} ) \cdot \val^{e}$ and
    $\chi' \cdot ((e,q,\symbval), \vec{n})$ is the sequence of
    configurations given by the induction hypothesis for $\sigma'
    \cdot \val^{e}_{1}$. If $\set{\phi' \in \Omega^{\phi}_{l} \mid
    \sigma' \cdot (\val^{e}_{1} \oplus \val^{s}_{1}),|\sigma'|+1-e
    \models \phi'} \ne \Omega_{\symbval}$, it corresponds to the case where
    \sys in the \pmainlogic$[\langle \previous, \since\rangle, \oblieqlocal,
    \leftarrow]$ game has already deviated from the strategy we have
    defined so far. So in this case, we define $\ts(\sigma\cdot
    \val^{e})$ and the sequence of configurations to be arbitrary.
    Otherwise, we have $\set{\phi' \in \Omega^{\phi}_{l} \mid
    \sigma' \cdot (\val^{e}_{1} \oplus \val^{s}_{1}),|\sigma'|+1-e
    \models \phi'} = \Omega_{\symbval}$. Let $V \subseteq \bvars^{e}$ be
    defined as $p \in V$ iff $\val^{e}(p)=\top$ and let $\ssys( \chi' \cdot (
    (e,q,\symbval),\vec{n}) \cdot  ((e,q,\symbval,V),\vec{n}))$ be the
    transition $(e,q,\symbval,V) \act{\inc(\symbval) - \dec(\symbval')} (\lceil e+1 \rceil l, q',
    \symbval')$. We define the sequence of configurations as
    $\chi'\cdot ((e,q,\symbval), \vec{n}) \cdot ( (e,q,\symbval,V)
    \vec{n}) \cdot ((\lceil e+1 \rceil l, q',\symbval'),
    \vec{n} + \inc(\symbval) - \dec(\symbval'))$. Since $\ssys$ is a
    winning strategy for \sys in the single-sided VASS
    game, $ \vec{n} + \inc(\symbval) - \dec(\symbval') \ge
    \vec{0}$. The valuation $\ts(\sigma \cdot
    \val^{e}): \dvars^{\phi} \to \Domain$ is defined as follows. The
    set $\dvars^{\phi}$ is partitioned into
    equivalence classes at level $\lceil e+1 \rceil l$ in
    $\symbval'$. For every such equivalence class ${[(x,\lceil e+1
    \rceil l)]}_{\symbval'}$, assign the data value $d'$ as defined
    below.
    \begin{enumerate}
        \item If there is a backward reference $\mynext^{\lceil e+1
            \rceil l}x \oblieqlocal \mynext^{\lceil e+1 \rceil l -j}y$ in
            $\symbval'$, let $d' = \sigma'\cdot (\val^{e}_{1} \oplus
            \val^{s}_{1})(|\sigma'|+2-j)(y)$.
        \item If there are no backward references from $(x,\lceil e+1
            \rceil l)$ in $\symbval'$ and the set
            $\pobli_{\symbval}(x,\lceil e+1 \rceil l)$ of past
            obligations of $x$ at level $\lceil e+1 \rceil l$ in
            $\symbval'$ is empty, let $d'$ be $d_{j}$, where $j$ is
            the smallest number such that $d_{j}$ is not assigned to
            any variable yet.
        \item If there are no backward references from $(x,\lceil e+1
            \rceil l)$ in $\symbval'$ and the set
            $O=\pobli_{\symbval}(x,\lceil e+1 \rceil l)$ of past
            obligations of $x$ at level $\lceil e+1 \rceil l$ in
            $\symbval'$ is non-empty, then ${[(x,\lceil e+1
            \rceil l)]}_{\symbval'}$ is a point of decrement for the
            repetition history
            $H=H_{\symbval'}(x,\lceil e+1 \rceil l)$ in
            $\symbval'$. Pair off this with a point of increment for
            $H_{\symbval'}(x,\lceil e+1 \rceil l)$ in a frame that occurs in $\chi' \cdot (
            (e,q,\symbval),\vec{n}) \cdot  ((e,q,\symbval,V),\vec{n})$
            that has not been paired off before. It is possible to do
            this for every point of decrement for $H$ in $\symbval'$,
            since $(\vec{n} + \inc(\symbval))(H)$ is the number of
            points of increment for $H$  occurring in $\chi' \cdot (
            (e,q,\symbval),\vec{n}) \cdot  ((e,q,\symbval,V),\vec{n})$
            that have not yet been paired off and $(\vec{n} +
            \inc(\symbval))(H) \ge \dec(\symbval')(H)$ for every
            repetition history $H$. Suppose we
            pair off ${[(x,\lceil e+1 \rceil l)]}_{\symbval'}$ with a
            point of increment ${[(y,0)]}_{\symbval_{i}}$ in the frame
            $\symbval_{i} = \Pi_{\symbvals}(\chi' \cdot (
            (e,q,\symbval),\vec{n}) \cdot
            ((e,q,\symbval,V),\vec{n}))(i)$, then let $d'$ be
            $\sigma'\cdot (\val^{e}_{1} \oplus \val^{s}_{1})(i)(y)$.
            From the definition of the repetition history $\rh$ matching
            the set $\po$ of past obligations
            (page~\pageref{matching-second-condition}), we infer that for every
            variable $z$ such that the data value $d'$ is assigned to
            $z$ at some past position that satisfies some nested
            formula $\psi \in \Phi$, $\Omega_{\symbval'}$ contains the
            formula $\mynext^{\lceil e+1 \rceil
            l}(\oblieqp{x}{z}{\psi})$. This ensures that
            $\Omega_{\symbval'}$ contains \emph{all} the past
            repetitions of $d'$, which is required to ensure that the
            concrete model we build realizes the sequence of frames
            that we identify next.
    \end{enumerate}
    Suppose \sys plays according to the strategy
    $\ts$ defined above, resulting in the model
    $\sigma = (\val^{e}_{1} \oplus \val^{s}_{1}) \cdot (\val^{e}_{2} \oplus
    \val^{s}_{2}) \cdots$. It is clear from the construction that
    there is a sequence of configurations
    \begin{align*}
        ((-1, q^{\phi}_{\init}, \bot), \vec{0}) ((-1,
        q^{\phi}_{\init}, \bot, V_{1}), \vec{0})\\ ((0, q^{\phi}_{\init},
        \symbval_{1}), \vec{n}_{1}) ((0, q^{\phi}_{\init},
        \symbval_{1}, V_{2}), \vec{n}_{1})\\
        ((1,q^{\phi}_{\init}, \symbval_{2}), \vec{n}_{2}) \cdots ((l,
        q, \symbval_{l+1}), \vec{n}_{l+1})\\
        ((l,q,\symbval_{l+1}, V_{l+2}), \vec{n}_{l+1})
        ((l,q',\symbval_{l+2}), \vec{n}_{l+2}) \cdots
    \end{align*}
    that is the result of \sys playing according to the
    strategy $\ssys$ in the single-sided VASS game such that the
    concrete model $\sigma$ realizes the symbolic model
    $\symbval_{l+1} \symbval_{l+2} \cdots$. Since $\ssys$ is a winning
    strategy for \sys, the sequence of configurations
    above satisfy the parity condition of the single-sided VASS game,
    so $\symbval_{l+1} \symbval_{l+2} \cdots$ symbolically satisfies
    $\phi$. From Lemma~\ref{lem:symbolicToConcreteSatNested} (symbolic
    vs.~concrete models), we conclude that $\sigma$ satisfies
    $\phi$.
\end{proof}


\section{Single-sided \pmainlogic[\texorpdfstring{$\langle \future \rangle$, $\oblieqlocal$, $\leftarrow$}{(F),≈,←}]  is undecidable}%
\label{undec-singlesided-nestedpast}
In this section, we will show that if nested formulas can use the
$\future$ modality, then the winning strategy existence problem is
undecidable, even if future constraints are not allowed. We prove the
undecidability for a fragment of \pmainlogic[$\langle \future \rangle$,
$\oblieqlocal$, $\leftarrow]$ in which the scope of the $\future$
operator in the nested formulas has only Boolean variables.

We prove the undecidability in this section by a reduction from a
problem associated with lossy counter machines.
A lossy counter machine is a counter machine as defined in Section~\ref{prelims}, with
additional \emph{lossy transitions}. Many types of lossy transitions are
considered in~\cite{Mayr2003}, of which we recall here \emph{reset
lossiness} that is useful for us. A reset step $(q,m_{1}, \ldots,
m_{n}) \rightarrow (q,m_{1}',\ldots, m_{n}')$ is possible iff for all
$i$, either
\begin{enumerate}
    \item $m_{i}' = m_{i}$ or
    \item $m_{i}' = 0$ and there is an instruction
        $(q:\mathrm{If}~c_{i}=0~ \mathrm{then}~
        \mathrm{goto}~q'~\mathrm{else}~c_{i}:=c_{i}-1;~\mathrm{goto}~q'')$.
\end{enumerate}
Intuitively, if a counter is tested for zero, then it can suddenly
become zero.

\begin{thmC}[{\cite[Theorem 10]{Mayr2003}}]%
\label{thm:LCM_undec_prob}
Given a reset lossy counter machine with four counters and initial
state $q_{0}$, it is undecidable to check if there exists an
$n \in \Nat$ such that starting from the configuration $(q_{0},
0,0,0,n)$, there is an infinite run.
\end{thmC}
There are more powerful results shown in~\cite{Mayr2003} by
considering other forms of lossiness and imposing more restrictions on
the lossy machine. For our purposes, the above result is enough.

We give a reduction from the problem mentioned in
Theorem~\ref{thm:LCM_undec_prob} to the winning strategy existence
problem for single-sided \pmainlogic[$\langle \future \rangle$,
$\oblieqlocal$, $\leftarrow]$ games. Given a reset lossy counter
machine, we add a special state $q_{s}$, make it the initial state and
add the following instructions:
\begin{itemize}
    \item $(q_{s}: c_{4}:=c_{4}+1; \mathrm{goto}~q_{s})$
    \item $(q_{s}:\mathrm{If}~c_{4}=0~ \mathrm{then}~
        \mathrm{goto}~q_{0}~\mathrm{else}~c_{4}:=c_{4}-1;~\mathrm{goto}~q_{0})$
\end{itemize}
The above modification will let us reach the configuration
$(q_{0},0,0,0,n)$ for any $n$ and start the reset lossy machine from
there. One pitfall is that there is an infinite run that stays in
$q_{s}$ for ever and keeps incrementing $c_{4}$. We avoid this by
specifying that the state $q_{0}$ should be visited at some time.

For convenience, we let
\sys{} use Boolean variables and start the play instead of
\env{}. This doesn't result in loss of generality since the Boolean
variables can be encoded by data variables and the positions of the
two players can be interchanged as done in
Section~\ref{undec-plrvgame}. In our reduction, \sys{} will simulate
the reset lossy machine and \env{} will catch any errors during the
simulation. There will be one Boolean variable
$b$ for \env{} to declare cheating. For \sys{}, there will be four
data variables $x_{1}, \ldots, x_{4}$ to simulate the four counters.
For every \review{instruction} $t$ of the reset lossy machine, there will be a
Boolean variable $p_{t}$ owned by \sys{}. For the purposes of this
reduction, we treat the instruction $(q:\mathrm{If}~c_{i}=0~
\mathrm{then}~
\mathrm{goto}~q'~\mathrm{else}~c_{i}:=c_{i}-1;~\mathrm{goto}~q'')$ as
two \review{instructions} $t_{1}$ and $t_{2}$ with source state being $q$. The
target state for $t_{1}$ is $q'$, which tests $c_{i}$ for zero. The
target state for $t_{2}$ is $q''$, which decrements $c_{i}$ (assuming
that $c_{i}$ was not zero). For
a set $S$ of positions
along a run, we denote by $S \restr{x_{k}++}$, the set of positions
$j$ in $S$ such that the $j$\textsuperscript{th} transition along the run
increments the counter $c_{k}$. Suppose the first $i$ rounds of the
\pmainlogic[$\langle \future \rangle$, $\oblieqlocal$, $\leftarrow]$
game have been played and the resulting sequence of valuations is
$\sigma$. It encodes the value of counter $c_{k}$ as the cardinality
of the set $\set{d \in \Domain \mid \exists j \in
    [1,i]\restr{x_{k}++},
    \sigma(j)(x_{k})=d, \forall j' \in [j+1,i],
    \sigma(j')(x_{k}) \ne d \text{ and } j'\text{\textsuperscript{th}
transition doesn't test } c_{k} \text{ for zero} }$. Intuitively, the
counter value is equal to the number of distinct data values that
have appeared in incrementing positions and not repeated in the future
such that there is no zero test for that counter in the future. For
incrementing a counter $c_{k}$, it is enough to ensure that a new data
value is assigned to $x_{k}$. For zero testing a counter, nothing
special is needed since by definition, a counter that is tested for
zero can suddenly become zero. For decrementing a counter $c_{k}$, we
ensure that the data value assigned $c_{k}$ repeats in the past in an
incrementing position. Here
cheating can happen in two ways.  First, the same data value can be
used in multiple decrementing transitions. We avoid this by using
nested formulas to ensure that no data value appears thrice anywhere
in the model. Second, the data value used in a decrementing transition
may have a matching repetition in a past incrementing position, but
there might be a zero testing transition between that past position
and the current decrementing position. We avoid this by getting
\env{} to declare such a cheating and using nested formulas to check
that \env{} correctly declared a cheating.

The following two formulas represent two possible mistakes
\env{} can make, in which case \sys{} wins immediately.
\begin{itemize}
    \item \env{} declares cheating at a position that doesn't
        decrement any counter.
        \begin{align*}
            \phi_{1} = F( \lnot b \land \bigvee_{t: t \text{ doesn't
            decrement any counter}}p_{t})
        \end{align*}
    \item \env{} declares cheating at a decrementing position that was
        properly performed.
        \begin{align*}
            \phi_{2} = F(\lnot b \land \bigvee_{t: t \text{ decrements
            }c_{k}} p_{t} \land
            \oblieqp{x_{k}}{x_{k}}{\bigvee_{t'': t'' \text{ increments
            } c_{k}} p_{t''} \land \lnot F(
                \bigvee_{t': t'
                \text{tests } c_{k} \text{ for zero}}p_{t'})})
        \end{align*}
\end{itemize}
The second formulas above says that there is a position in the future
where $b$ is false (which is the position where \env{} declared a
cheating). At that position, \review{instruction} $t$ is fired, which decrements
$c_{k}$. Also at that position, the data value assigned to
$x_{k}$ repeats in the past in $x_{k}$ satisfying two nested formulas.
The First nested formula says that the past position fired some \review{instruction}
$t''$ that incremented $c_{k}$. The second nested formula says that
starting from that past position, no \review{instruction} ever tests
$c_{k}$ for zero.

If the environment doesn't make any of the above two mistakes, then
\sys{} has to satisfy all the following formulas in order to win.

\begin{itemize}
    \item Exactly one \review{instruction} must be fired at every position:
        \begin{align*}
            \phi_{3} = G(
            (\bigvee_{t}p_{t} \land \bigwedge_{t \ne t'}(\lnot
            p_{t} \lor \lnot p_{t'})))
        \end{align*}
    \item The first \review{instruction} must be from the initial state:
        \begin{align*}
            \phi_{4} = \bigvee_{t: t \text{ is from the initial
            state}}p_{t}
        \end{align*}
    \item Consecutive \review{instruction}s must be compatible:
        \begin{align*}
            \phi_{5} = G(
            \bigvee_{t' \text{ can fire after } t}(p_{t} \land \nxt
            p_{t'}))
        \end{align*}
    \item The state $q_{0}$ must be visited some time:
        \begin{align*}
            \phi_{6} = F(\bigvee_{t: \text{ target of } t \text{ is }
            q_{0}}p_{t})
        \end{align*}
    \item At every incrementing \review{instruction}, the data value must be
        new:
        \begin{align*}
            \phi_{7} = \bigwedge_{k \in [1,4], t \text{ increments } c_{k}}
            G(p_{t} \Rightarrow
            \lnot\oblieqp{x_{k}}{x_{k}}{\top})
        \end{align*}
    \item At every decrementing \review{instruction}, the data value must repeat
        in the past at a position with an incrementing \review{instruction}:
        \begin{align*}
            \phi_{8} = \bigwedge_{k\in [1,4], t \text{ decrements } c_{k}}
            G(p_{t} \Rightarrow
        \oblieqp{x_{k}}{x_{k}}{\bigvee_{t': t' \text{ increments }
        c_{k}}p_{t'}})
        \end{align*}
    \item No data value should appear thrice:
        \begin{align*}
            \phi_{9} = \lnot \bigvee_{k \in [1,4]}
            F(\oblieqp{x_{k}}{x_{k}}{\oblieqp{x_{k}}{x_{k}}{\top}})
        \end{align*}
    \item If \env{} sets $b$ to $\bot$ at a position that decrements
        $c_{i}$, then the data value in $x_{i}$ should repeat in the
        past at a position such that after that past position, there
        is no zero testing for $c_{i}$.
        \begin{align*}
            \phi_{10} = \bigwedge_{k \in [1,4]} G(\lnot b ~\land ~ \bigvee_{t: t \text{ decrements
            }c_{k}}p_{t}~\Rightarrow~ \oblieqp{x_{k}}{x_{k}}{\lnot F
                \bigvee_{t': t' \text{ tests } c_{k} \text{ for
                zero}}p_{t'}})
        \end{align*}
    \item The run should be infinite:
        \begin{align*}
            \phi_{11} = G(\nxt \top)
        \end{align*}
\end{itemize}
For \sys{} to win, either \env{} should make a mistake setting
$\phi_{1}$ or $\phi_{2}$ to true or \sys{} should satisfy all the
formulas $\phi_{3}, \ldots, \phi_{11}$. The formula defining the
single-sided \pmainlogic[$\langle \future \rangle$, $\oblieqlocal$,
$\leftarrow]$ game is $\phi_{1} \lor \phi_{2} \lor
(\bigwedge_{i=3}^{11}\phi_{i})$.

\begin{lem}%
    \label{lem:undec-singlesided-nestedfuture}
    Given a reset lossy machine with four counters and initial state
    $q_{0}$, consider the single-sided \pmainlogic[$\langle \future
    \rangle$, $\oblieqlocal$, $\leftarrow]$ game with winning
    condition given by the formula written above. There exists an
    $n \in \Nat$ such that there is an infinite computation from
    $(q_{0},0,0,0,n)$ iff \sys{} has a winning strategy in the
    corresponding game of repeating values.
\end{lem}
\begin{proof}
    Suppose there exists an $n \in \Nat$ such that there is an
    infinite computation from $(q_{0},0,0,0,n)$. Then \sys{} will
    first simulate $(n+1)$ incrementing \review{instructions} at the special state
    $q_{s}$. Then \sys{} simulates the decrementing \review{instruction} from
    $q_{s}$ to $q_{0}$. From then onwards, \sys{} faithfully simulates
    the infinite run starting from $(q_{0},0,0,0,n)$. If at any stage,
    \env{} sets $b$ to $\bot$, either $\phi_{1}$ or $\phi_{2}$ becomes
    true and \sys{} wins immediately. Otherwise, \sys{} will
    faithfully simulate the infinite run for ever, all the formulas
    $\phi_{3}, \ldots, \phi_{11}$ are satisfied and \sys{} wins.

    Conversely, suppose \sys{} has a winning strategy in the
    corresponding game of repeating values. Consider the game in which
    \env{} doesn't set $b$ to $\bot$ if \sys{} has faithfully
    simulated the reset lossy machine so far. Thus, the formulas
    $\phi_{1}$ and $\phi_{2}$ will never become true. So for
    \sys{} to win, all the formulas $\phi_{3}, \ldots, \phi_{11}$
    should be true. To make $\phi_{6}$ true, $p_{t}$ must be set to
    true at some time for some \review{instruction} $t$ whose target is
    $q_{0}$. Let $n=0$ if this \review{instruction} $t$ tests $c_{4}$ for zero.
    Otherwise, let $n$ be one less than the number of times
    $p_{t'}$ is set to true, where $t'$ is the \review{instruction} $(q_{s}:
    c_{4}:=c_{4}+1; \mathrm{goto}~q_{s})$. We will now prove that
    there is an infinite computation starting from $(q_{0}, 0,0,0,n)$.
    Indeed, the only way to satisfy all the formulas $\phi_{3},
    \ldots, \phi_{11}$ is to simulate a run faithfully for ever, which
    proves the result.
\end{proof}


\section{Conclusion}%
\label{conclusion}
It remains open whether the \textsc{3Exptime} upper bound given in
Corollary~\ref{cor:upperBound} is optimal.
Another open question is the
decidability status of single-sided games with future obligations restricted to only two data variables;  the reduction we have in
Section~\ref{undec-singlesided-plrvgame} needs three.

Some future directions for research on this topic include finding
restrictions other than single-sidedness to get decidability. For the
decidable cases, the structure of winning strategies can be studied,
e.g., whether memory is needed and if yes, how much.


\subsection*{Acknowledgements}
The work
reported was carried out in the framework of ReLaX, UMI2000 (CNRS,
Univ. Bordeaux, ENS Paris-Saclay, CMI, IMSc).
The authors thank St\'{e}phane Demri and Prakash Saivasan for useful
discussions, and anonymous reviewers for their
constructive feedback, which helped us correct some omissions and
improve the presentation.

\bibliographystyle{plain}
\bibliography{references}

%




\end{document}